\documentclass[aps,pre,twocolumn,eqsecnum]{revtex4}

\usepackage{amsmath,bm,epsfig,here}

\def \ed {\end{document}}
\def\Fbox#1{\vskip1ex\hbox to 8.5cm{\hfil\fboxsep0.3cm\fbox{%
  \parbox{8.0cm}{#1}}\hfil}\vskip1ex\noindent}  %%  {TEXT} in BOX
\let \nn  \nonumber
\newcommand{\br}{\\ \nn}
\let\*\cdot
\def\<{\left\langle} \def\>{\right\rangle} \def\({\left(} \def\){\right)}
\let\p\partial \let\~\widetilde \let\^\widehat \def\ort#1{\^{\bf{#1}}}
 \def\x{\ort x} 
\def\z{\ort z}  \def\1{\bm1} 
 
\newcommand{\B}[1]{{\bm{#1}}}%% Bold Roman & Greek Lower & Upper Case
\newcommand{\C}[1]{{\mathcal{#1}}}    %%   Calligrapfic Upper case
\newcommand{\BC}[1]{\bm{\mathcal{#1}}}%% Bold Calligrapfic Upper case
\def\BE{\begin{equation}}\def\EE{\end{equation}}
\def\BEA{\begin{eqnarray}}\def\EEA{\end{eqnarray}}
\def\BSE{\begin{subequations}}\def\ESE{\end{subequations}}

\renewcommand{\sb}[1]{_{\text {#1}}}  %% sub-   for lower case
\newcommand{\Sp}[1]{^{^{\text {#1}}}} %% Super- for Upper case
  \def\Sb#1{_{\scriptscriptstyle\rm{#1}}}

\newcommand{\eq}[1]{(\ref{#1})}%%  requires \eq{label}
\newcommand{\Eq}[1]{Eq.~(\ref{#1})}%%  requires \eq{label}
\newcommand{\Eqs}[1]{Eqs.~(\ref{#1})}%%  requires \eq{label}
\newcommand{\Fig}[1]{Figure~\ref{#1}}%%  requires \Fef{label}
\renewcommand{\a}{\alpha}\renewcommand{\b}{\beta}\newcommand{\g}{\gamma}
\newcommand{\G} {\Gamma}\renewcommand{\d}{\delta}
\newcommand{\e}{\epsilon}\newcommand{\ve}{\varepsilon}

\renewcommand{\t}{\tau}
\def\r{\rho}\def\k{\kappa}
\def\t{\theta } 
\def\R{\bm R}

\def\R{\mathcal R}
\let\p\partial 
\def\x{\ort x}   \def\z{\ort z}
  \def\P{\mathcal P}

\def\ld {\ell^\ddag}

 \def \vK {von-K\'arm\'an~}

\let \= \equiv

 \let\*\cdot
 \def\sb#1{_{\rm{#1}}}

 \def\({\left(} \def\){\right)}
 \def \[ {\left [} \def \] {\right ]}
   \def\Sp#1{^{\scriptscriptstyle\rm{#1}}}
    \let\^\widehat
  \let\-\overline

  \def\Eq#1{Eq.~\eq{#1}}
   \def\Ref#1{(\ref{#1})}  \def\<{\left\langle}
   \def\>{\right\rangle}
   \def\ort#1{\^{\bf{#1}}}

   \let\~\widetilde

\def\Rig{\mbox{Ri}\sb{grad}}
\def\Rif{\mbox{Ri}\sb{flux}}
\def\ld {{\ell^\ddag}}  \def\lp {{\ell^+}}
\begin{document}

\title{ Energy Conservation and Second-Order Statistics \\ in Stably
Stratified Turbulent Boundary Layers }
\author{Victor S. L'vov, Itamar Procaccia and Oleksii Rudenko}
\email{oleksii@wisemail.weizmann.ac.il}%%
\affiliation{Dept. of Chemical Physics, The Weizmann Institute of
Science, Rehovot 76100, Israel}

\begin{abstract}
We address the dynamical and statistical description of stably
stratified turbulent boundary layers with the important example of
the atmospheric boundary layer with a stable temperature
stratification in mind. Traditional approaches to this problem,
based on the profiles of mean quantities, velocity second-order
correlations, and dimensional estimates of the turbulent thermal
flux run into a well known difficulty,  predicting the suppression
of turbulence at a small critical value of the Richardson number, in
contradiction with observations. Phenomenological attempts to
overcome this problem suffer from various theoretical
inconsistencies. Here we present an approach taking into full
account all the second-order statistics, which allows us to respect
the conservation of total mechanical energy. The analysis culminates
in an analytic solution of the profiles of all mean quantities and
all second-order correlations  removing the unphysical predictions
of previous theories. We propose that the approach taken here is
sufficient to describe the lower parts of the atmospheric boundary
layer, as long as the Richardson number does not exceed an order of
unity. For much higher Richardson numbers the physics may change
qualitatively, requiring careful consideration of the potential
Kelvin-Helmoholtz waves and their interaction with the vortical
turbulence.
\end{abstract}


\maketitle

%%%%%%%%%%% End of FrontMater %%%%%%%%%%%%%%%%%%%

%\begin{table*}
\begin{widetext}

{{%%
\textbf{Nomenclature}~\\ ~\\
\begin{tabular}{l c l c l c l}
  $\BC A$ & ~~ & Thermal flux production vector, (\ref{defOfA})& ~~~~ & %%
  $\beta$ & ~~ & Buoyancy parameter, $\B g \~\beta$ \\
  %%%%%%%%%%%%%%%%%%%%%%%%%%%%%%%%%%%%%%%%
  $\BC B$ & ~~ & Pressure-temperature-gradient-vector, (\ref{defs1f})& ~~~~ & %%
  $\~\beta$ & ~~ & Thermal expansion coefficient, ($\ref{beta}$) \\ %%
  %%%%%%%%%%%%%%%%%%%%%%%%%%%%%%%%%%%%%%%%
  $\C C_{ij}$ & ~~ & Energy conversion tensor, (\ref{defs1B})& ~~~~ & %%
  $\gamma\Sb{RI}$ & ~~ & Relaxation frequency of $\tau_{ij}$, $i = j$ \\ %%
  %%%%%%%%%%%%%%%%%%%%%%%%%%%%%%%%%%%%%%%%
  $\C D / \C D t$ & ~~ & Substantial derivative, $\partial / \partial t + \BC U \!\cdot\! \B \nabla$ & ~~~~ & %%
  $\~\gamma\Sb{RI}$ & ~~ & Relaxation frequency of $\tau_{ij}$, $i \neq j$ \\ %%
  %%%%%%%%%%%%%%%%%%%%%%%%%%%%%%%%%%%%%%%%
  $D / D t$ & ~~ & Mean substantial derivative, $\partial / \partial t + \B U \!\cdot\! \B \nabla$ & ~~~~ & %%
  $\gamma\Sb{RD}$ & ~~ & Relaxation frequency of $\B F$ \\ %%
  %%%%%%%%%%%%%%%%%%%%%%%%%%%%%%%%%%%%%%%%
  $E\Sb K$ & ~~ & Turbulent kinetic energy per unit mass, $\frac12 |\B u|^2$& ~~~~ & %%
  $\gamma_{uu}$ & ~~ & Relaxation frequency of $E\Sb K$ \\ %%
  %%%%%%%%%%%%%%%%%%%%%%%%%%%%%%%%%%%%%%%%
  $E\Sb \Theta$ & ~~ & "Temperature energy" per unit mass, $\frac12 \theta^2$ & ~~~~ & %%
  $\gamma_{\theta\theta}$ & ~~ & Relaxation frequency of $E\Sb \Theta$ \\ %%
  %%%%%%%%%%%%%%%%%%%%%%%%%%%%%%%%%%%%%%%%
  $\B F$ & ~~ & Turbulent thermal flux per unit mass, $\<\B u \theta\>$ & ~~~~ & %%
  $\varepsilon_{ij}$ & ~~ & Dissipation tensor of $\tau_{ij}$, ($\ref{diss}$)\\ %%
  %%%%%%%%%%%%%%%%%%%%%%%%%%%%%%%%%%%%%%%%
  $ F_*$ & ~~ & Thermal flux at zero elevation $z = 0$ & ~~~~ & %%
  $\B \epsilon$ & ~~ & Dissipation vector of $\B F$, ($\ref{diss}$)\\ %%
  %%%%%%%%%%%%%%%%%%%%%%%%%%%%%%%%%%%%%%%%
  $\B g$ & ~~ & Gravity acceleration, $\B g = -g\, \z$  & ~~~~ & %%
  $\varepsilon$ & ~~ & Dissipation of $E\Sb \Theta$, ($\ref{diss}$)\\ %%
  %%%%%%%%%%%%%%%%%%%%%%%%%%%%%%%%%%%%%%%%
  $L$ & ~~ & Monin-Obukhov length, $u_*^3/\beta F_*$ & ~~~~ & %%
  $\Theta$ & ~~ & Total potential temperature \\ %%
  %%%%%%%%%%%%%%%%%%%%%%%%%%%%%%%%%%%%%%%%
  $\ell$ & ~~ & Outer scale of turbulence, external parameter & ~~~~ & %%
  $\Theta_d$ & ~~ & Deviation of $\Theta$ from BRS\\ %%
  %%%%%%%%%%%%%%%%%%%%%%%%%%%%%%%%%%%%%%%%
  $\C P_{ij}$ & ~~ & Rate of Reynolds stress production, ($\ref{defs1A}$) & ~~~~ & %%
  $\overline{\Theta}$ & ~~ & Mean potential temperature, $\<\Theta_d\>$ \\ %%
  %%%%%%%%%%%%%%%%%%%%%%%%%%%%%%%%%%%%%%%%
  $p$, $\~p$, $p_*$ & ~~ & Total, fluctuating and zero level pressures & ~~~~ & %%
  $\theta$ & ~~ & Fluctuating potential temperature \\ %%
  %%%%%%%%%%%%%%%%%%%%%%%%%%%%%%%%%%%%%%%%
  Pr$\Sb T$ & ~~ & Turbulent Prandtl number, $\nu\Sb T / \chi\Sb T$ & ~~~~ & %%
  $\theta_*$ & ~~ & Potential temperature \\ %%
  %%%%%%%%%%%%%%%%%%%%%%%%%%%%%%%%%%%%%%%%
  Ri$\sb{flux}$ & ~~ & Flux Richardson number, $\beta F_z / \tau_{xz} S\Sb U$ & ~~~~ & %%
    & ~~ & at zero elevation, $F_* / u_*$\\ %%
  %%%%%%%%%%%%%%%%%%%%%%%%%%%%%%%%%%%%%%%%
  Ri$\sb{grad}$ & ~~ & Gradient Richardson number, $\beta S\Sb \Theta / S^2\Sb U$ & ~~~~ & %%
  $\lambda_*$ & ~~ & Viscous lengthscale, $\nu / u_*$ \\ %%
  %%%%%%%%%%%%%%%%%%%%%%%%%%%%%%%%%%%%%%%%
  $\C R_{ij}$ & ~~ & Pressure-rate-of-strain-tensor, ($\ref{defs1C}$) & ~~~~ & %%
  $\nu$ & ~~ & Kinematic viscosity \\ %%
  %%%%%%%%%%%%%%%%%%%%%%%%%%%%%%%%%%%%%%%%
  $S\Sb U$ & ~~ & Mean velocity gradient, $dU/dz$ & ~~~~ & %%
  $\nu\Sb T$ & ~~ & Turbulent viscosity \\ %%
  %%%%%%%%%%%%%%%%%%%%%%%%%%%%%%%%%%%%%%%%
  $S\Sb \Theta$ & ~~ & Mean potential temperature gradient, $d\overline{\Theta}/dz$ & ~~~~ & %%
  $\rho$ & ~~ & density of the fluid \\ %%
  %%%%%%%%%%%%%%%%%%%%%%%%%%%%%%%%%%%%%%%%
  $T$ & ~~ & Molecular temperature & ~~~~ & %%
  $\tau_{ij}$ & ~~ & Reynolds stress tensor, $\<u_i u_j\>$ \\ %%
  %%%%%%%%%%%%%%%%%%%%%%%%%%%%%%%%%%%%%%%%
  $\BC U$ & ~~ & Velocity field & ~~~~ & %%
  $\tau_*$ & ~~ & Mechanical momentum flux \\ %%
  %%%%%%%%%%%%%%%%%%%%%%%%%%%%%%%%%%%%%%%%
  $\B U$ & ~~ & Mean velocity, $\<\BC U\>$ & ~~~~ & %%
    & ~~ & at zero elevation (at the graund) \\ %%
  %%%%%%%%%%%%%%%%%%%%%%%%%%%%%%%%%%%%%%%%
  $\B u$ & ~~ & Fluctuating velocity, $\BC U - \B U$ & ~~~~ & %%
  $\chi$ & ~~ & dynamical thermal conductivity \\ %%
  %%%%%%%%%%%%%%%%%%%%%%%%%%%%%%%%%%%%%%%%
  $u_*$ & ~~ & (Wall) friction velocity, $\sqrt{\tau_*}$ & ~~~~ & %%
  $\chi\Sb T$ & ~~ & Turbulent thermal conductivity \\ %%
  %%%%%%%%%%%%%%%%%%%%%%%%%%%%%%%%%%%%%%%%
  $\x$, $\z$ & ~~ & stream-vise and vertical (wall-normal) directions & ~~~~ & %%
  BRS & ~~ & Basic Reference State \\
\end{tabular}
}}\\
%\end{table*}
~\\
\end{widetext}

\newpage

\section*{\label{s:intro}Introduction}

The lower levels of the atmosphere are usually strongly influenced
by the Earth's surface. Known as the atmospheric boundary layer,
this is the part of the atmosphere where the surface influences the
temperature, moisture, and velocity of the air above through the
turbulent transfer of mass.

The stability of the atmospheric boundary layer depends on the
profiles of the density and the temperature as a function of the
height above the ground. During normal summer days the land mass
warms up and the temperature is higher at lower elevations. If it
were not for the decrease in density of the air as a function of the
height, such a situation of heating from below would have been
always highly unstable. In fact, the boundary layer is considered
stable as long as the temperature decreases at the dry adiabatic
lapse rate ($T'\approx - 9.8^\circ C$ per kilometer) throughout most
of the boundary layer. With such a rate of cooling one balances out
the decrease in density. With a higher degree of cooling one refers
to the atmospheric boundary layer as unstably stratified, whereas
with a lower degree of cooling the situation is stably
stratified. Stably stratified boundary layer occurs typically
during clear, calm nights. In extreme cases  turbulence tends to
cease, and radiational cooling from the surface results in a
temperature that increases with height above the surface.

The tendency of the atmosphere to be turbulent does not depend only
on the rate of cooling but also on the mean shear in the vertical
direction. The commonly used parameter to describe the tendency of
the atmosphere to be turbulent is the ``gradient" Richardson number (Richardson, 1920),
defined as
\begin{equation}\label{Rig}
\Rig \= \frac{ \beta \, d\Theta (z)\big / d z}{ [d U_x/dz]^2} \,,
\end{equation}
where  $x$  is the stream-wise direction, $z$ is the height above
the ground, $\overline{\Theta}(z)$ is the  the ``mean potential temperature"
profile, (which differs from the mean temperature profile $T(z)$ by
accounting for the adiabatic cooling of the air during its
expansion: $ d\overline{\Theta} (z)\big / d z=  d T (z)\big / d z + |T'|$), $
\beta=\~\b g$ is the buoyancy parameter in which $\tilde \b$ is the
adiabatic thermal expansion coefficient, [defined below in
\Eq{beta};  for an ideal gas $\~\beta= 1/ T$], and $g$ is the
gravitational acceleration. The mean shear $d U_x /d z$ is defined
in terms of the mean velocity $\B U $, which in the simplest case of
flat geometry depends only on the vertical coordinate $z$.  The
parameter $\Rig$ represents the ratio of the generation or
suppression of turbulence by buoyant production of energy to the
mechanical generation of energy by wind shear.

In this paper we consider the description of stably stratified
turbulent  boundary layers (TBL), taking as an example the case of
stable thermal stratification. Since the 50's of twentieth century,
traditional models of stratified TBL generalize models of
unstratified TBL, based on the budget equations for the kinetic
energy and mechanical momentum; see reviews of Umlauf and Burchard (2005),
Weng and Taylor (2003). The main difficulty is that the budget
equations are not closed; they involve turbulent fluxes of
mechanical moments $\tau_{ij}$  (known as the ``Reynolds stress"
tensor) and a thermal flux $\B F$ (for the case of thermal
stratification): \begin{equation} \label{def3}%%
\tau_{ij} \= \langle u_i u_j \rangle\,,\quad \B F  \=   \<  \B u
\, \theta \> \,,
\end{equation} %%
where $\B u$ and $\theta$ stand for the turbulent
fluctuating velocity and the potential temperature with zero mean.
The nature of the averaging procedure behind the symbol
$\langle\cdots\rangle$ will be specified below.

Earlier  estimates of the fluxes~\eq{def3} are based on the concept
of the down-gradient turbulent transport, in which, similarly to the
case of molecular transport, the flux is taken proportional to the
gradient of transported property times a corresponding (turbulent)
transport coefficient:%%
 \begin{subequations}\label{dt}%%
 \begin{eqnarray}\label{dtM}
 \tau_{xz}&=& -\nu\Sb T  {d U_x}\big /{dz}\,, \quad
\nu\Sb T \approx C_\nu \,  \ell _z \sqrt {\tau_{zz}}\,, \\
\label{dtH} %%
F_z&=&- \chi \Sb T
  {d\Theta}\big /{dz}\,,\quad    ~  \chi \Sb T \approx C_\chi \, \ell _z
 \sqrt {\tau_{zz}}\,, \quad \mbox{etc.}~~~~
\end{eqnarray} \end{subequations} %%
Here the turbulent-eddy viscosity $\nu \Sb T$ and turbulent thermal
conductivity $\chi \Sb T$ are estimated by dimensional reasoning via
the vertical turbulent velocity $\sqrt{\tau_{zz}}$ and a scale
$\ell_z$ (which in the simplest case is determined by the elevation
$z$). The dimensionless coefficients $C_\nu$ and $C_\chi$ are
assumed to be of the order of unity.

This approach meets serious difficulties (Zeman, 1981), in
particular, it predicts a full suppression of turbulence when the
stratification exceeds a critical level, for which  $\Rig
=$Ri$\sb{cr}\approx 0.25$. On the other hand, in observations of
the atmospheric turbulent boundary layer   turbulence exists for
much larger values than $\Rig=0.25$: experimentally above
$\Rig=10$ and even more (for detailed discussion see Zilitinkevich
\emph{et al}., submitted to \emph{Science}). In models for weather
predictions this problem is ``fixed" by introducing fit functions
$C_\nu (\Rig)$ and $C_\chi(\Rig)$ instead of the constant $C_\nu $
and $C_\chi$ in the model parametrization \eq{dt}. This technical
``solution" is not based on a physical derivation and just masks
the shortcomings of the model. To really solve the problem one has
to understand its physical origin, even though from a purely
formal viewpoint it is indeed possible that a dimensionless
coefficient like $C_\chi$ can be any function of $\Rig$.

To expose the physical reason for the failure of the down-gradient
approach, recall that in a stratified flow, in the presence of
gravity,  the turbulent kinetic energy is {\em not}  an integral
of motion. Only the total mechanical energy, the sum of the
kinetic and the potential energy, is conserved in the inviscid
limit. As it was shown already by Richardson, the difficult point
is that an important contribution to the potential energy comes
not just from the mean density profile, but from the density
fluctuations. Clearly, any reasonable model of  the turbulent
boundary layer must obey the conservations laws.

 The physical requirement of conserving the total mechanical
 energy calls for an explicit   consideration not only of the mean
profiles, but also of {\it all} the relevant second-order,
one-point, simultaneous correlation functions of  {\it all} the
fluctuating fields together with some closure procedure for the
appearing third order moments. First of all, in order to account for
the important effect of stratification on the anisotropy, we must
write explicit equations for the entire Reynolds stress tensor,
$\tau _{ij}=\< u_iu_j\>$ . Next, in the case of the temperature
stratified turbulent boundary layers we follow tradition [see, e.g.
see e.g.
 Zeeman, (1981), Hunt et al. (1988), %%
Schumann and Gerz (1995),  %%
Hanazaki and Hunt (2004), %%
Keller and van Atta (2000), %%
Stretch et al. (2001), %%
Elperin et al. (2002), %%
 Cheng et al. (2002)
Luyten et al. (2002),  and %%
Rehmann and Hwang (2005)] and  account for  the turbulent potential
energy  which is proportional to the  variance of the potential
temperature deviation, $\langle \theta^2\rangle$. And last but not
least, we have to consider explicitly equations for the the vertical
fluxes, $\tau_{xz}$ and $F_z$, which include the down gradient terms
proportional to the velocity and temperature gradients, and
counter-gradient terms, proportional to $F_x$ (in the equation for
$\tau_{xz}$) and to $\< \theta^2\>$ (in the equation for $F_z$) .

Unfortunately, the resulting second order closure seems to be
inconsistent with the variety of boundary-layer data, and many
authors took the liberty to introduce additional fitting parameters
and sometimes fitting functions to achieve a better agreement with
the data (see reviews of
 Umlauf and Burchard (2005),
 Weng and Taylor (2003),
 Zeeman, (1981),
 Melor and Yamada (1974),
  and references therein). Moreover, in the second order
closures the problem of critical Richardson number, Ri$\sb{cr}$,
seems to persists (Cheng et al., 2002; Canuto, 2002).

An interesting attempt to improve the modeling of stably
stratified flows with the aim to remove Ri$\sb{cr}$ is reported in
a paper  by Zilitinkevich, Elperin, Kleeorin and Rogachevskii
(2007) in this issue, referred hereafter to as the ZEKR-paper.
Following the traditional second-order scheme (see e.g. Zeeman,
1981), they removed the unwarranted approximation~\eq{dtH}, and
replace it with an exact equation for the vertical thermal flux
$F_z$ and for the ``temperature energy" $\<\theta^2\>/2$,   using
dimensional reasoning to estimate required third-order
correlations. To account for the important effect of
stratification on the anisotropy of turbulence, they, in agreement
with Cheng et al. (2002), explicitly considered budget equations
for the partial kinetic energies $\tau_{ii}/2$. Nevertheless ZEKR
do not close rigorously the momentum budget equation. They combine
three terms in this equations into, what they called ``effective
dissipation rate", arriving actually again at the down-gradient
approximation~\eq{dtM} for the vertical flux of $x$-component of
the mechanical moment, $\tau_{xz}$.  In our paper   we treat these
three terms separately, considered also explicitly the horizontal
heat flux $F_x$.

Notice that in spite of obvious inconsistency of the first-order
schemes, ``most of the practically used turbulent models are based
on the concept of the down-gradient transport". One of the  reasons
is that in the second-order schemes instead of two down-gradient
equations~\eq{dt} one needs to take into account eight nonlinear
coupled additional equations i.e. four equations for the Reynolds
stresses, three equations for the heat fluxes and equation for the
temperature variance. As the result, second-order schemes have
seemed to be rather cumbersome for comprehensive analytical
treatment and have allowed to find only some relationships between
correlation functions (see, e.g., Cheng et al., 2002).
Unfortunately, the numerical solutions to the complete set of the
second-order schemes equations which involve too many fitting
parameters are much less informative in clarification of physical
picture of the phenomenon than desired analytical ones.

In this paper we suggest a relatively simple second-order closure
model of turbulent boundary layer with stable temperature
stratification that, from the one hand, accounts (as we believe) for
main relevant physics in the stratified TBL and, from the other
hand, is simple enough to allow complete analytical treatment
including the problem of critical Ri$\sb{grad}$. To reach  this goal
we approximate the third order correlations via the first- and
second-order ones, accounting only for  the most physically
important terms. We will try to expose the approximations in a clear
and logical way, providing the physical justification as we go
along. Resulting second-order model consist of nine coupled
equations for the mean velocity and temperature gradients, four
components of the Reynolds stresses, two components of the
temperature fluxes and the temperature variance. Thanks to the
achieved simplicity of the model  we found an approximate analytical
solution of these equations, expressing all nine correlations as
functions of only one governing parameter, $\ell(z)/L$, where
$\ell(z)$ is the outer scale of turbulence (depending on the
elevation $z$ and also known as the ``dissipation scale") and $L$ --
is the Monin-Obuckhov length.

 We would like also to stress, that in our approach $\ell(z)/L$
  is an external
 parameter of the problem. For small elevations
$z \ll L$, it is well accepted that $\ell(z)$ is proportional to
$z$, while the $\ell(z)$ dependence is still under debate for $z$
comparable or exceeding $L$. For $z  \gtrsim L$ the assignment and
discussion of the actual dependence of the outer scale of
turbulence, $\ell(z)$, which is manifested in the nature is out of
the scope of this paper, and is remained for future work.  At   time
being, we can analyze consequences  of our approach for  the
following versions of $\ell(z)$ dependence at $z\gg L$:  \\
\textbullet~ function $\ell(z)$ is saturated  at some level  of the
order of $L$. For concreteness we   can  take
 \BE \label{sless}
1/\ell(z) = \sqrt{(d_1z)^{-2}+(d_2 L)^{-2}}\,,
 \quad d_1\sim d_2\sim 1\ .
 \EE
\noindent \textbullet~  $\ell(z)$ is again proportional to $z$ for
elevations much larger than $L$: $\ell(z)=d_3 z$ but with the
proportionality constant $d_3< d_1$. If so,    we can also study
the case $\ell(z) \gg L$ even though such a condition may not be
realizable in nature. In that case our analysis of the  limit
$\ell(z) \gg L$  has only a methodological character: it allows to
derive an approximate analytic solution for all the objects of
interest as functions of $\ell(z)/L$ that, of course, is also
valid for the outer scale of turbulence not exceeding a value of
the order of $L$.

It should be noticed that  traditional turbulent closures
(including ours) cannot be applied for strongly stratified flows
with $\Rig\gtrsim 1$ (may be even at  $\Rig\sim 1$). The main
reason is that these closures are roughly justified for developed
\emph{vortical} turbulence, in which the eddy-turnover time is of
the order of its life time; in other words, there are no well
defined ``quasi-particles" or waves. This is NOT the case for
stable stratification with $\Rig\gtrsim 1$, in which the
Brunt-V\"ais\"al\"a frequency %%
\BE \label{BVf} N\=\sqrt{\b d \Theta(z)/d\,z }\,,
 \EE%%
 is   larger then the eddy-turnover frequency $\gamma$.  It means that for
 $\Rig\gtrsim  1$  there are weakly decaying Kelvin-Helmholts internal
 gravity waves (with characteristic frequency $N$ and decay time above
$1/\g$), propagating on large distances, essentially effecting on
TBL, as pointed out by  Zilitinkevich, (2002). In contrast to
ambitious ZEKR attempt to describe entire TBL without limitation in
values of $\Rig$, and without explicit accounting for the internal
gravity waves, we concentrate in our paper on self-consistent
description of the lower part of the atmospheric TBL, in which
turbulence has vortical character and consequently,  large values of
$\Rig$ do not appear. We relate large values of $\Rig$ in the upper
part of TBL with contributions of the internal gravity waves  to the
energy and the energy flux in  TBL, to the momentum flux, and to the
production of (vortical) turbulent energy. Due to their instability
in a shear flow, the waves can break and create turbulent kinetic
energy. All these effects are beyond the  framework of   our
paper.  Their description in the upper  ``potential-wave" TBL and
intermediate region with the combined ``vortical-potential"
turbulent velocity field is in our nearest agenda.

To make the paper more transparent for wide audience, not
necessarily experts in atmospheric TBL, we attempt to present the
material in a self-contained manner, and organized it as follows.

In Sect.~\ref{s:balance}A we used the Oberbeck-Boussinesq
approximation and applied the standard Reynolds decomposition (into mean
values and turbulent zero-mean fluctuations of the velocity and
temperature fields) to derive equations for the mean values and
balance equations for all the relevant second-order correlation
functions. In Sect. \ref{s:balance}B we demonstrate that the resulting
balance equations exactly preserve (in the non-dissipative limit)
the total mechanical energy of the system, that consists of the
kinetic energy of the mean flow, kinetic energy and potential energy
of the turbulent subsystem.

In Sect.~\ref{s:closure}   we describe the proposed closure
procedure that results in a model of stably stratified TBL, that
accounts explicitly  for all the relevant second-order correlations.
The third order correlations which appear in the theory are modeled
in terms of second-order correlations in Sects.~\ref{s:closure}A and
B. Further simplifications are presented  in Sects.~\ref{s:closure}C
and D for stationary  turbulent flows in a plane geometry outside
the viscous and buffer layers.  In Sect.~\ref{s:closure}E we suggest
some generalization of the standard ``wall-normalization" to obtain
the model equations in dimensionless form with only one governing
parameter, $\ell(z)/L$.

Section~\ref{ss:strat41} contains approximate analytical solution of
the model. It is shown that the analytical solution deviates from
the numerical counterpart in less than a few percent in the entire
interval $0\le (\ell/L)<\infty$.

The last Sect.~\ref{s:res} is devoted to a detailed description of
our results: profiles of the mean velocity and potential temperature
(Sect.~\ref{s:res}A), profiles of the turbulent kinetic and
``temperature" energies, profiles of the anisotropy of partial
kinetic energies (Sect.~\ref{s:res}B),  profiles of the turbulent
transport parameters $\nu\Sb T$ and $\chi\Sb T$, profiles of the
gradient- and flux-Richardson numbers $\Rig$ and $\Rif$, and the
dependence of the turbulent Prandl number Pr$\Sb T$ vs. $\ell/L$ and
$\Rig$, Sect.~\ref{s:res}C. In conclusion Sect.~\ref{s:res}D, we
consider the  validity of the down-gradient transport
concept~\eq{dt} and explain why it is violated in the upper part of
TBL. The problem of critical $\Rig$ is also discussed.

 In Appendix \ref{s:approx} we revisit the derivation of the
Oberbeck-Boussinesq approximation. The reason for doing so stems
from the fact that usual derivations are too specialized either to
fluids whose density variation is very small, or to ideal gases.
Thus it cannot be directly applied, for example, to humid air.

The detailed procedure of obtaining an approximate analytical
solutions for all the objects of interest as functions of $\ell/L$
is presented in Appendix \ref{ss:gen}. Exact solutions are obtained
for two cases: neutral stratification, $\ell/L = 0$, and
(methodological) limit $\ell/L \rightarrow \infty$. Both solutions
are corrected up to the first order in corresponding small
parameters. The desired approximate analytical solutions interpolate
these two limits over all values of $\ell/L$.

Appendix \ref{ss:onClosure} contains a  discussion of  the
applicability for the stratified flows of applied closure of the
triple correlations via second order ones with the help of
dimensional reasoning.

Appendixes A-C are available in the online version of this paper or
at Los-Alamos archive \# nlin.CD/0612018

\section{\label{s:balance} Simplified dynamics in a stably
 temperature-stratified TBL and their conservation laws}

The aim of this section is to consider the simplified dynamics of
a stably temperature-stratified turbulent boundary layer, aiming
finally at an explicit description of the height dependence of
important quantities like the mean velocity, mean temperature,
turbulent kinetic and potential energies, etc. In general one
expects very different profiles from those known in standard
(unstratified) wall-bounded turbulence. We want to focus on these
differences and propose that they occur already relatively close
to the ground allowing us to neglect (to the leading order) the
dependence of the density on height and the Coriolis force. We
thus begin by simplifying the hydrodynamic equations which are
used in this section.

%%%%%%%%%%%%%%%%%%%%%%%%%%%%%%%%%%
\subsection{\label{ss:wall-units}Simplified hydrodynamic
equations and Reynolds decomposition}%%
First we briefly overview the derivation of the governing
equations in the Boussinesq approximation. For the extensive
details the reader is referred to the Appendix A.

The system (\ref{full}) of hydrodynamic equations describing a
fluid in which the temperature is not uniform  consists of the
Navier-Stokes equations for the fluid velocity, $\BC U(\B r,t)$, a
continuity equation for the space and time dependent (total)
density of the fluid, $\rho (\B r,t)$, and of the heat balance
equation for the (total) entropy per unit mass, $\C S(\B
r,t)$,~Landau and Lifshitz, 1987.

These equations are considered with boundary conditions that
maintain the solution far from the equilibrium state, where $\BC
U=\C S=0$. These boundary conditions are $\BC U=0$ at zero
elevation, $\BC U=const$ at a high elevation of a few kilometers.
This reflects the existence of a wind at high elevation, but we do
not attempt to model the physical origin of this wind in any
detail. The only important condition with regards to this wind is
that it maintains a momentum flux towards the ground that is
prescribed as a function of the elevation. Similarly, we assume
that a stable temperature stratification is maintained such that
the heat flux towards the ground is prescribed as well.

We neglect the viscous entropy production term assuming that the
temperature gradients are large enough such that the thermal
entropy production term dominates. For simplicity of the
presentation we restrict ourselves by relatively small elevations
and disregard the Coriolis force (for more details, see Wyngaard,
1992). On the other hand we assume that the temperature and
density gradients in the entire turbulent boundary layer are
sufficiently small to allow employment of local thermodynamic
equilibrium. In other words, we assume the validity of the
equation of state.

As a ``basic reference state" (BRS) denoted hereafter by a
subscript ``$\sb b$" we use the isentropic  model of the
atmosphere, where the entropy is considered space homogeneous. Now
assuming a smallness of deviations of the density and pressure
from their BRS values and exploiting the equation of state, one
obtains a simplified equation (\ref{B6a}) from the complete system
(\ref{full}), which is already very close to the standard
Navier-Stoks equation in the Boussinesq approximation. Introducing
(generalized, see eq. (\ref{defTheta})) potential temperature, one
results in the well-known system (\ref{FS}) of hydrodynamic
equations in the Boussinesq approximation.

We confine ourselves to regions which are close to the ground
hence neglect the dependence of the density on height, thus we
replace $\B \nabla (p/\rho\sb{\, b})\Rightarrow (\B \nabla p)
/\rho\sb{\, b}$, and $\B \nabla(\rho\sb{\, b}\BC U)\Rightarrow
\rho\sb{\, b} \B \nabla\cdot \BC U$. The resulting equations read
\begin{equation}\label{GovEq} %%
 \frac{\C D\, \BC U }{\C D t}   =
-\frac{\B {\nabla} p}{ \rho\sb b} - \B  \beta\, \Theta\sb {\, d}
 + \nu\, \Delta\, \BC U \,,  \  \frac{\C D\,
 \Theta\sb {\, d}}{\C D t}  = \chi\,
\Delta\, \Theta\sb {\, d} \ .
\end{equation}%%
Here $ {\C D}/{C Dt}\=  {\partial}/{\partial{t}}+ \BC U \cdot \B
\nabla$ is the convection time derivative, $p$ -- pressure,
$\rho\sb b$ is the density in BRS, $\beta = g \~\beta$ is the
buoyancy parameter ($g$ is the gravity acceleration and $\~\beta$
is the thermal expansion coefficient Eq. (\ref{beta}), which is
equal to $1/T$, reciprocal molecular temperature, for an ideal
gas), $\Theta\sb {\, d}$ is the deviation of the potential
temperature from its BRS value, $\nu$ -- dynamical viscosity and
$\chi$ is the dynamical thermal conductivity. To develop equations
for mean quantities and correlation functions one applies the
Reynolds decomposition $ \BC U   =  \B U + \B u\,, \ \langle \BC
U\rangle = \B U\,, \  \<\B u\>=0\,,
 \Theta\sb{\,d} = \overline \Theta+\theta\,, \
 \langle\Theta\sb{\,d}\rangle=\overline\Theta\,, \  \langle \theta
 \rangle= 0\,, \ p =  \langle p\rangle+\~p\,, \quad \langle \~p \rangle
 = 0$. Here the average $\langle \cdots \rangle$ stands for
  averaging over a horizontal plane
at a constant elevation. This leaves the average quantities
 with a $z,t$ dependence only.
 Substituting in Eqs.~\eq{GovEq} one gets  equations of motion
 for the mean velocity and
mean temperature profiles
 \begin{equation}\label{mean}
\frac{D\, U_i}{Dt}  + \nabla\!_j\, \~ \tau_{ij}=
  - \frac{ \nabla\!_i  \langle p\rangle}{\rho_b} -
\beta_i \,\overline\Theta\,,\
\frac{D\, \overline\Theta }{Dt}+ \B\nabla\cdot \~ {\B F}=0\ .%%%%
\end{equation}
Here $ {D}/{Dt}\=  {\partial}/{\partial{t}}+ \B U \cdot \B \nabla$
is the mean convection derivative.  The total (molecular and
turbulent) momentum
  and thermal fluxes are
\begin{equation} \label{def-tau}\~ \tau_{ij}
 \=  - \nu \nabla\!_j\, U_i+\tau_{ij}\,,\quad
 \~ {\B F}  \=  -\chi \,\B\nabla \Theta +\B F\,,
\end{equation}  %%
where $\tau_{ij} = \langle u_i u_j \rangle$ is the Reynolds stress
tensor describing the turbulent momentum flux, and $\B F = \langle
\B u \theta \rangle$ is the turbulent thermal flux.  In order to
derive equations for these correlation functions, one considers
the equations of motion for the fluctuating velocity and
temperature: %%
\begin{subequations}\label{fluct}%%
\begin{eqnarray}   %%===================================
\label{NSEFluct}%%
 {D\,\B u}/{D\,t}
 &\!=\!& -\B u \cdot\B  \nabla \B U -\B u \cdot\B  \nabla \B u
 +\left\langle \B u \cdot\B  \nabla \B u\right\rangle \\ \nn%%
&& -({\B \nabla \~p}/{\rho_b}) + \nu\, \Delta \B u - \B
\beta\,\theta\,,\\ %%
\label{fluctb}
 {D\,\theta}/{D\,t} &\!=\!& -\B
u \cdot \B \nabla \Theta -\B u \cdot \B \nabla \theta +\chi\,\Delta
\theta +\left\langle \B u \cdot \B \nabla \theta \right\rangle
.~~~~~~~~
\end{eqnarray}\end{subequations} %%

The whole set of the second order correlation functions includes the
Reynolds stress, $\tau_{ij}$, the turbulent thermal flux, $\B F$,
 and the ``temperature energy"
$
 E_{\theta} \=
\<\theta^2\>/2$,  which is denoted and named by analogy with the
turbulent kinetic energy density (per unit mass and unit volume),
 $
 E\Sb K = \frac12\,\langle |\B u|^2\rangle/2=   \mbox{Tr}
 \{\tau_{ij}\}/2$.
Using \eq{fluct} one gets the following ``balance equations": %%
\begin{subequations} \label{corr}%%
\begin{eqnarray} \label{corra}%%%% Reynplds balance
    \frac{D\,\tau_{ij}}{D\,t} +\varepsilon_{ij}
     + \frac{\partial}{\partial{x_k}}T _{ijk}
    &=&
   \C P_{ij}- \C C_{ij} +\C \R_{ij}
    \,,~~~~~~~~~~~\\
\label{corrb}
  \frac{D\,F_i}{D\,t} +  \epsilon _i+
    \frac{\partial}{\partial{x_j}}T_{ij} &=& \C A_i+\C B_i
     \,,  ~~~~\\
\label{corrc}%%
   \frac{D\, E_{\theta}}{D\,t}+   \varepsilon+\B \nabla
   \cdot \B T&=& -\B F \cdot
    \B\nabla \Theta  \ . %%
\end{eqnarray}%%
\end{subequations}
Here we denoted the dissipations of the Reynolds-stress, heat-flux
and the temperature energy by
  \begin{eqnarray}\nn %%
\varepsilon_{ij}&\!\!\=\!\!&  2\,\nu\left\langle
\frac{\partial{u_i}}{\partial{x_k}}\,
\frac{\partial{u_j}}{\partial{x_k}} \right\rangle,
 \epsilon_{i} \=  \(\nu +\chi\)\left\langle
\frac{\partial{\theta}}{\partial{x_k}}\,
\frac{\partial{u_i}}{\partial{x_k}} \right\rangle, \\
 \label{diss}
\varepsilon & \= &  \chi\,\<|\B\nabla\theta|^2\>,  %%
\end{eqnarray} %%
The last term on the LHS of each of Eqs.~\eq{corr} describes  {\em
spatial} flux of the corresponding quantity. In models of wall
bounded unstratified turbulence it is known that these terms are
very small almost everywhere. We do not have sufficient experience
with the stratified counterpart to be able to assert that the same
is true here. Nevertheless, for simplicity we are going to  neglect
these terms. It is possible to show that the accounting for these
terms does not influence much the results. Note that keeping these
terms turns the model into a set of differential equations which are
very cumbersome to analyze. This is a serious uncontrolled step in
our development, so we cross our fingers and proceed with caution.
Since these terms are neglected we do not provide here the explicit
expressions for $T_{ijk}$, $T_{ij}$, and $\B T$.

The first term on the RHS of the balance Eq.~\eq{corra} for the
Reynolds stresses is the Energy Production tensor $\C P_{ij}$,
describing the production of the turbulent kinetic energy from the
kinetic energy of the mean flow, proportional to the gradient of the
mean velocity: %%
\begin{subequations} \label{defs1} \begin{equation} \label{defs1A}%%
\C P_{ij} \= - \tau_{ik}\, {\partial{U_j}}/{\partial{x_k}}
-\tau_{jk}\, {\partial{U_i}}/{\partial{x_k}} \ . \end{equation}%%
 The second term on the RHS of Eq.~\eq{corra}, $\C C_{ij} $, will be
 referred hereafter to as the
 ``Energy conversion tensor". It
 describes the conversion of the kinetic  turbulent energy into
 potential energy. This term is proportional to the buoyancy parameter
 $ \b$ and the turbulent thermal flux $\B F$:
\begin{equation}\label{defs1B} \C C_{ij} \=
 - \beta\big (F_i\,\delta_{j\,z}+
F_j\,\delta_{i\,z}\big)\ .\end{equation}%%
The next term in the RHS of Eq.~\eq{corra} is known as the
``Pressure-rate-of-strain" tensor: %%
\begin{equation}\label{defs1C}  \C R_{ij} \=  \left\langle   {\~p}
\,s_{ij}/{\rho\sb b}\right\rangle, \quad s_{ij}\=
 {\partial{u_i}}/{\partial{x_j}} + {\partial{
u_j}}/{\partial{x_i}} \ . \end{equation}%%
In incompressible turbulence its trace vanishes, therefore $ \C
R_{ij}$ does not contribute to the balance of the kinetic energy. As
we will show in Sec.~\ref{sss:PRS}, this tensor can be presented as
the sum of three contributions (Zeman, 1981),
\begin{equation}\label{rof}%%
\C R_{ij}= {R_{ij}\Sp{\, RI}} +{R_{ij}\Sp {\,IP}} +{R_{ij}\Sp{\,IC}}
\,,%%
\end{equation} %%
in which ${R_{ij}\Sp{\, RI}}$ is  responsible for the nonlinear
process of isotropization of turbulence and is traditionally called
the ``Return-to-Isotropy", ${R_{ij}\Sp {\,IP}}$ is similar to the
energy production tensor~\eq{defs1A} and is called ``Isotropization
of production".  New term, appearing in the stratified flow
${R_{ij}\Sp{\,IC}}$,  is similar to the energy conversion
tensor~\eq{defs1B} and will be refereed to as the ``Isotropization
of conversion".

Consider  the balance of the turbulent thermal flux $\B F$,
\Eq{corrb}. The first term in the RHS, $\BC A$ describes the source
of $\B F$ and, by analogy with the energy-production tensor, $\C P
_{ij}$, is called ``Thermal-flux production vector". Like $\C P
_{ij}$, \Eq{defs1B}, it has the contribution, $A_i^{^{SU}} $,
proportional to the mean velocity gradient:
 \begin{eqnarray}\label{defOfA}\nonumber
 {\C A}_i& \=\ & { A_i^{^{SU}}}
 + { A_i^{^{S\Theta}}} + { A_i^{^{E\theta}}} \,,\quad
  { A_i^{^{SU}}}  \= -\B F \cdot \B \nabla\, U_i \,,
  \\
 { A_i^{^{S\Theta}}} & \= & -\tau_{ij}\,
 {\partial \Theta}/{\partial x_j}\,,
\quad { A_i^{^{E\theta}}} \= 2\,\beta\,E_{\theta}\,\delta_{i\,z}\,,
\end{eqnarray}%%
 and two additional contributions, related to the temperature
gradient and to the ``temperature energy", $E_\theta$, and the
buoyancy parameter. One sees, that in contrary to the oversimplified
assumption~\eq{dtH} the thermal flux in turbulent media cannot be
considered  as proportional to the temperature gradient. It has also
a contribution, proportional to the velocity gradient and even to
the square of the temperature fluctuations. Moreover, the RHS of the
flux-balance \Eq{corrb} has an additional term, the
``Pressure-temperature-gradient vector" which, similarly to the
pressure-rate-of-strain tensor~\eq{rof}, can be divided into three
parts (Zeman, 1981): %%
\begin{equation}\label{defs1f} %%
\BC B  \=   \left\langle {\~p }\, \B \nabla \theta /\rho_b
\right\rangle  = \B B \Sp {RD}  + \B B \Sp{SU} +  \B B \Sp{E\theta}\
.
\end{equation}%%
As we will show in Sec.~\ref{sss:PRS} the first contribution, $
B_i\Sp {RD}\propto \< u\, u\, \nabla\!_i\, \theta\>$ is responsible
for the nonlinear flux of  $\B F$ in the space of scales, toward
smaller scales, similarly to the correlation $\< u\, u\, u\>$, which
is responsible for the flux of kinetic energy $\< u^2\>$ toward
smaller scales. The correlation $ B_i\Sp {RD}
 \propto \< u\, u\, \nabla\!_i\, \theta\>$ may be understood as the nonlinear contribution to the dissipation of the thermal flux. Correspondingly
we will call it  ``Renormalization of the thermal-flux Dissipation"
and will supply it with a superscript ``~$\Sp {RD}$~". The next  two terms  in the
decomposition~\eq{defs1f} are  ${B_i^{^{SU}}}\propto S\Sb U$ and
${B_i^{^{E\theta}}}\propto E_\theta$.  They describe the
renormalization of the thermal-flux production terms
${A_i^{^{SU}}}\propto S\Sb U$ and ${A_i^{^{E\theta}}}\propto
E_\theta$, accordingly. %%
\end{subequations}%%
%%%%%%%%%%%%%%%%%%%%%%%%%%%%%%%%%%%%%%%%%%%%%%
\subsection{\label{ss:cons}Conservation of total mechanical energy
in the exact balance equations}

The total mechanical energy of temperature stratified turbulent
flows consists of three parts with densities (per unit mass): $E=
E_{\C K}+ E\Sb K+E\Sb P$, where $E_{\C K}=|\B U|^2/2$ is the
density of kinetic energy of the mean flow, $E\Sb K=\tau_{ii}/2$
is the density of turbulent kinetic energy and $E\Sb P={\beta}
E_\theta/{S_\Theta}$ is the density of potential energy,
associated with turbulent density fluctuation $ \~\rho= \~\beta\,
\theta \rho_b$, caused by the (potential) temperature fluctuations
$\theta$, and $S_\Theta = d\, \overline{\Theta}/d z$.  Note that
the formula for $E\Sb P$ appears different from the plane average
of the $\C E\Sb P$ in Eq. (\ref{TotMechEn}). In Appendix \ref{a:1}
we show that in fact the difference between the two objects is a
time independent total potential energy in the basics reference
state, and therefore  it can be considered as the zero level of
the potential energy.

The balance Eq. for  $E_{\C K}$ follows   from \Eq{mean}:
\begin{subequations}\label{balE}%%
\begin{equation}\label{balETm} %%
 {D E_{\C K}}/{D\,  t}+ \nu \(\nabla_{\!\! j}  U_i\)^2+
 \nabla_{\!\! j}\,
(U_i   \tau_{ij})= [\mbox{source } E_{\C K}]+ \tau_{ij}\nabla_{\!\!
j} \, U_i\,,
 \end{equation}
with the help of identity: $ U_i\nabla_{\!\! j}\, \tau_{ij}\=
\nabla_{\!\! j}\, (U_i \tau_{ij})-\tau_{ij} \nabla_{\!\! j}\, U_i$
and definition~\eq{def-tau}. The terms on the LHS of this Eq.,
proportional to $\nu$ and $ \tilde \tau_{ij}$ respectively, describe
the dissipation and the spatial flux of $ E_{\C K}$. The term
[source $E_{\C K} $]  on the RHS of \Eq{balETm} describes the
external source of energy, originating from the boundary conditions
described above and $\tau_{ij}\nabla_{\!\! j} \, U_i$, describes the
kinetic energy out-flux from the mean flow to turbulent subsystem.

The balance Eq. for the turbulent kinetic energy follows directly
from~\Eq{corra}:  %%
\begin{equation}\label{balETt} %%
 {D\, E\Sb K}/{D\,  t}+ \big[  \ve_{ii}+\nabla_{\!\! j}\,
T_{iij}\big]/2=- \tau_{ij} \nabla_{\!\! j} \, U_i - \b F_z\ . %%
\end{equation}%%
On the LHS of \Eq{balETt} one sees the dissipation and spatial flux
terms. The first term on the RHS originates from the energy
production, $\frac12 \, \C P_{ii}$, defined by \Eq{defs1A}. This
term has an opposite sign to the last term on the RHS of \Eq{balETm}
and describes the production of the turbulent kinetic energy from
the kinetic energy of the mean flow. The last term on the RHS of
\Eq{balETt} originates from the energy conversion tensor $\frac12 \,
\C C_{ii}$, \Eq{defs1B}, and describes the conversion of the
turbulent kinetic energy into potential one.

According to the last of Eqs.~\eq{corr}, one gets the balance
equation for the potential energy $E\Sb P$; multiplying \Eq{corrc}
for $E_\theta$ by $\b/S_\Theta$:%%
\begin{equation}\label{balEP}
  {D\, E\Sb P}/{D\,  t}+  \beta \Big [ \epsilon +
 \nabla_{\!\! j} T_j \Big ]/S_\Theta  = \beta F_z\ .\end{equation}
 \end{subequations}
 The RHS of this Eq. [coinciding up to a sign with the last term
  on the RHS of \Eq{balETt}] is the source of potential energy
(from the kinetic one).

In the sum of the three balance equations the conversion terms (of the
kinetic energy from the mean to turbulent flows and of the turbulent
kinetic to potential one) cancel and one gets the total mechanical
energy balance: %%
\begin{equation}\label{bal-tot}
 {D\, E\ }/{D\,  t}+ [\mbox{diss }E] + \B \nabla\,
 [\mbox{flux} E]= [\mbox{source } E_{\C K}]\ .
 \end{equation}
This equation exactly respects the conservation of total mechanical
energy in the dissipation-less limit,  irrespective of the closure
approximations. This is because the energy production and conversion
terms are exact and do not require any closures, while the
pressure-rate-of-strain tensor, that requires some closure, does not
contribute to the total energy balance.

%%%%%%%%%%%%%%%%%%%%%%%%%%%%%%%%%%%%%%%%%%%%%%%%%
\section{\label{s:closure} The Closure Procedure and the resulting model}
In this section we describe the proposed closure procedure that
results in a model of  stably stratified TBL. In developing this
model we rely strongly on the analogous well developed modeling of
standard (unstratified) TBL. The final justification of this
approach can be only done in comparison to data from experiments
and simulations. We will do below what we can to use the existing
data, but we propose at this point that much more experimental and
simulational work is necessary to solidify all the steps taken in
this section.

\subsection{\label{sss:PRS} Pressure-Rate-of-Strain tensor $\C R_{ij}$ and Pressure-Temperature-Gradient vector $\BC B$} %%
The correlation functions $\C R_{ij}$   and $\BC B$, defined by
Eqs.~\eq{defs1C} and \eq{defs1f}, include fluctuating part of the
pressure $\~ p$.  The Poisson's equation for $\~ p$
 follows from  \Eq{fluct}: $
\Delta \~p = \rho \sb b\Big[ -\nabla _i \nabla _j \( u _i u _j -\<u
_i u _j\>
 +U_i u_j +U_j u_i\) +\beta\nabla_z\theta\Big]
$.
 The solution of this equation includes a harmonic part, $\Delta \~p =
 0$, which is  responsible for sound propagation and does not
 contribute to  turbulent dynamics at small Mach numbers. Thus this
 contribution can be neglected. the inhomogeneous solution
 includes  three parts $\~{p}=\rho \sb b[p_{uu}+p_{Uu}+p_\theta]$,
 where
  \begin{eqnarray}\label{pr1}
  p_{uu}&=&
 \Delta^{-1}\nabla _i \nabla _j\(   \<u _i u _j\>- u _i u _j\)\,,\br
p_{Uu}&=& \Delta^{-1}\nabla _i \nabla _j\(  U_iu_j+ U_ju_i\)\,, \
p_\theta= \beta\Delta^{-1}\nabla_z\ \theta \ ,
\end{eqnarray}
and the inverse Laplace operator $ \Delta^{-1}$ is defined as usual
in terms of an integral over the Green's function.

Correspondingly the correlations  $\C R_{ij}$   and $\BC B$ consist
of three terms, Eqs.~\eq{rof} and \eq{defs1f}, in which %%
 \begin{eqnarray}\label{decP}   &&R_{ij}\Sp{RI} = \left\langle
 p_{uu} s_{ij} \right\rangle\,, \       R_{ij}\Sp{IP}
 \=  \left\langle  p_{Uu}\, s_{ij}\right\rangle , \
R_{ij}\Sp{IC} \= \left\langle  p_\theta s_{ij} \right\rangle,~\\
&&  \B B_i\Sp {RD}  =\< p _{uu} \B \nabla \theta \>,\ \B B ^{^{SU}}
   \=   \< p _{Uu} \B \nabla \theta \>,\ \B B_i^{^{E\theta}} \=
   \< p _\theta \B \nabla \theta \>\nonumber \ .
   \end{eqnarray} %%
All of those terms  originating from $p_{uu}$ are the most
problematic because they introduce coupling to triple correlation
functions: ${R_{ij}\Sp{\, RI}}\propto \<u_iu_ju_k\>$ and  $\B
B\Sp{RD}\propto \< u^2 \B \nabla \theta \>$. Thus they require
closure procedures whose justification can be only tested
a-posteriori against the data.

Having in mind to simplify the model in most possible manner, we
adopt for the diagonal part of the Return-to-Isotropy tensor, the
simplest Rota form (Rotta, 1951)%%
\begin{subequations}\label{ROF}%%
\begin{equation}\label{RIa} %%
{R_{ii}\Sp{\, RI}} \simeq   -\g\Sb {RI}\(  \tau_{ii} - 2\,E\Sb
K / 3\) \,, %%
\end{equation} %%
in which  $\g\Sb{RI}$ is the relaxation frequency of diagonal
components of the Reynolds-stress  tensor toward its isotropic form,
$2 E\Sb K/3$.  The parametrization of  $\g\Sb{RI}$    will be
discussed later.  The tensor ${R_{ij}\Sp{\, RI}}$ is traceless,
therefore the frequency $\g\Sb{RI}$ must be the same for all the
diagonal components of ${R_{ii}\Sp{\, RI}}$. On the other hand there
are no reasons to assume that off-diagonal terms have the same
relaxation frequency.   Therefore, following~L'vov~et~al.~(2006) we
assume that %%
\begin{equation}\label{RIb} %%
{R_{ij}\Sp{\, RI}} \simeq   -   \~\g\Sb {RI} \tau_{ij}  \,, \quad
i\ne j\,,%%
\end{equation}%%
with, generally speaking, $\~\g\Sb {RI}\ne \g\Sb {RI}$. Moreover, on
intuitive level,  we  can expect, that off-diagonal terms should
decay faster then the diagonal ones, i.e. $\~\g\Sb {RI}> \g\Sb
{RI}$. Indeed, our analysis of DNS results, see Eq.~\Ref{est} shows
that $\~\g\Sb {RI}/\g\Sb {RI}\simeq 1.46$.

The term $\B B\Sp{RD}$ also describes return-to-isotropy due to
nonlinear turbulence self interactions  (Zeman, 1981), and may be
modeled as: %
\begin{equation}\label{RD1} {B_i\Sp {RD}} =-\g \Sb {RD} F_i\ .
\end{equation} %%
 This equation dictates the vectorial structure of $
{B_i\Sp {RD}} \propto F_i$, which will be confirmed below. The rest
can be understood as the definition of the $\g \Sb {RD}$ as  the
relaxation frequency of the thermal flux. Its parametrization  is
the subject of further discussion in Sec. \ref{ss:closure}.

The traceless ``Isotropization-of-Production" tensor, ${R_{ij}\Sp
{\,IP}}$, has a very similar structure to the production tensor, $\C
P_{ij}$, \Eq{defs1},  and thus is traditionally  modeled in terms of
$\C P_{ij}$~(Pope, 2001):%%
\begin{equation} \label{IP}%%
{R_{ij}\Sp {\,IP}} \simeq  -C\Sb {IP}\(   \C \P_{ij} - \delta_{ij}\,{\C P}/3
\)\,,  \quad \C P\=  \mbox{Tr} \,\{ \C \P_{ij}\}\ .%%
\end{equation}%%
The accepted value of the numerical constant $ C\Sb {IP} = \frac35\ $
(Pope, 2001).

The traceless ``Isotropization-of-Conversion" tensor, ${R_{ij}\Sp
{\,IC}}$ does not exist in unstratified TBL. Its structure is very
similar to the conversion tensor, $\C C_{ij}$, \Eq{defs1}. Therefore
it is reasonable to  model it in the same way in terms of $\C
C_{ij}$ (Zeman, 1981): \begin{equation} \label{IC}%%
{R_{ij}\Sp {\,IC}} \simeq  -C\Sb {IC}\(
  \C C_{ij} -\delta_{ij}\, {\C C}/3
\)\,,  \quad \C C\=  \mbox{Tr} \,\{ \C C_{ij}\}\,,%%
\end{equation}%%
with some new constant $ C\Sb {IC}$ . %%

The renormalization of production terms ${B_i^{^{SU}}}$ and
${B_i^{^{E\theta}}}$ are very similar to the corresponding thermal flux
production terms, ${A_i^{^{SU}}}$ and ${A_i^{^{E\theta}}}$, defined
by Eqs.~\eq{defs1}. Therefore, in the spirit of Eqs.~\eq{IP} and
\eq{IC}, they are modelled as follows:%%
\begin{eqnarray} %%
{B_i^{^{SU}}}&=& (C_{_{SU}}-1){A_i^{^{SU}}}= (1-C_{_{SU}})
(\B F \cdot \B \nabla\,) U_i\,, \\ %%
{B_i^{^{E\theta}}} &=& -(C_{_{E\theta}}+1) {A_i^{^{E\theta}}} =
 -2\,\beta\,(C_{_{E\theta}}+1)E_{\theta}\,\delta_{i\,z}\ .~~~~~~~~ %%
\end{eqnarray} %%
\end{subequations} %%
Using this and \eq{pr1} one finds the sign of $C_{_{E\theta}}$:
 \begin{eqnarray}\nonumber%%
&& -\beta\(C_{_{E\theta}} +1\) E_{\theta} = \<\~p_{\theta} \nabla_z
\theta\> = \beta\langle (\nabla_z\theta) \Delta^{-1}
(\nabla_z\theta)\rangle,~~~\\ %%
&& \label{CT}%%
C_{_{E\theta}} = -\(1 + {\langle (\nabla_z\theta) \Delta^{-1}
(\nabla_z\theta)\rangle}/{\langle \theta^2\rangle}\) < 0\ .
\end{eqnarray}%%
To estimate $C_{_{E\theta}}$ we assume that on the gradient scales
the temperature fluctuations are roughly isotropic (ZEKR-paper),
and therefore we can estimate
 $\Delta= \nabla_x^2+\nabla_y^2+\nabla_z^2 \approx 3\nabla_z^2$.
  Introducing this
estimate and integrating by parts leads to $ C_{_{E\theta}}\approx
-2/3$.

\subsection{\label{sss:dis}Reynolds-stress-, thermal-flux-, and
 thermal-dissipation} %%
Far away from the wall and for large Reynolds numbers the
dissipation tensors are dominated by the  viscous scale motions,
at which turbulence can be considered as isotropic. Therefore, the
vector $\B \epsilon$ should vanish, while the tensor $\ve_{ij}$,
\Eq{defs1}, should be diagonal:  %%
\begin{subequations}\label{distau1}
\begin{equation}\label{distau1a} \epsilon_i=0\,, \quad \ve_{ij}=
2\, \g_{uu} \,E\Sb K \, \d_{ij}/3\,, %%
\end{equation}%%
where  the numerical prefactor $\frac23 $ is chosen such that
$\g_{uu}$ becomes the relaxation frequency of the turbulent kinetic
energy. Under stationary conditions the rate of turbulent kinetic
energy dissipation is equal to the energy flux through scales, that
can be estimated as $\< u u u \>/ \ell$, where $\ell$ is the outer
scale of turbulence. Therefore, the natural estimate of $\g_{uu}$
involves the tripple-velocity correlator, $ \g_{uu}\sim \( \< u u u
\>/ \ell \< uu \>\)$, %%
exactly in the same manner, as the Return-to-Isotropy frequencies,
$\g  \Sb{RI}$ and $ \~{\g}\Sb{RI}$ in Eqs.~\eq{RIa} and \eq{RIb}.
Similarly,
\begin{equation}\label{distau1b}
\varepsilon  =\g _{\theta\theta}\, E_\theta  \,,    \quad \g_
{\theta\theta} \sim  \<\theta \theta u \>\big / \ell \<\theta
\theta\> \ . \end{equation}\end{subequations} %%
%%
%%%%%%%%%%%%%%%%%%%%%%%%%%%%%%%%%%%%%%%%%%
\subsection{\label{ss:balance}Stationary balance equations
 in plain geometry}%%

In the plane geometry, the equations simplify further. The mean
velocity is oriented in the (streamwise) $\x$ direction and all
mean values depend  on the vertical (wall-normal) coordinate $z$
only: %%
$\B U  =  U (z)\,\x $, $\overline{\Theta}=\overline{\Theta}(z)$, $
\tau_{ij}=\tau_{ij}(z)$, $ ~\B F=\B F(z)$, $E_\t =E_\t(z)$.
Therefore     $\(\B U\cdot\B \nabla\)\< \dots\> = 0$, and in the
stationary case, when $\p\ /\p t= 0$,  the mean convective
derivative  vanishes: $D\ /D\, t=0$.  Moreover due to the $y\to -y$
symmetry of the problem the following correlations vanish:
 $
\~\tau_{xy}=\~\tau_{yz}=\~F_y=0$.
 The only non-zero components of the mean velocity and
temperature gradients are:
 \begin{equation}\label{Shears} S_{_U}\= {d U}/{d z}\,, \quad
S_{_\Theta}\=  {d \Theta}/{d z}\ . \end{equation} %%
Finally, the acceleration due to gravity $\B g$ is directed vertically and
thus the buoyancy parameter has just one component:
 $ \B  \beta = \z \, \beta $.

%%%%%%%%%%%%%%%%%%%%%%%%%%%%%%%%%%%%%%%%%%%%
\subsubsection{Equations for the mean velocity and temperature profiles}
 Having in mind Eqs. of Sec.~\ref{ss:balance} and integrating    Eqs.
\eq{mean} for $U_x$ and $\Theta$   over $z$,   one gets equations
for the total (turbulent and molecular) mechanical-momentum flux,
$\~\tau (z)$, and thermal flux, $\~F$, toward the wall %%
\begin{subequations} \label{TBL-Sim}%%
  \begin{eqnarray}  \label{Sim-U-x}%%
 \~\tau_{xz}(z)=-\nu\,S_{_U} +\tau_{xz}   \Rightarrow
 \~\tau_{xz}(0)\=- \tau_* \,, \\
\label{Sim-Teta-Avg}%%
  \~ F_z(z)=-\chi\,S_{_\Theta} + F_z  \Rightarrow \~F_z(0)\= - F_*\ . %%
\end{eqnarray}\end{subequations} %% ===============================
The total flux of the $x$-component of the
mechanical moment in $z$-direction is %%
$ \rho_b\~\tau_{xz} (z)  \=   \int dz ({\partial \< p\>/\p x}) +
\mbox{const}$.  Generally speaking, $\~\tau_{xz} (z)$ depends on
$z$. For example, for the pressure driven planar channel flow (of
the half-wight $\delta$) $\rho_b\~\tau_{xz} (z)   =
 (\partial \< p\>/\p x)(\delta-z)<0$.

Relatively close to the ground, where $z\ll \delta$, the $z$
dependence of $\~\tau_{xz}(z)$ can be neglected. In the absence of
the mean horizontal pressure drop and spatial distributed heat
sources $\~\tau$ and $\~F$ are $z$-independent, and thus equal to
their values at zero elevation, as indicated in Eqs.~\eq{TBL-Sim}
after ``$\Rightarrow$"-sign. Notice, that in our case of stable
stratification both vertical fluxes, the $x$-component of the
mechanical momentum, $\~\tau_{xz}$, and the thermal flux, $\~F_z$,
are directed toward the ground, i.e. negative. For the sake of
convenience, we introduce in Eqs.~\eq{TBL-Sim} notations for their
(positive) zero level modulus: $\tau_*$ and $F_*$.

Recall that in the plain geometry $U_z=0$. Nevertheless one can
write the equation for $U_z$:  %%
\begin{equation} \label{Sim-U-z}%%
 d \(\tau_{zz} +{\< p \>/\rho_b}\)/d z   = \beta\,\Theta \,,%%
\end{equation}
 which describes a turbulent correction ($\propto \tau_{zz}$) to the
 hydrostatic equilibrium~\eq{heq}. Actually, this equation determines
 the profile of $\< p\>$, that does not appear in the system of
balance equations ~\eq{TBL-Sim}.
%%%%%%%%%%%%%%%%%%%%%%%%%%%%%%%%%%%%%
\subsubsection{Equations for the pair (cross)-correlation functions}
  Consider first
  the balance Eqs.~\eq{corra} for the
diagonal components of the Reynolds-stress tensor in algebraic
model (which arises when we neglect the spatial fluxes): %%
\begin{eqnarray}\nn%%
&&   \G  E\Sb K +3 \gamma\Sb{RI} \tau_{xx} /2 = - \Big(3-
2\,C\Sb{IP} \Big)\tau_{xz}S_{_U} - C\Sb{IC}\,\beta \,F_{z}\,,~~~~ \\
\label{taup}  %%
&& \G E\Sb K +3\gamma\Sb{RI} \tau_{yy}/2   = -
C\Sb{IP}\tau_{xz}S_{_U} - \,C\Sb{IC}\,\beta \,F_{z}\,,\br
 &&  \G E\Sb K + 3\gamma\Sb{RI} \tau_{zz} /2 =
-C\Sb{IP}\tau_{xz}S_{_U} +\Big(3 +2\,C\Sb{IC} \Big)\beta \,F_{z}\
.%%
\end{eqnarray}%%
where $\G\=\g_{uu}-\g \Sb{RI}$.
 The LHS of these equations include the dissipation and
Return-to-isotropy terms. On the RHS  we have the    kinetic energy
production and isotropization of production terms (both proportional to
$S_{_U}$) together with the conversion and isotropization of
conversion terms, that are proportional to the vertical thermal flux
$F_z$. The horizontal component of the thermal flux,  $F_x$, does
not appear in these equations.

Equations~\eq{taup} allow to find anisotropy of the
turbulent-velocity fluctuations and to get  the balance Eqs. for the
turbulent kinetic energy with the energy production and conversion
terms on the RHS: %%
 \begin{subequations}\label{tau1} \begin{eqnarray}%%
3\tau_{xx} &=& 2 \big\{ [2(1 -C\Sb{IP})\Gamma_{uu}/\gamma\Sb{RI} +1
] E\Sb K \br  &&  -(3 -2C\Sb{IP} +C\Sb{IC})\, \beta F_z /
\gamma\Sb{RI} \big\} \,,
\\ %%
%%%%%%%%%%%%%%%%%%%
3\tau_{yy} &=& 2 \big\{ [(C\Sb{IP}-1)\Gamma_{uu}/\gamma\Sb{RI}  +1 ]
E\Sb K \br  &&
-(C\Sb{IP} +C\Sb{IC}) \beta F_z/\gamma\Sb{RI}  \big\} \,, \\
%%
%%%%%%%%%%%%%%%%%%%
3\tau_{zz} &=& 2 \big\{ [\(C\Sb{IP}-1\)
 \Gamma_{uu}/\gamma\Sb{RI}  +1 ]E\Sb K \br  &&   -
(C\Sb{IP} -2C\Sb{IC} -3 ) \, \beta F_z/\gamma\Sb{RI}\big\} \,, %%
 \\ \label{bale}%%
 \Gamma_{uu}E\Sb K &=&   -\tau_{xz}S_{_U}+ \beta F_z \,,%%
\end{eqnarray} %%
Equation~\eq{bale} includes   the only non-vanishing tangential
(off-diagonal) Reynolds stress $\tau_{xz}$ and has to be accompanied
with an equation for this object:
  \begin{equation}
\label{txz}%%
\~ \g\Sb {RI} \tau_{xz}  =  \big(C\Sb{IP} -1\big)\tau_{zz}\,S_{_U}
+\big(1 +C\Sb{IC}\big)\beta\,F_x \ . \end{equation} %%
\end{subequations}%%
\begin{subequations}%%
This equation manifests  that the tangential Reynolds stress
$\tau_{xz}$, that determines the energy production [according to
\Eq{bale}], influences, in its turn, on the value of the
streamwise thermal flux $F_x$,  which therefore  effects on the
turbulent kinetic energy production.%%

 As we mentioned, in the plain geometry $\~ F_y=0$.
  Equations~\eq{corrb} for the  $F_x$ and $F_z$  in this case take the form: %%
\begin{eqnarray}  \label{Sim-F-x}%%
\gamma\Sb{RD} F_x &=& -\(\tau_{xz} S_{_\Theta} +C\Sb{SU} F_z
S_{_U}\)\,, \\ %%
\label{Sim-F-z}%%
\gamma\Sb{RD} F_z &=& -\(\tau_{zz} S_{_\Theta} +2\,C\Sb{E\Theta}\,
\beta E_{\theta}\)\,,  %%
\end{eqnarray} %%
in which the RHS describes the thermal-flux production, corrected by
the isotropization of production terms.

The last   \Eq{corrc} for $E_{\theta}$,   represents the balance
between the dissipation (LHS) and  production (RHS):
\begin{equation} \label{Sim-ET}%%
 \gamma_{\theta\theta}\,E_\theta = -F_z\,S_{_\Theta}\ .%%
\end{equation}%%
\end{subequations} %%

\subsection{\label{ss:closure} Dimensional closure of time-scales and
the balance equations in the turbulent region}

At this point we follow tradition in modeling of all the nonlinear
inverse time-scales by dimensional estimates (Kolmogorov, 1941): %%
\begin{eqnarray} \label{freqs}%%
\g_{uu}&=&c_{uu} \sqrt{E\Sb K} \big / \ell\,, \quad \g\Sb {RI}= C\Sb
{RI} \g_{uu}\,,  \br \~ \gamma\Sb {RI}& = &  \~ C\Sb {RI} \gamma\Sb
{RI}\,, \gamma_{\theta\theta}  = C_{\theta\theta} \g_{uu}\,, \quad
\gamma\Sb{RD}= C_{u\theta}\g_{uu}\ .%%
\end{eqnarray}%%
Remember that  $\ell $ is the ``outer  scale of turbulence". This
scale equals to $z$ for $z< L$, where $L$ is the Monin-Obukhov
length (definition is found below).

Detailed analysis of experimental, DNS and LES data (see
L'vov~et~al.,~2006, and references therein) shows that for
unstratified flows, $\textrm{g}=0$, the anisotropic boundary
layers exhibits values of the Reynolds stress tensor that can be
well approximated by the values $\tau_{xx}=E\Sb K$,
$\tau_{yy}=\tau_{zz}=E\Sb K/2$. In our approach this dictates the
choice $C\Sb{RI} = 4(1-C\Sb{IP})$. Also we can expect that
$\tau_{yy}$ is almost not affected by buoyancy. This gives simply
$ C\Sb{IC} = -C\Sb{IP}$.  If so, Eqs.~\eq{tau1} with the
parametrization~\eq{freqs} can be identically rewritten as
follows: %%
\begin{subequations}\label{simple}
\begin{eqnarray} \nn %%
&&   \tau_{xx} =   E\Sb K -\frac{\beta F_z}{2\,\gamma_{uu}}\,,
\quad%%
\tau_{yy}  = \frac{E\Sb K}2\,,\quad  %%
\tau_{zz}  = \frac{E\Sb K}2 +\frac{\beta F_z}{2\,\gamma_{uu}}\,, \\
\label{simpleA}
 &&  \gamma_{uu}E\Sb K = \beta F_z -\tau_{xz}S_{_U} \,, \quad
\g_{uu}=c_{uu} \sqrt{E\Sb K} \big / \ell \,, \br &&  ~~~~~~~~~~~
4\,\~C\Sb{RI}\,\gamma_{uu} \tau_{xz} = \beta\,F_x -\tau_{zz}\,S_{_U}
\  .%%
\end{eqnarray}  %%
For  completeness we also repeated here the
parametrization~\eq{freqs} of $\g_{uu}$. Finally we present the
version of the balance Eqs. for the thermal flux \eq{Sim-F-x},
\eq{Sim-F-z}, and for the ``temperature energy", \eq{Sim-ET}, after
all the simplified assumptions:
\begin{eqnarray} \nn
C_{\theta\theta}\,\gamma_{uu}E_\theta &=& -F_z\,S_{_\Theta}\,, \\
\label{simpleB} %%
C_{u\theta}\,\gamma_{uu} F_x &=& -\(\tau_{xz}S_{_\Theta}
 +C\Sb{SU}F_z\, S_{_U}\)\,,  \br%%
C_{u\theta}\,\gamma_{uu} F_z &=& -\(\tau_{zz}  S_{_\Theta}
 +2\,C\Sb{E\Theta}\,\beta  E_{\theta}\)\ .%%
\end{eqnarray} \end{subequations}  %%

%%%%%%%%%%%%%%%%%%%%%%%%%%%%%%%%%%%%%%%%%%%
\subsection{\label{ss:wall-units}Generalized wall normalization}%%
The analysis of the balance Eqs.~\eq{simple}   is drastically
simplified if they are presented in dimensionless form.
Traditionally, the conventional ``wall units" are introduced via the
wall friction velocity, $u_*$, and the viscous length-scale,
$\lambda_*$, defined by $u_*  \=  \sqrt{ \tau_*}$, $ \lambda_*
 \=   {\nu}/{u_*}$.
In addition to  $u_*$, and   $\lambda_*$, we introduce a new wall
unit for the temperature, $\theta_*\= {F_*}/{u_*}$, defined via
the thermal flux at the wall  and friction velocity. Subsequently,
$ \B r^+ \= { \B r}/{\lambda_*}$,     $t^+ \=  {t\, \lambda_*}/{
u_*}$, $\BC U^+ \=   {\BC U}/{ u_*}$, $p^+ \= p/\rho_b\,u_*^2$, $
\overline{\Theta}^+   \=   {\overline{\Theta}}/{\theta_*}$, $
\theta^+ \=  {\theta}/{\theta_*}$,  etc. Then the governing
Eqs.~\eq{GovEq} take the form:%%
\begin{eqnarray}\nn %%
 {\C D^+ \, \BC U^+}/{\C D t ^+ } + \B {\nabla}^+ p^+  &=& %%
 { \z}\,\Theta^+\sb {\,d}/{L^+} \ +\Delta^+\, \BC U^+ \,, \\
 \label{NSE-dim}
 {\C D^+ \,   \Theta^+\sb {\,d}}/{\C D t ^+ } &=&  %%
  \Delta^+\, \Theta^+\sb {\,d}/{\textrm{Pr}} \ .%%
\end{eqnarray}  %%
%%%
These Eqs. include two dimensionless parameters: the conventional
Prandtl number Pr$=\nu\big / \kappa$, and $L^+$ -- the
Monin-Obukhov length $L$ measured in wall units: $L\= u_*^3\big /
\beta F_*$, $L^+\=L\big / \lambda_*$.
 Following~ZEKR-paper, we used here the modern
definition of the Monin-Obukhov length, which differs from the old
definition by the absence of the \vK constant $\kappa\approx 0.436$
in its denominator (Monin and Obukhov, 1954).

 Outside of the viscous sub-layer, where the  kinematic viscosity
and kinematic thermal conductivity can be ignored,  $L^+$ is the
only dimensionless parameter in the problem, which separates the
region of weak stratification, $z^+< L^+$, and the region of strong
stratification, where $z^+>L^+$.

Given the generalized wall normalization we introduce objects
with a superscript  ``~$^+$"
 in the usual manner:
 \begin{eqnarray}\nn
  S_{_U}^+ &\=&  t_*\, S_{_U}\,, \
S_{_\Theta}^+\= {\lambda_* \, S_{_\Theta}}/{\theta_*}\,, \
\gamma^+\= t_* \gamma\,, \ \tau_{ij}^+ \=   {\tau_{ij}}/{u_*^2}\,,\\
 \label{plus-m}\B F^+ &\=&
 {\B F}/{u_* \theta_*}\,,\ E_\theta^+ \=
 {E_\theta}/{\theta_*^2}\ .\end{eqnarray}
In the turbulent region, governed by $L^+$ only,   Eqs.~\eq{TBL-Sim}
simplify to $ \tau^+_{xz}  = -1$, $F^+_z = -1$. Then, in the wall
units Eqs.~\eq{simple} can be reduced  to Eqs.~\eq{eqs-plus} for the
six profiles $S^+_{_U}$, $S^+_{_\Theta}$, $E^+\Sb K$, $E^+_{\theta}$
and $ F^+_x$ and $\tau^+_{zz}$, presented in the
Appendix~\ref{ss:plus}. This equation   can be effectively analyzed,
see next Sec.~\ref{ss:strat41}.  %%
%%
%%%%%%%%%%%%%%%%%%%%%%%%%%%%
\subsection{\label{sss:resc}Rescaling symmetry and
$\ddag$-representation} Outside of the viscous region, where
Eqs.~\eq{simple} were derived, the problem has only one
characteristic length, the Monin-Obukhov scale  $ L$.
Correspondingly, one expects that the only dimensionless parameter
that governs the turbulent statistics in this region should be the
ratio of the outer scale of turbulence, $\ell$, to the Monin-Obukhov
length-scale $L$, which we denote as $ \ell^\ddag\=
 {\ell}/{L}=   \ell^+ / L^+  $.
Indeed, introducing ``$\ddag$-objects": %%
\begin{equation}\label{ddagb} %%
 \ell^\ddag\=
 {\ell}/{L}\,, \quad
 S_{_U}^\ddag\= S_{_U}^+\ell^+\,, \  S_{_\Theta}^\ddag\=
S_{_\Theta}^+\ell^+\,,  \end{equation} %%
and using Eqs.~\eq{plus-m} one rewrites the balance Eqs.~\eq{simple}
as follows: %%
\Fbox{
\begin{subequations} \label{MM}%%
\begin{eqnarray} \label{9-eqA}%%
&& \hskip -1.3 cm \tau^+_{xx} =  E^+\Sb K + {\ell^\ddag  } /2
c_{uu}
\sqrt{E^+\Sb K}\,, \quad \tau^+_{yy}  =  {E^+\Sb K}/2\,,\\
\label{9-eqB}%%
&&  \hskip  -1.3 cm 2\, \tau^+_{zz}= {E^+\Sb K}  -{\ell^\ddag
 } / c_{uu} \sqrt{E^+\Sb K}\,,~~~~ \\ \label{9-eqC}
 &&  \hskip  -1.3 cm c_{uu}  { E^+\Sb K}^{3/2} =  \ell^\ddag F^+_z -\tau^+_{xz}
S^\ddag_{_U} \,, \\ \label{9-eqD}%%
\label{not-used}%%
&&  \hskip  -1.3 cm 4\,\~C\Sb{RI}\,c_{uu} \sqrt{E^+\Sb K}
\tau^+_{xz} =\ell^\ddag F^+_x -\tau^+_{zz}\,S^\ddag_{_U} \,, \\
\label{9-eqE}%%
&&  \hskip  -1.3 cm  C_{\theta\theta}\,c_{uu} \sqrt{E^+\Sb K}
E^+_\theta = -F^+_z\,S^\ddag_{_\Theta}\,, \\ \label{9-eqF}%%
&&  \hskip  -1.3 cm C_{u\theta}\,c_{uu} \sqrt{E^+\Sb K} F^+_x = -
\tau^+_{xz}S^\ddag_{_\Theta} -C\Sb{SU}F^+_z\, S^\ddag_{_U} \,,
\\ \label{9-eqG}%%
&&  \hskip  -1.3 cm  C_{u\theta}\,c_{uu} \sqrt{E^+\Sb K} F^+_z = -
\tau^+_{zz}S^\ddag_{_\Theta} - 2\,C\Sb{E\Theta}\,\ell^\ddag
E^+_{\theta} \,,%%
\end{eqnarray}\end{subequations}}%%
These equations are the main result of  current
Sec.~\ref{s:closure}. It may be considered as ``Minimal Model" for
stably stratified TBL, that respects the conservation of energy,
describes anisotropy of turbulence and  all relevant fluxes
explicitly and, nevertheless is still simple enough to allow
comprehensive analytical analysis, that results in approximate
analytical solution (with reasonable accuracy) for the mean
velocity and temperature gradients $S\Sb U$ and $S\Sb \Theta$,
and all second-order (cross)-correlation functions.%%

As expected, the only parameter appearing in the Minimal
Model~\eq{MM}   is  $\ell^\ddag$. The outer scale of turbulence,
$\ell$, does not appear by itself, only via the definition of
$\ell^\ddag$ ~\eq{ddagb}. Therefore our goal now is to solve
Eqs.~\eq{MM} in order to find five functions of only one argument $
\ell^\ddag $: $S^\ddag _{_U}$, $S_{_\Theta}^\ddag$, $E^+\Sb K$,
$E_{\theta}^+$ and $ F^+_x$. After that we can specify the
dependence $\ell^+(z^+)$ and then reconstruct the $z^+$-dependence
of these five objects. %%

\section{\label{s:res}  Results and discussion}

%%%%%%%%%%%%%%%%%%%%%%%%%%%%%%%%%
\subsection{\label{ss:strat41} Analytical solution of the Minimal-Model
balance equations~\eq{MM}}
\begin{figure*}
 \includegraphics[width=0.495  \textwidth]{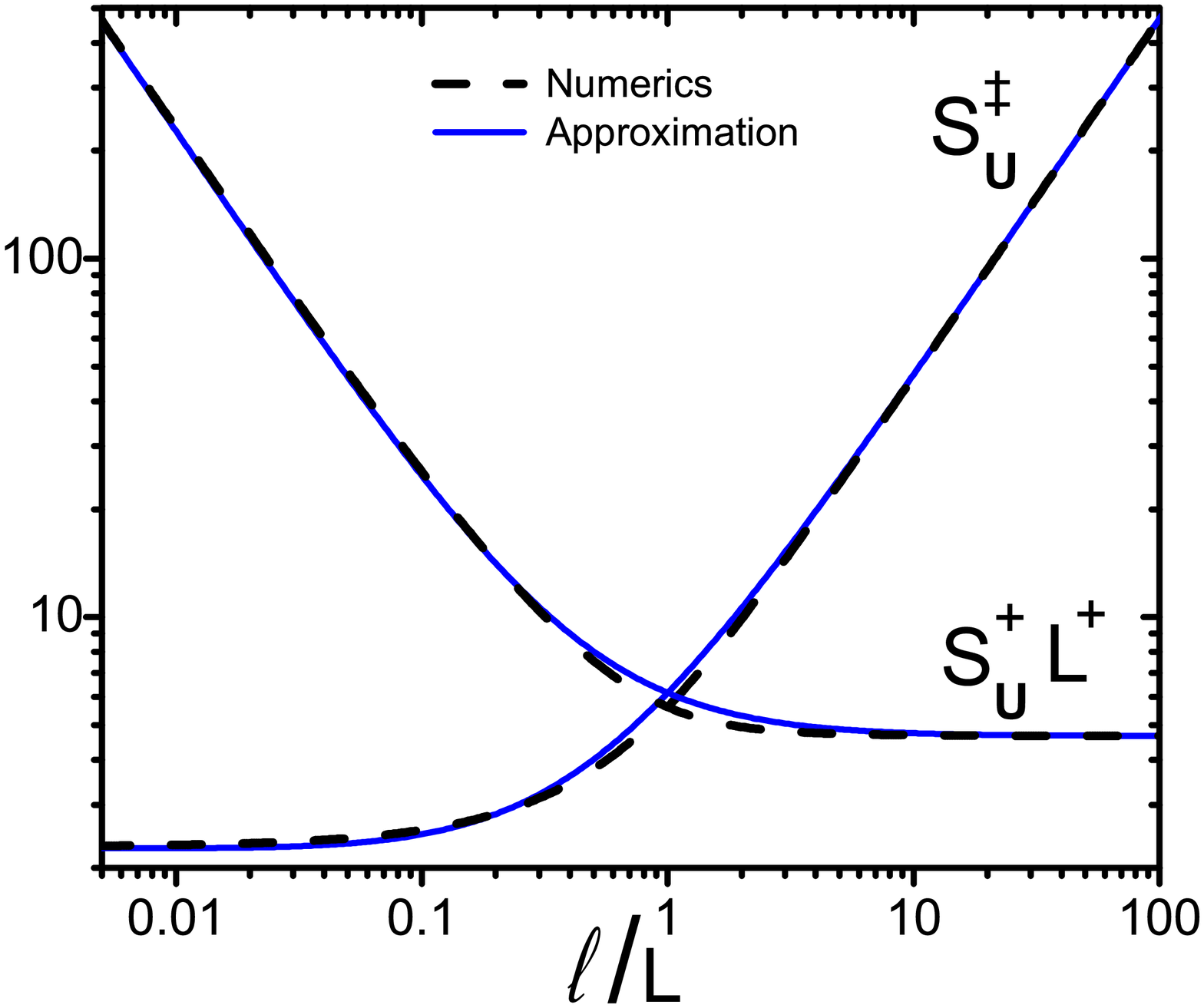}
     \includegraphics[width=0.495  \textwidth]{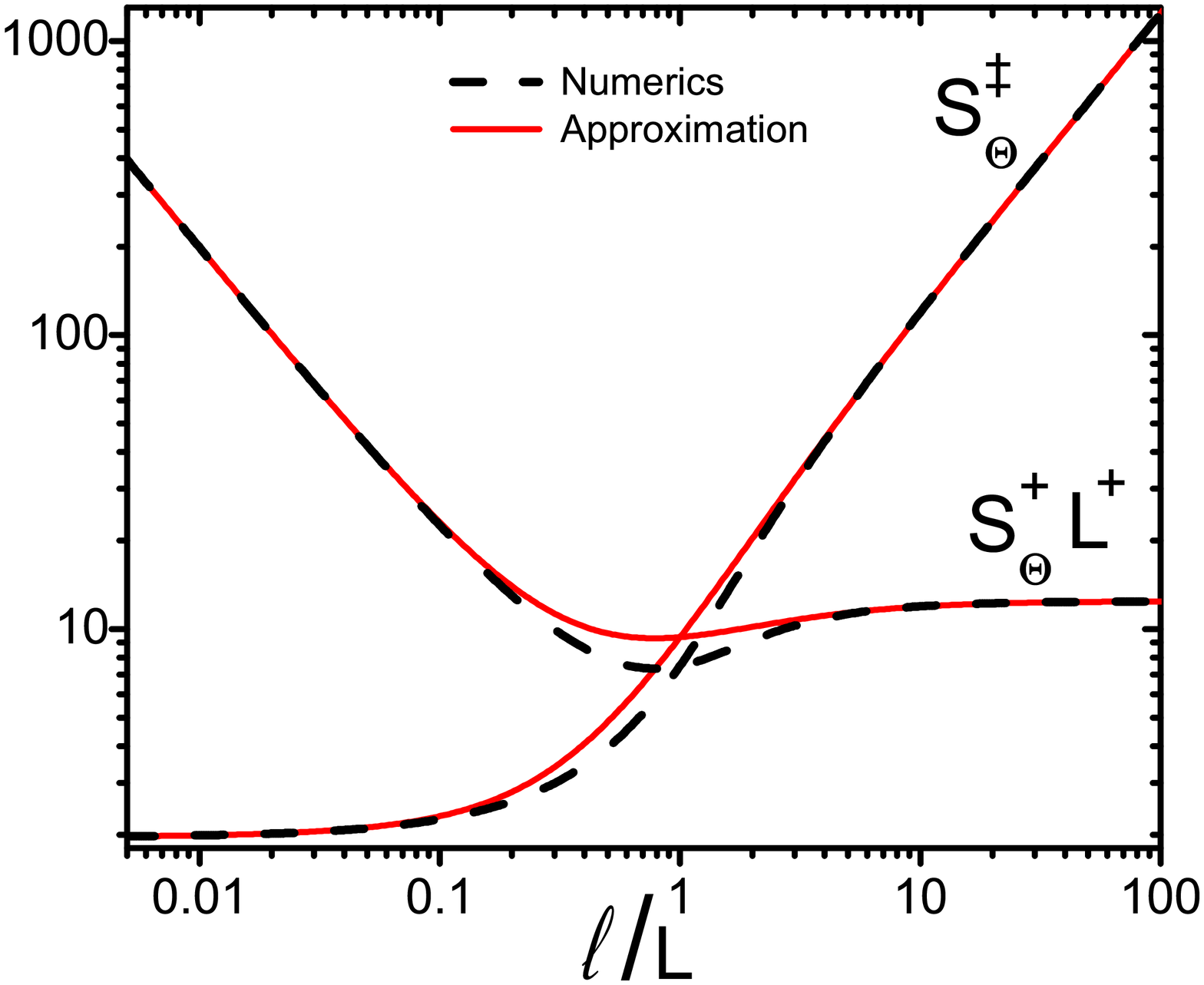} \vskip -0.6cm
      \includegraphics[width=0.495  \textwidth]{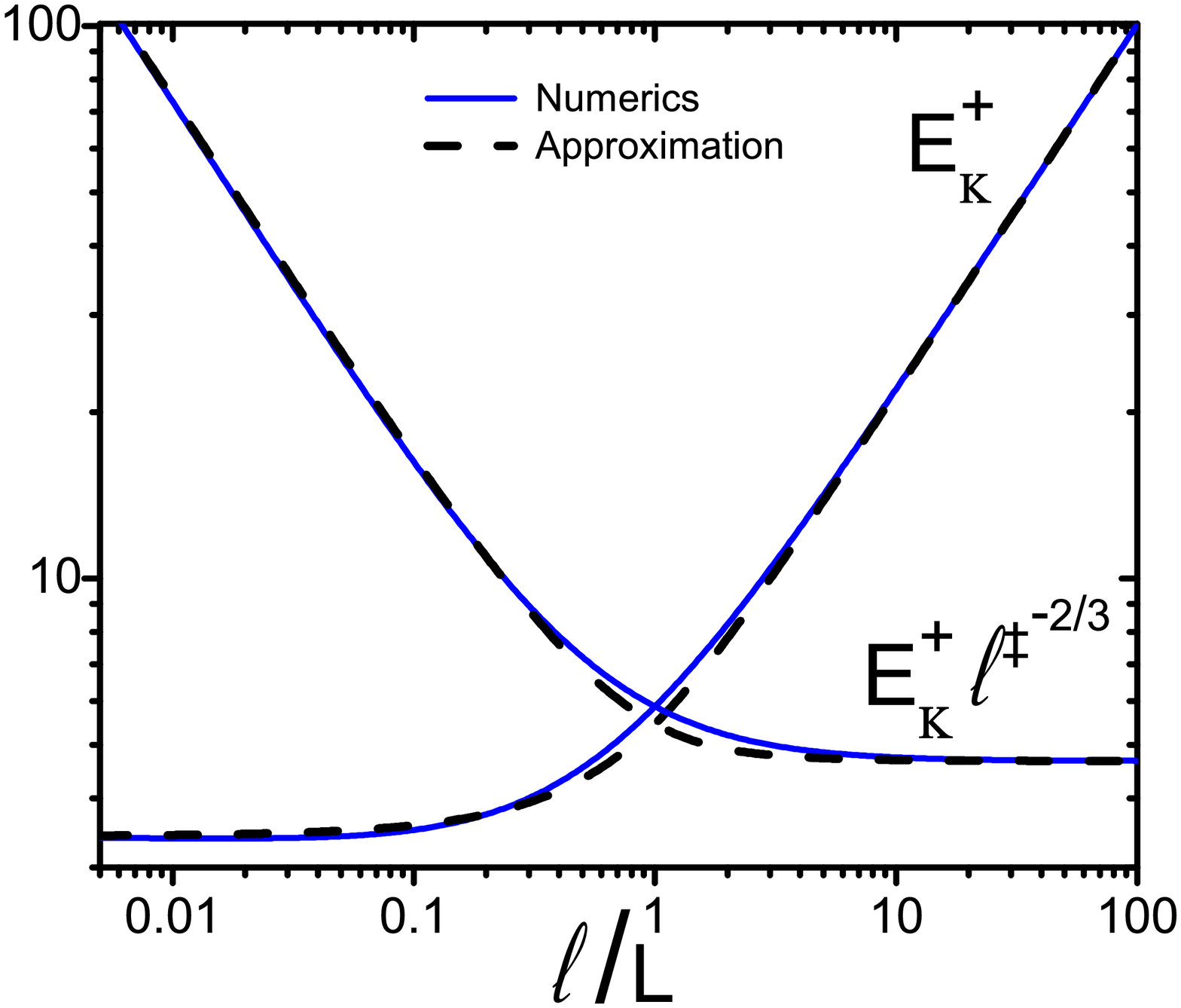}
 \includegraphics[width=0.475  \textwidth]{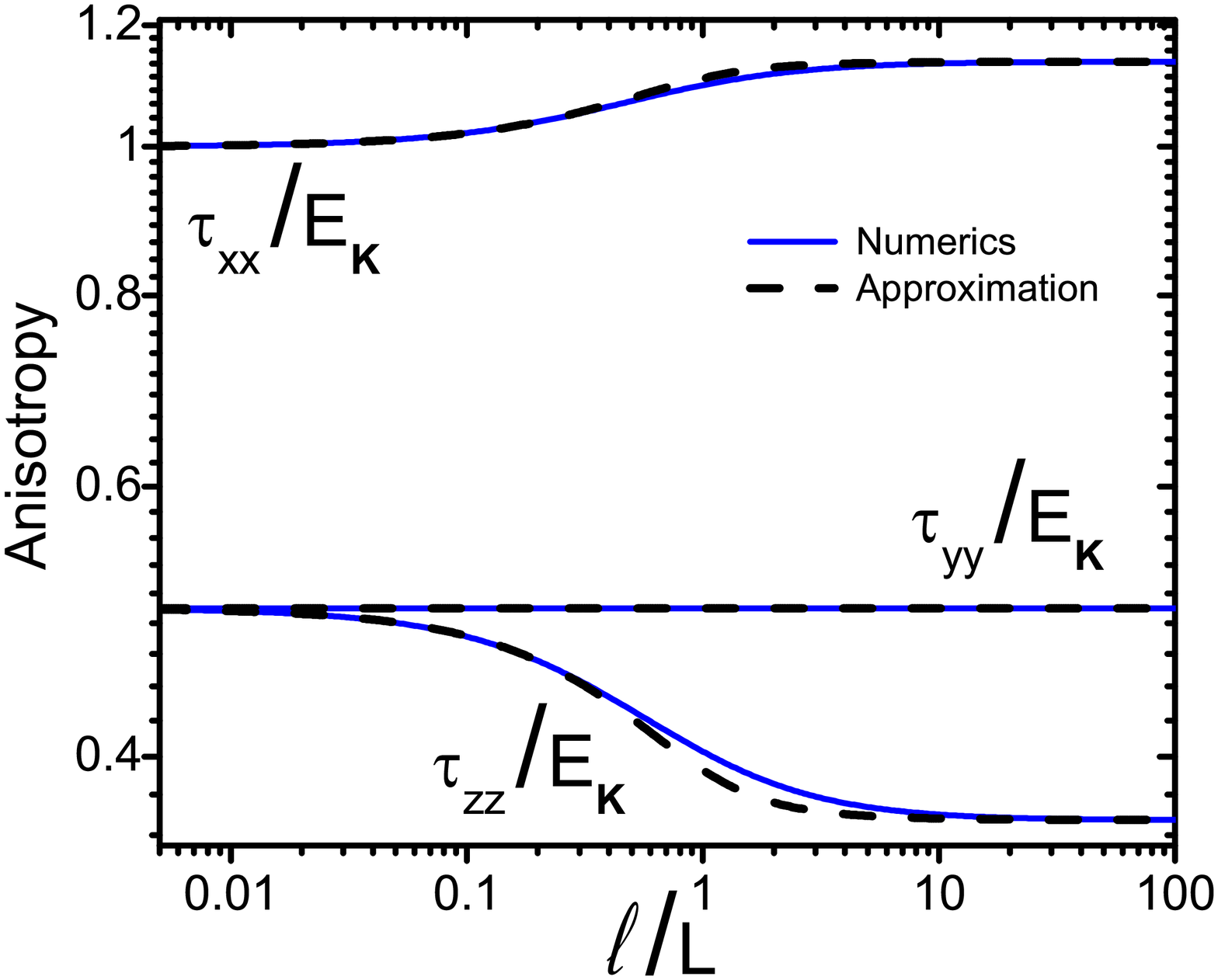} \vskip -0.6cm
  \centering\includegraphics[width=0.495  \textwidth]{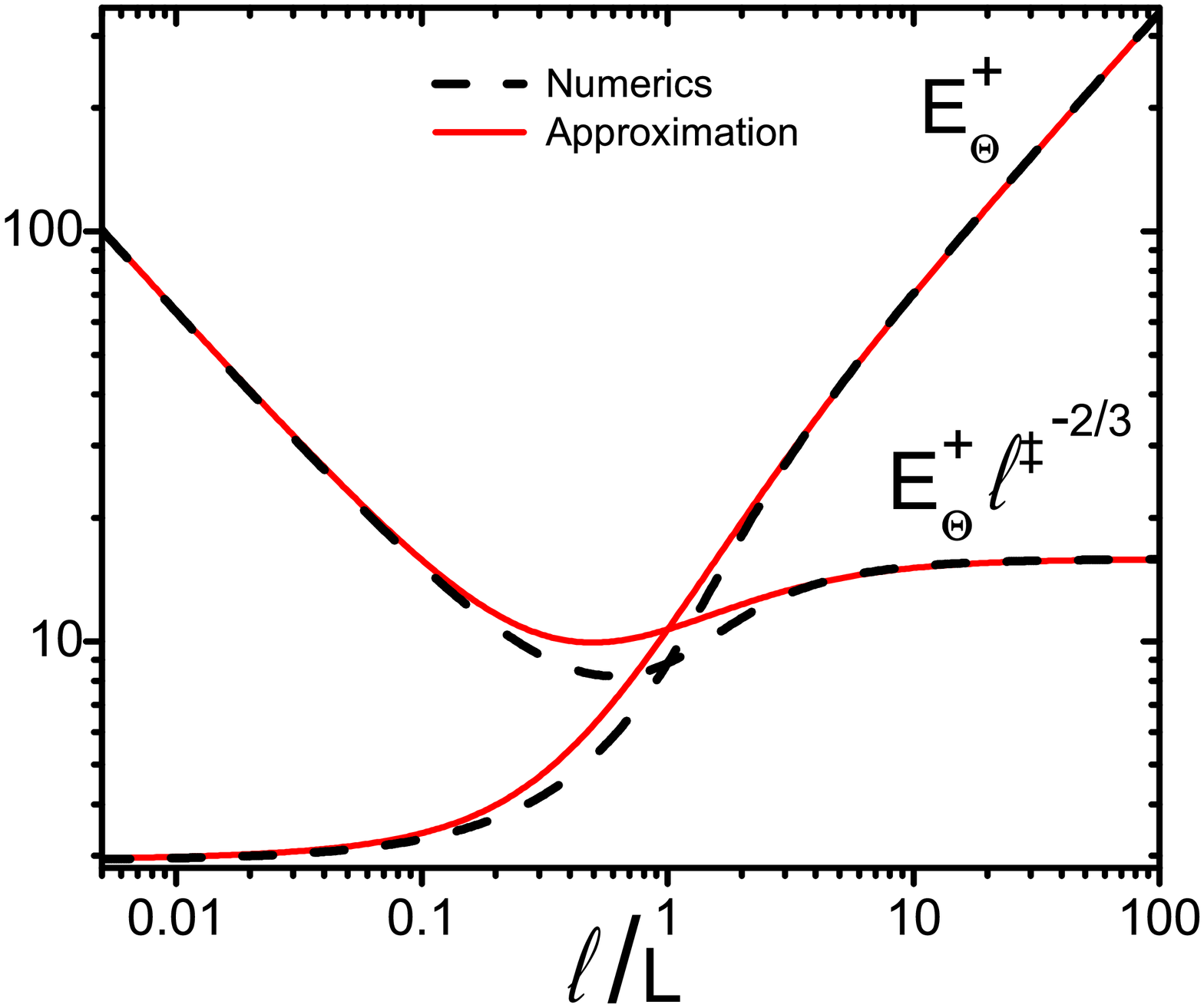}
     \centering\includegraphics[width=0.495 \textwidth]{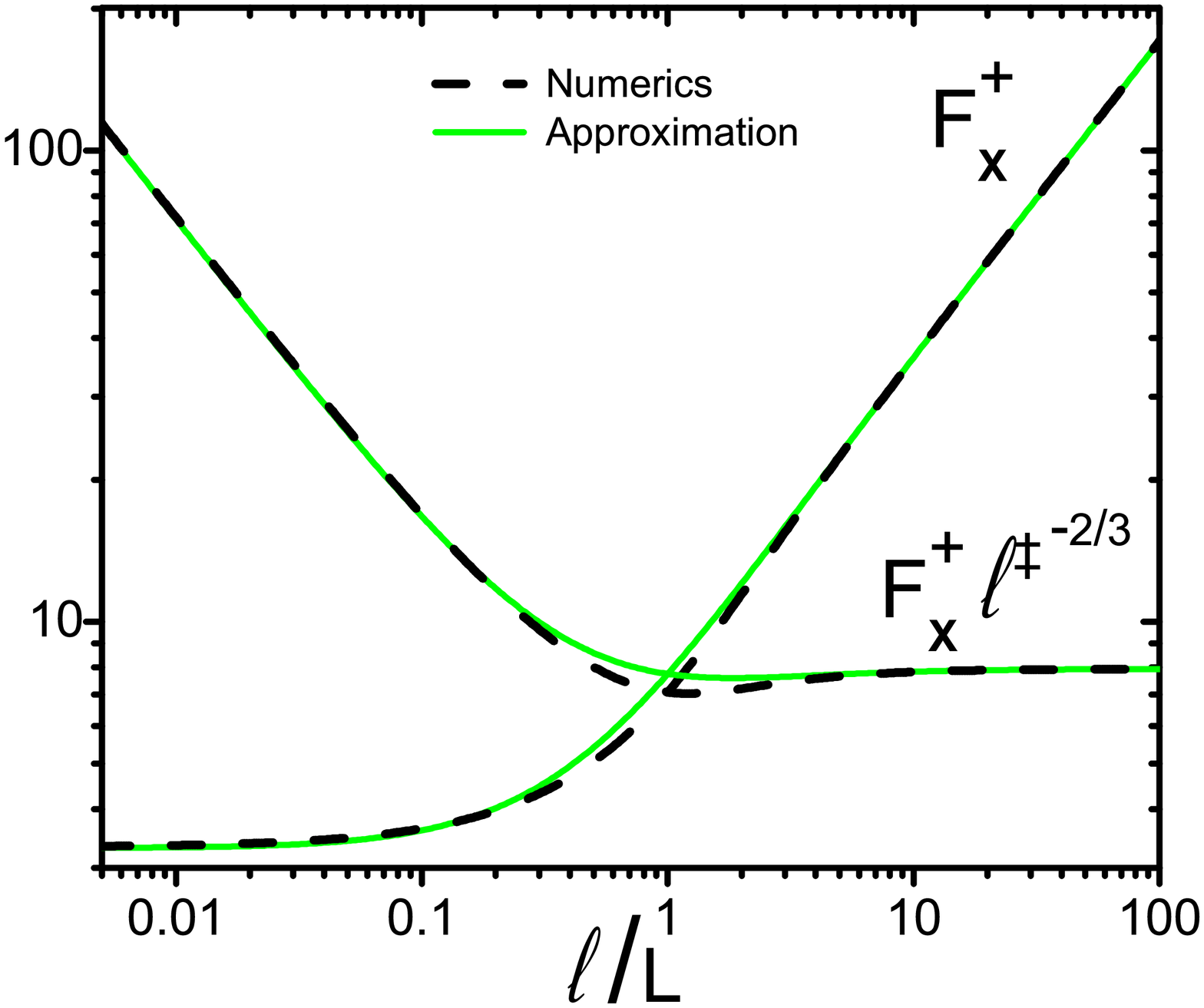}

  \caption{\label{f:S} Color online. Log-log plots of
  the normalized velocity mean shears $  S\Sb U
  ^\ddag \equiv  \lp\, S\Sb U^+$ and $L^+ S\Sb U^+$
  (left upper panel),  normalized mean-temperature
  gradients $ S\Sb \Theta ^\ddag \equiv   \lp\,S\Sb
  \Theta ^+ $ and  $L^+ S\Sb \Theta ^+$ (right upper
  panel), the turbulent kinetic energy  $E\Sb K^+$
  and $E\Sb K^+  /{\ell^\ddag}^{2/3} $ (left middle
  panel), partial kinetic energies $\tau_{ii}/E\Sb K$
  (left middle panel),   temperature energy $E_
  \theta ^+ $ and   $E_ \theta ^+/{\ell^\ddag}^{2/3}$
  (left lower panel) and horizontal thermal flux
  $F_x^+$ and   $F_x^+  /{\ell^\ddag}^{2/3} $ (right
  lower panel) vs. $\ell^\ddag=\ell/L=\lp /L^+$.
  Red and blue solid lines -- exact numerical
  solutions before and after normalization by the
  large $\ell^\ddag$ asymptotics, black dashed and
  dot-dashed lines -- approximate analytical
  solutions.  The region $\ell \gtrsim L$ may not be
  realized in the Nature. In this case it has only
  methodological character.    }
\end{figure*}

 This Subsec. is devoted to analytical and numerical analysis of the
Minimal-Model~\eq{MM}. Example of numerical solution of Eqs.~\eq{MM}
(with some reasonable choice of the phenomenological parameters) is
shown in Fig.~\ref{f:S}. Nevertheless,  it would be much more
instructive to have approximate analytical solutions for all
correlations that will describe their $\ell^\ddag$-dependence with
reasonable accuracy. The detailed procedure of finding these
solutions is presented in Appendix~\ref{ss:gen}. A brief overview is
as follows.

 Appendix~\ref{ss:plus} presents the
balance Eqs.~\eq{simple} in generalized  wall units~\eq{plus-m}.
Analysis of the resulting Eqs.~\eq{eqs-plus} allows us to clarify
the rescaling symmetry and to suggest ``~$^\ddag$~"
normalization~\eq{ddagb}, in which the balance Eqs.~\eq{simple} take
final even simple form~\eq{MM}, which is the basis for further
analysis.

In Ap.~\ref{ss:poly} we show that Eqs.~\eq{MM} can be reformulated
as polynomial Eq.~\eq{poly} of ninth order for the only unknown
$\sqrt{E^+}$. Analysis of its structure helps to formulate effective
interpolation formula~\eq{inter}, discussed below.

Next,  in Ap.~\ref{ss:Ri0} we find the solutions~\eq{sol1} of
Eqs.~\eq{MM} at neutral stratification, $\ell^\ddag=0$, corrected by
Eqs.~\eq{small-Ri} up to linear order in $\ell^\ddag$. Its
comparison with the existing DNS data results in an estimate for the
constants $\~ C\Sb{RI}\approx 1.46$, and $c_{uu}\approx 0.36$, see
L'vov at el. (2006) and references therein.

 Then, in Ap.~\ref{ss:Ri-infty} we
considered the region $\ell^\ddag \to \infty$. Even though such a
condition may not be realizable in nature, from a methodological
point of view, as we will see below, it enables to obtain the
desired analytical approximation. The $\ell^\ddag \to \infty$
asymptotic solution~\eq{sol2} with corrections~\eq{pert}, linear in
the small parameter $\d =2(c_{uu}/\ell^\ddag)^{4/3}$ is found. Now
we are armed to suggest an interpolation formula %%
\BSE\label{inter} \Fbox{\begin{equation} \label{interA}%%
  {E^{+}\Sb K}(\ell^\ddag)^{3/2}\simeq \frac{11
\ell^\ddag}{3 \, c_{uu}} +\frac{8\,  \~C\Sb {RI}}{\sqrt{
\big(11\ell^\ddag/3\, c_{uu}\big)^{2/3}
+\big(8\,\~C\Sb{RI}\big)^{1/2}}}\,, %%
\end{equation} }%%
 that coincide
with the exact solutions for $\ell^\ddag=0$, ~\Eqs{sol1}, and for
$\ell^\ddag\to \infty$, ~\Eq{sol2},  including the leading
corrections to both asymptotics, linear in $\ell^\ddag$,
\Eqs{small-Ri} and $1/{\ell^\ddag}^{4/3}$, \Eqs{pert}. Moreover,
in the  region  $\ell^\ddag\sim 1$, \Eq{interA}  accounts for the
structure of exact polynomial \Eq{poly}. As a result, the
interpolation formula~\eq{interA} is close to the numerical solution
with deviations smaller than
 3\% in the entire region
$ 0\le \ell^\ddag  < \infty$, see left middle panel on
Fig.~\ref{f:S}.  Together with \Eq{sols-a} it produces a solution
for $S_{_U}^+$, that can be written as %%
\Fbox{\BE%%
 \label{interB}%
  S^+\Sb{U}(\lp) \simeq  \(L_1^+\)^{\!-1}+ \Big({\kappa\,
  \ell^+\sqrt{1+(\lp/L_2^+)^{2/3}}}\,\Big)^{\!-1}\,, %%
 \EE%%
} %%
where $ L_1^+ \=  3 L^+/14\,, \quad L_2^+\= 3L^+/{11\,\kappa} $ and
$\kappa$ is the \vK constant. This formula gives   the same accuracy
$ \sim3\%$, see left upper panel in Fig.~\ref{f:S}. We demonstrate
below that the proposed interpolation formulae describe the
$\ell^\ddag$-dependence of the correlations with a very reasonable
accuracy, about $  10\%$,  for any $0\leq\ell^\ddag<\infty$,    see
black dashed lines in Figs.~\ref{f:S}.

Unfortunately, a direct substitution of the interpolation
formula~\eq{inter} into the  exact relation for $S^\ddag\Sb \Theta$
obtained from the system (\ref{num}) (see~\eq{sols-b}) works well
only for small $ \ell^\ddag$, in spite of the fact that the
interpolation formula is rather accurate in the whole region.
 The reason is a small denominator in \Eq{sols-b} for large $
\ell^\ddag $. We need therefore to derive an independent
interpolation formula for $S^\ddag\Sb \Theta$. Using the expansions
\eq{exp} for small $ \ell^\ddag  \ll 1$ and \eq{pert} for large $
\ell^\ddag  \gg 1$ we suggest   %%
\Fbox{ \begin{eqnarray}\label{inter1a}%%
  S\Sb \Theta ^+ (\ell^+)  \simeq  {S\Sb \Theta ^+}^\infty
\!\! + \frac{S^+\Sb {\Theta,{\scriptstyle 0}}\!+\! 6
(c_{uu}\a)^{4/3}S\Sb {\Theta,1}^{+\infty}} {\(1 +\a\,\ell^+/L^+
\)^{4/3}}\,,~~~~~~%%
\end{eqnarray}}%%
in which
\begin{eqnarray}\nonumber
S^+ \Sb {\Theta,\scriptstyle 0}  &=& 2^{1/4}c_{uu} C\Sb{U\Theta}/
\~C\Sb{RI}^{1/4 }\ell ^+,\\ \nonumber%%
S^{+\infty}\Sb {\Theta,1} &=& -{C_{u\theta}}(2\~C\Sb{RI}
-{(11\,C_{u\theta} -3\,C\Sb{SU})}/{{3\,S\Sb \Theta ^{+\infty}}L^+}
)/{L^+},\\ \nonumber%%
S\Sb \Theta ^{ + \infty} &=& -14 (C\Sb{SU}- 4 C\Sb{U\Theta}/3)/3
L^+,%%
 \end{eqnarray}%%
 and $\alpha$ satisfies
 \begin{eqnarray}\nonumber%%
S^+\Sb {\Theta,{\scriptstyle 1}}\ell^+ &=& {S\Sb \Theta ^+}^\infty
L^+ + 6S\Sb {\Theta,1}^{+\infty}L^+ (c_{uu}\a)^{4/3} -{4 \a S\Sb
{\Theta,{\scriptstyle 0}}}/3 \,,%%
\end{eqnarray}%%
with
 \begin{eqnarray}\nonumber%%
 S^+\Sb {\Theta,
{\scriptstyle{1}}} \ell^+ &=& -{C_{u\theta}} \(3/{4\~C\Sb{RI}} -22
+3 {C\Sb{SU}}/{C_{u\theta}}\)/ {24\~C\Sb{RI}}\ .
\end{eqnarray}
\end{subequations}%%
 Equation~\eq{inter1a} is constructed such that the leading and
 sub-leading asymptotics for small and large $ \ell^\ddag $ coincide
 with  first two terms in the exact expansions "almost" neural stratification, \eq{exp},
 and extremely strong stratification, \eq{pert}. As a result \Eq{inter1a} approximates the exact solution
 with errors smaller then 5\% for $ \ell^\ddag < 1$ and $ \ell^\ddag > 50$
 and with errors smaller than 10\% for any $ \ell^\ddag $, see
 left upper  panel in \Fig{f:S}.

 Substituting the approximate Eqs. \eq{inter}   into
 the exact relations  \eq{sols-d} and \eq{sols-e} and one gets
 approximate solutions $E^+_\theta$ and
 $F_x^+$ with errors smaller than 10\%, see  lower panels
 in \Fig{f:S}.

%%%%%%%%%%%%%%%%%%%%%%%%%%%%%%%%%%%%%%%%%%%%%%%%%%%%%%%%%

%%%%%%%%%%%%%%%%%%%%%%%%%%%%%%%%%%%%%%%%%%%%%%%%%%%%%%%%%%%%%%%%%%%%%%%%%%%%%%

%%%%%%%%%%%%%%%%%%%%%%%%%%%%%%%%%%%%%%%%%%%%%%%%%%

\subsection{\label{ss:one} Mean velocity and temperature profiles}

In principle, integrating the mean shear $S\Sb U^+$ and the mean
temperature gradient $S\Sb \Theta ^+$, one can find the mean
velocity and temperature profiles. Unfortunately, to do so we need
to know $S\Sb U^+ $ and $S\Sb \Theta ^+$ as functions of the
elevation $z$, while in our approach they are found as functions of
$\ell/L$. Remember, that the external parameter $\ell$ is the outer
scale of turbulence that depends on the elevation $z$:
$\ell=\ell(z)$. For $z\ll L$ we can safely take $\ell=z$, however
for $z>L$  the function $\ell(z)$ is not found theoretically
although it was discussed phenomenologically with support of
observational, experimental and numerical data. It is traditionally
believed that for $z \gtrsim L$ the scale $\ell$  saturates at some
level of the order of $L$ [see, e.g. Eq. \Ref{sless}].

\begin{figure*}
 \includegraphics[width=0.475  \textwidth]{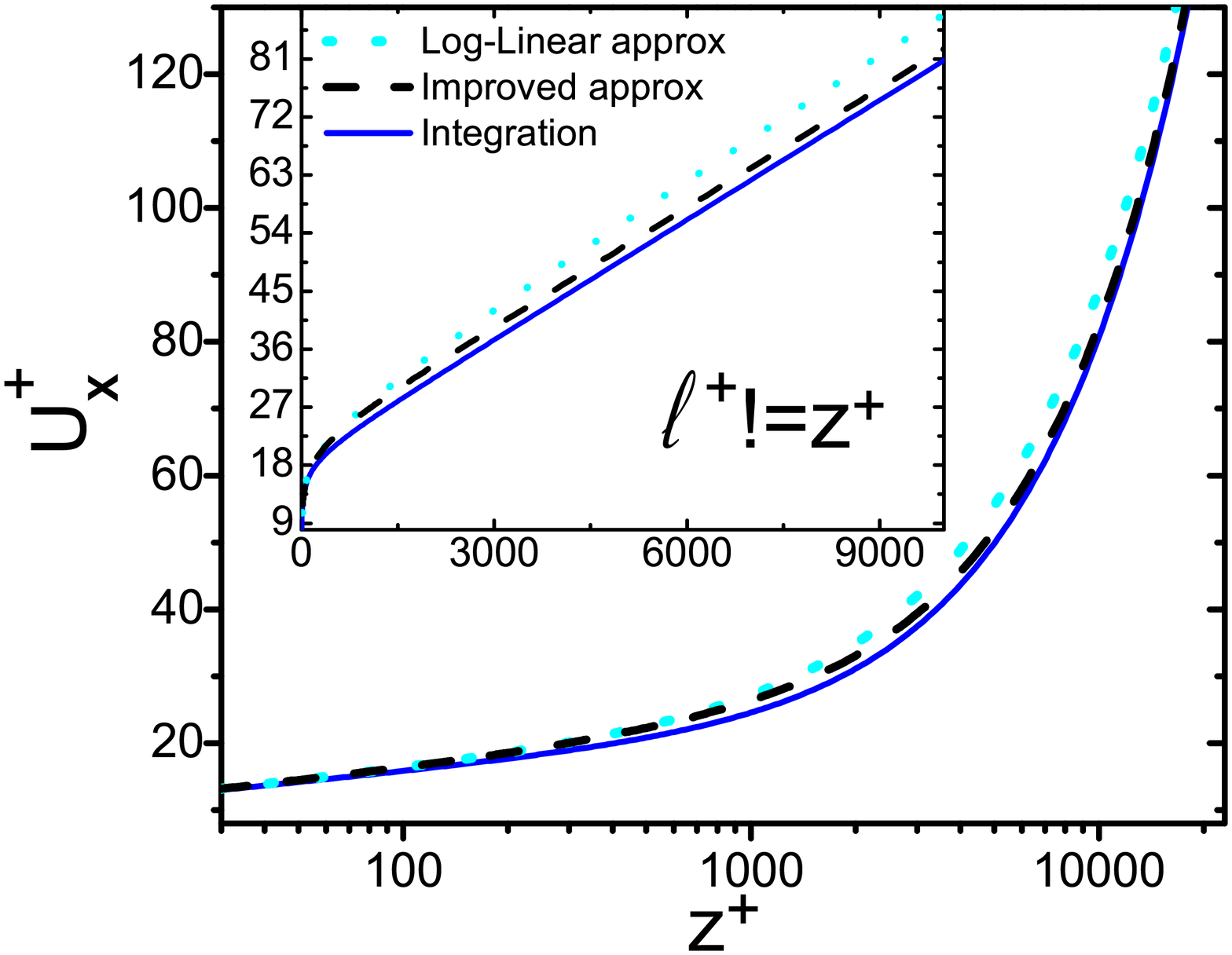}
 \includegraphics[width=0.45  \textwidth]{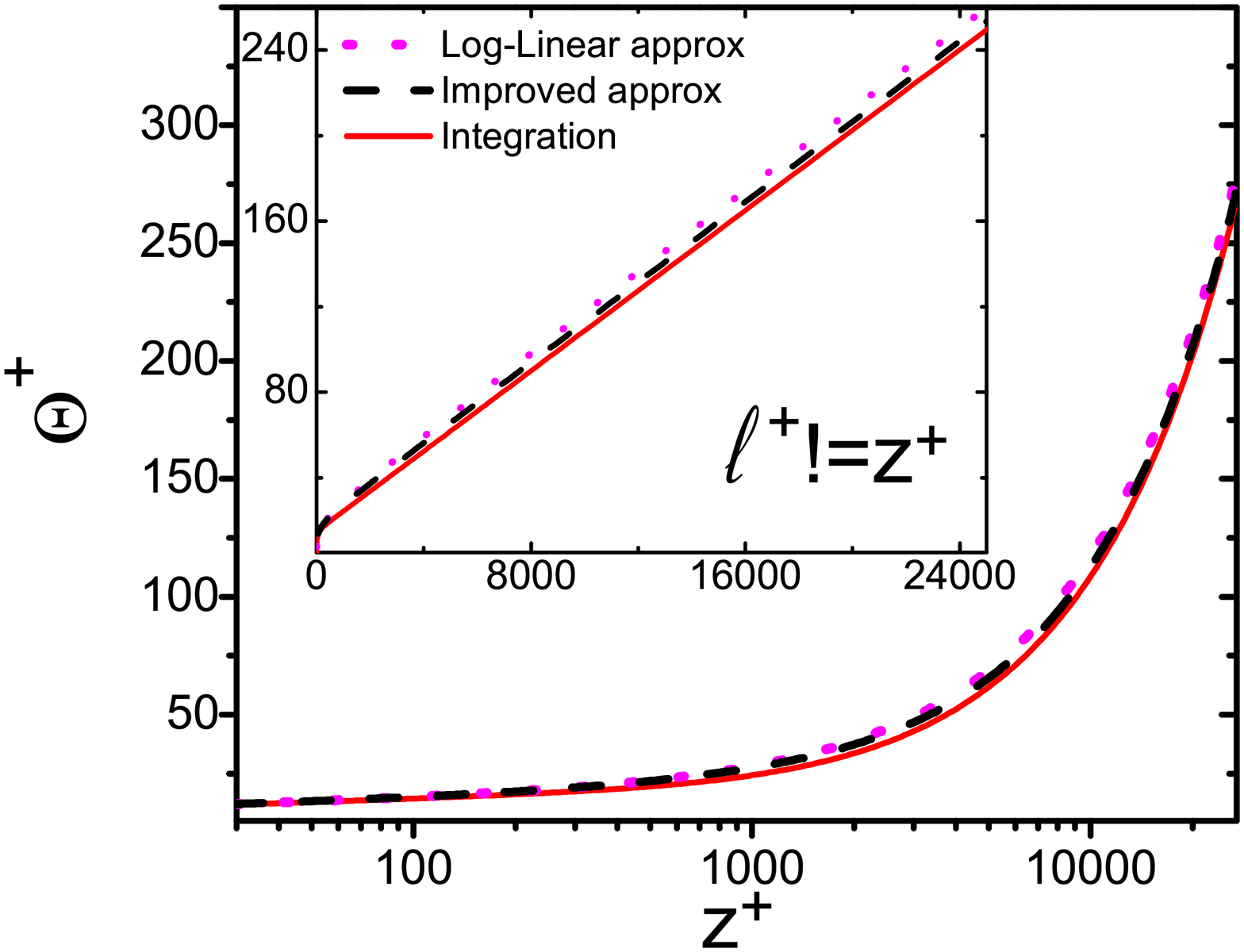}
 \includegraphics[width=0.475  \textwidth]{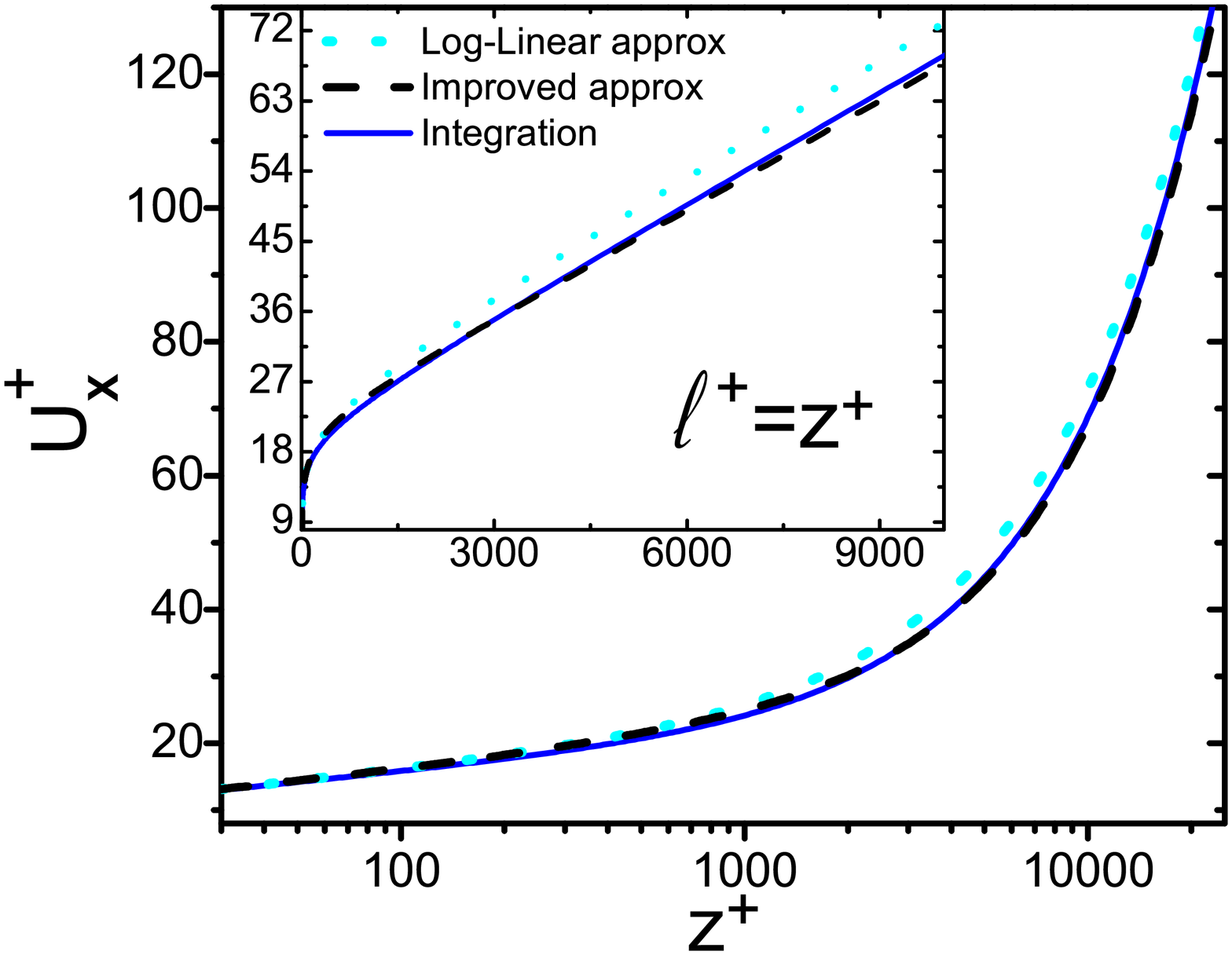}
 \includegraphics[width=0.45  \textwidth]{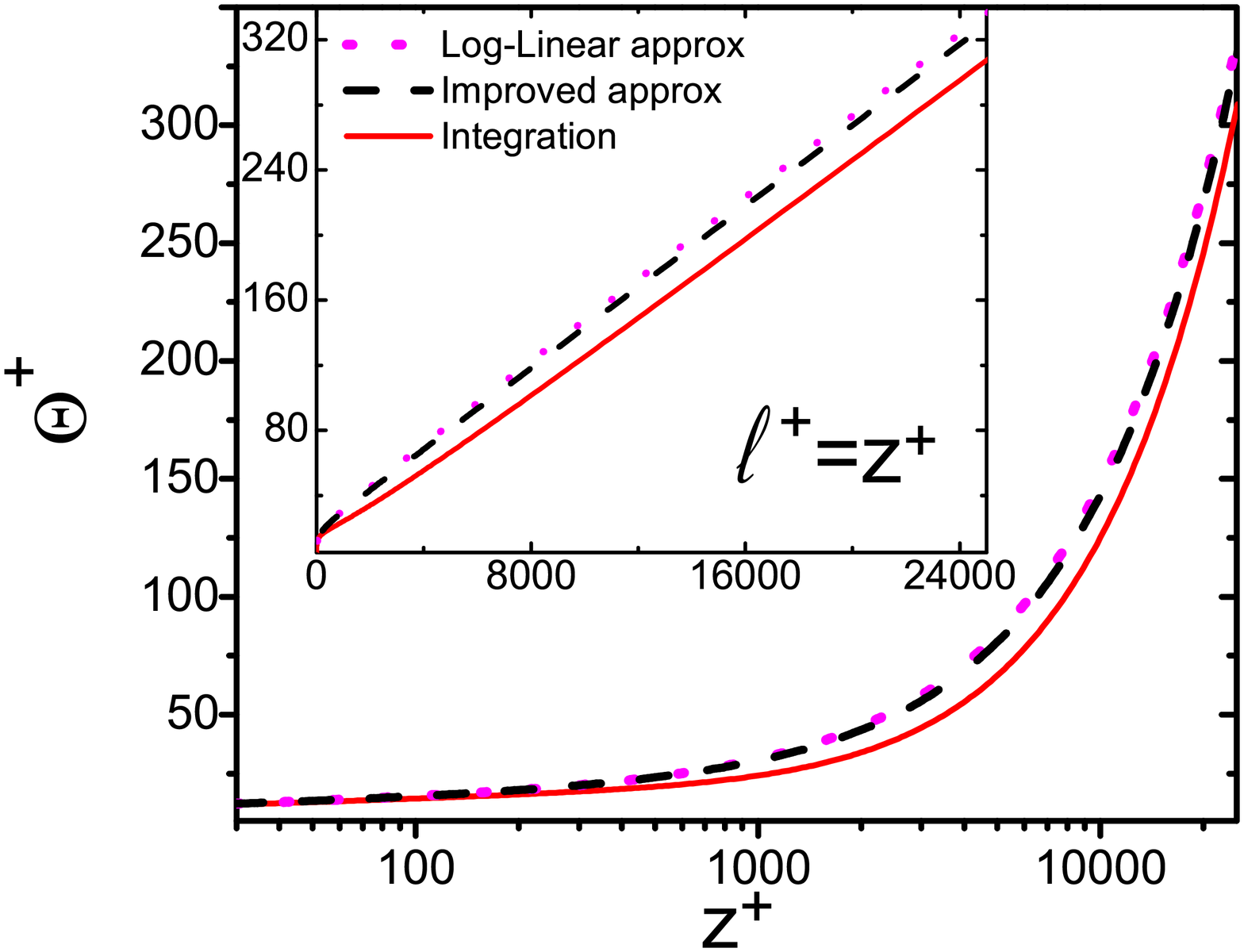}
 \caption{\label{f:means}
 Computed with \Eq{sless} (for $d_1=d_2=1$) plots of $ U^+ $
 (blue solid lines on left panels) and
 $ \overline{\Theta}^{\,+}$ (red solid lines on right panels)
 vs $\ln(z/L)$  and vs. $ z/L $ (in inserts) for $L^+=1000$. In
 upper panels $\ell(z)$ is taken from \Eq{sless}, while in lower
 panels  take $\ell(z)=z$.   Log-linear approximation~\eq{ll-aprV}
 to all profiles is shown by dotted lines, its improved
 version~\eq{ll-aprV2} by dashed lines.  The region $\ell \gtrsim L$ may not
  be realized in the Nature. In this case it has only
  methodological character.  }
\end{figure*}

The resulting plots of $U^+$  are  shown on \Fig{f:means}, upper
panel. Even taking $\ell(z)=z$ one gets very similar velocity
profile, see \Fig{f:means}, lower panel. With $\ell(z)=z$ we find an
analytical expression for the mean-velocity profile,  using the
interpolation \Eq{interB} for $S\Sb U^+$: %%
\BE \label{vel} %%
U^+(z)=\frac{1}{\kappa}\ln\Bigg[\frac{z}{z_{u0} \(1+\sqrt{1+ \(z /L
_2\)^{2/3}}\)^3} \Bigg]+\frac{z }{L_1 }  \ . %%
\EE %%
Here $z_{u0}$ is the roughness length. The ratios $L _j/L $ are
given with our choice of fitting constants.
 %%%%%%%%%%%%%%%%%%%%
%%%%%%%%%%%%%%%%%%%%%%%%%%%%%%%%%%%%%%%%%%%%%

The resulting mean velocity profiles have  logarithmic asymptotic
for $z<L$ and a linear behavior for $z>L$ in agreement with
meteorological observations. Usually the observations are
parameterized by a so-called log-linear approximation:
\BSE\label{ll-apr} \BE \label{ll-aprV}  %%
U^+ =  {\kappa}^{-1}  \ln ({z}/{z_{u0}})+ {z}/{L_1}\,,
 \EE
which is plotted in \Fig{f:means} by dotted lines. One sees some
deviation in the region of intermediate $z$. The reason is that the
real profile [see, e.g. \Eq{vel}] has a logarithmic term that
saturates for $z\gg L$, while in the approximation~\eq{ll-aprV} this
term continues to grow. To fix this one can use \Eq{vel} (with
$L_2=L_1$ for simplicity), or even its simplified version \BE
\label{ll-aprV2}  %%
U^+ = \frac{1}{\kappa}  \ln \frac{z}{z_{u0}\sqrt{1+(z/L_1)^2}}+
 \frac{z}{L_1}\ .
 \EE\ESE
This approximation is plotted as
a dashed line on \Fig{f:means} for comparison.
 One sees that our  approximation~\eq{ll-aprV2} works
much better than the traditional one. Thus we suggest \Eq{ll-aprV2}
for parameterizing meteorological observations.

The temperature profiles in our approach look  similar to the velocity
ones, see \Fig{f:means}, lower panel. They have logarithmic
asymptotic for $\ell < L$ and linear behavior for $\ell >L$.
Correspondingly, they can be fitted by the log-linear
approximation, like~\eq{ll-aprV}, or even better, by improved version of
it, like \Eq{ll-aprV2}. Clearly, the values of constant will be
different:  $\kappa \Rightarrow \kappa\Sb T$, $L_1 \Rightarrow
L_{1,\rm T}$, etc.
%%%%%%%%%%%%%%%%%%%%%%%%%%%%%
\subsection{\label{ss:one}  Profiles of second-order correlations}

The computed profiles of the turbulent kinetic and
 temperature energies, horizontal thermal flux profile and the
 anisotropy profiles are shown on \Fig{f:S} in the middle and lower
 panels. The anisotropy profiles, right middle panel, saturate at
 $\ell/L\approx 2$, therefore they are not sensitive to the $z$-dependence of $\ell(z)$;
 even quantitatively one can think of these profiles as if they were plotted as a function of $z/L$.

 Another issue is the profiles of $E\Sb K^+$ (left middle panel) and
 of $E\Sb \Theta$ and $F_x^+$ (lower panels), that are $\propto (\ell/L)^{2/3}$ for
 $\ell\gg L$ (if available). With the interpolation formula~\eq{sless} the
 profiles of the second order correlations have to saturate at
 levels corresponding to $\ld= 1$. This sensitivity to the $z$-dependence of
 $\ell(z)$ makes a comparison of the prediction with
 available data very desirable.

 \subsection{\label{ss:nums} Turbulent transport, Richardson and Prandtl numbers}

In our notations the turbulent viscosity and thermal conductivity,
turbulent Prandtl number, the gradient- and flux-Richardson numbers are %%
\begin{subequations} \label{num}
  \begin{eqnarray} \label{nu} %%
  \nu \Sb T&\= & - \frac{\tau_{xz}}{S\Sb U}= \frac1 {S\Sb U ^+}\= C_\nu (\ld )
  \frac{\tau_{zz}^+}{\g _{uu}^+}\,,\\
   \label{chi}%%
    \chi\Sb T &\=& -\frac{F_z}{S\Sb \Theta}= \frac 1 {S\Sb \Theta ^+}
    \= C_\chi  (\ld )
  \frac{\tau_{zz}^+}{\g _{uu}^+}\,,
     \\  \label{numPr}
\mbox{Pr}\Sb T& \= &\frac{\nu\Sb T }{\chi\Sb T }
=\frac{S_{_\Theta}^+}{S_{_U}^+}
=\frac{S_{_\Theta}^\ddag}{S_{_U}^\ddag}\,,  \\
\Rig  &\=& \frac{\b S_{_\Theta}}{S_{_U}^2}=  \frac{
S_{_\Theta}^+}{L^+\, {S_{_U}^+}^2 }=\frac{\ld
S_{_\Theta}^\ddag}{{S_{_U}^\ddag}^ 2}\,, \\
\Rif  &\=& \frac{\b  F_z}{\tau_{xy}S_{_U}}=  \frac1{ L^+\, S_{_U}^+
}=\frac{\ld  }{ S_{_U}^\ddag }\,, \label{numRig}\\
\label{rel4} \Rig&=&\Rif \, \mbox{Pr}\Sb T\ .
\end{eqnarray}
 \end{subequations} %%
With \Eqs{nu} and \eq{chi} we introduce also two dimensionless
functions $C_\nu (\ld )$  and $C_\chi  (\ld )$ that are taken as
$\ld$-independent constants in the down-gradient transport
approximation~\eq{dt} described in the Introduction.  We will show, however,  that the functions
have a strong dependence on $\ld$, going to zero in the limit
$\ld\to\infty$ as $1/ \ld^{4/3}$. Therefore this approximation is
not valid for large $\ld$ even qualitatively.

\subsubsection{\label{sss:dt} Approximation of down-gradient transport
and its violation in stably stratified TBL}
%%%%%%%%%%%%%%%%%%%%%%%%%%%%%%%%%%%%%%%%%%%%%%%%%%%%%%%%%%%%%%%%%%%%%%%%%%%%
\begin{figure*}
 \includegraphics[width=0.45\textwidth]{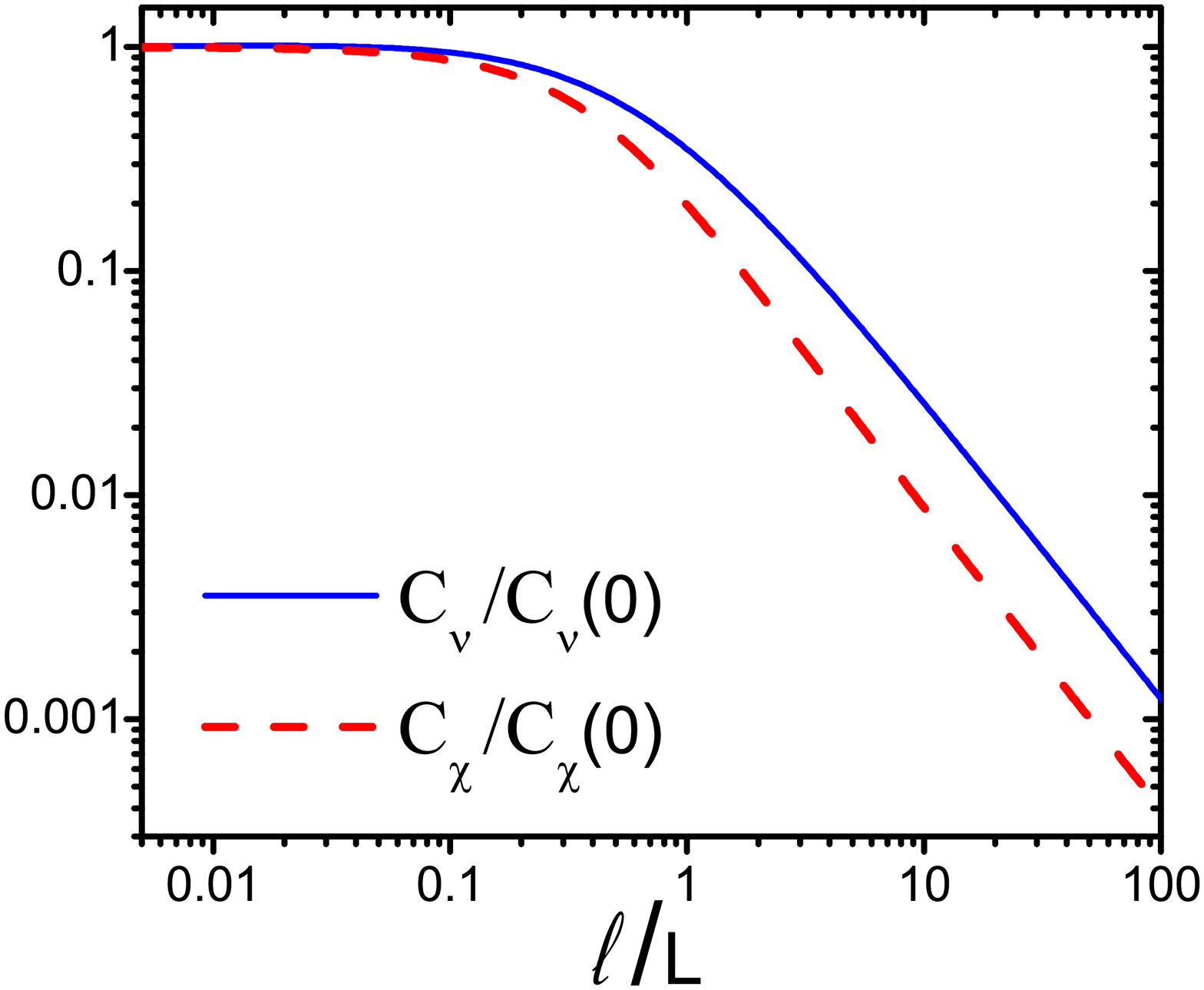}
 \includegraphics[width=0.45  \textwidth]{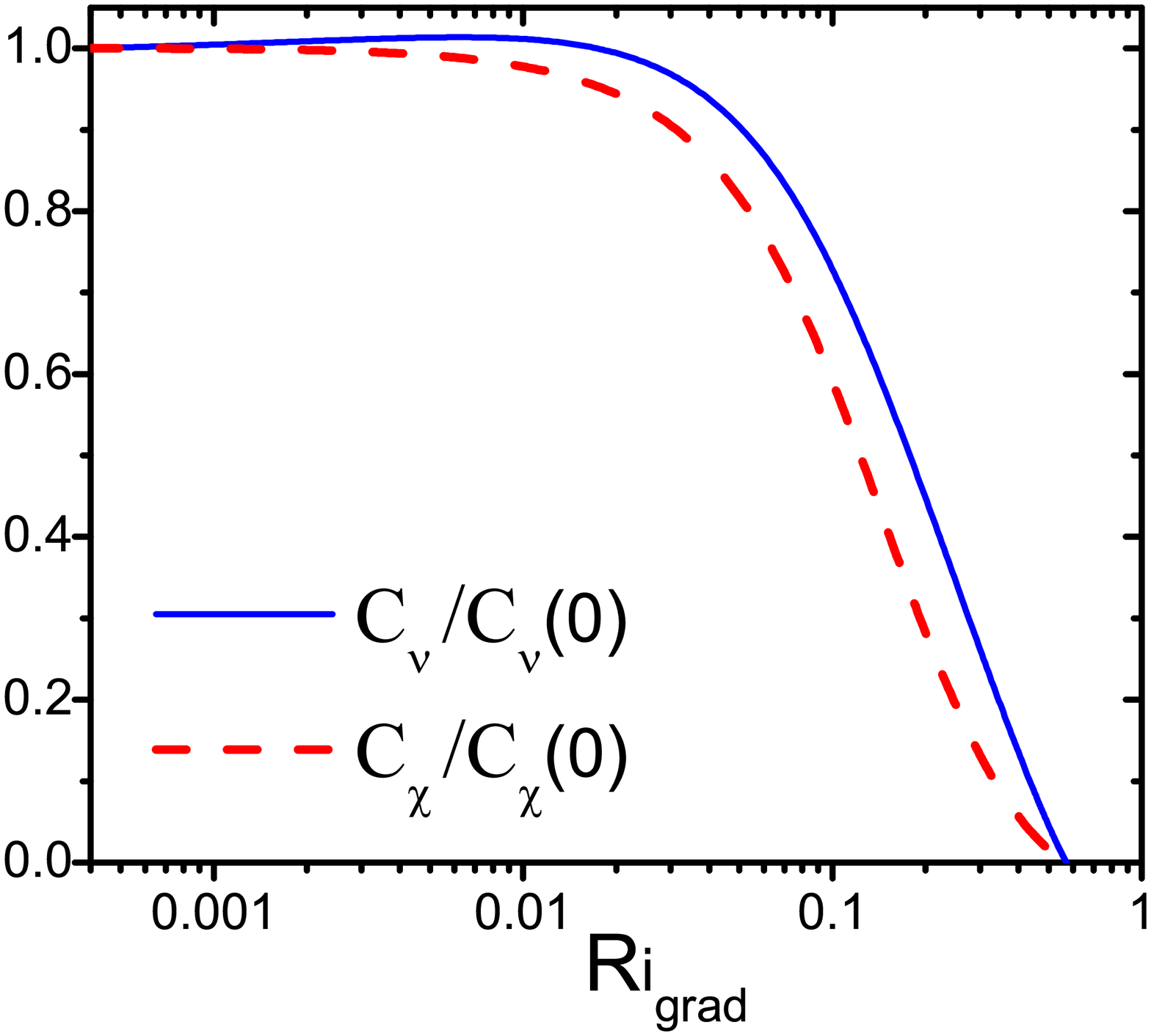}
 \includegraphics[width=0.46  \textwidth]{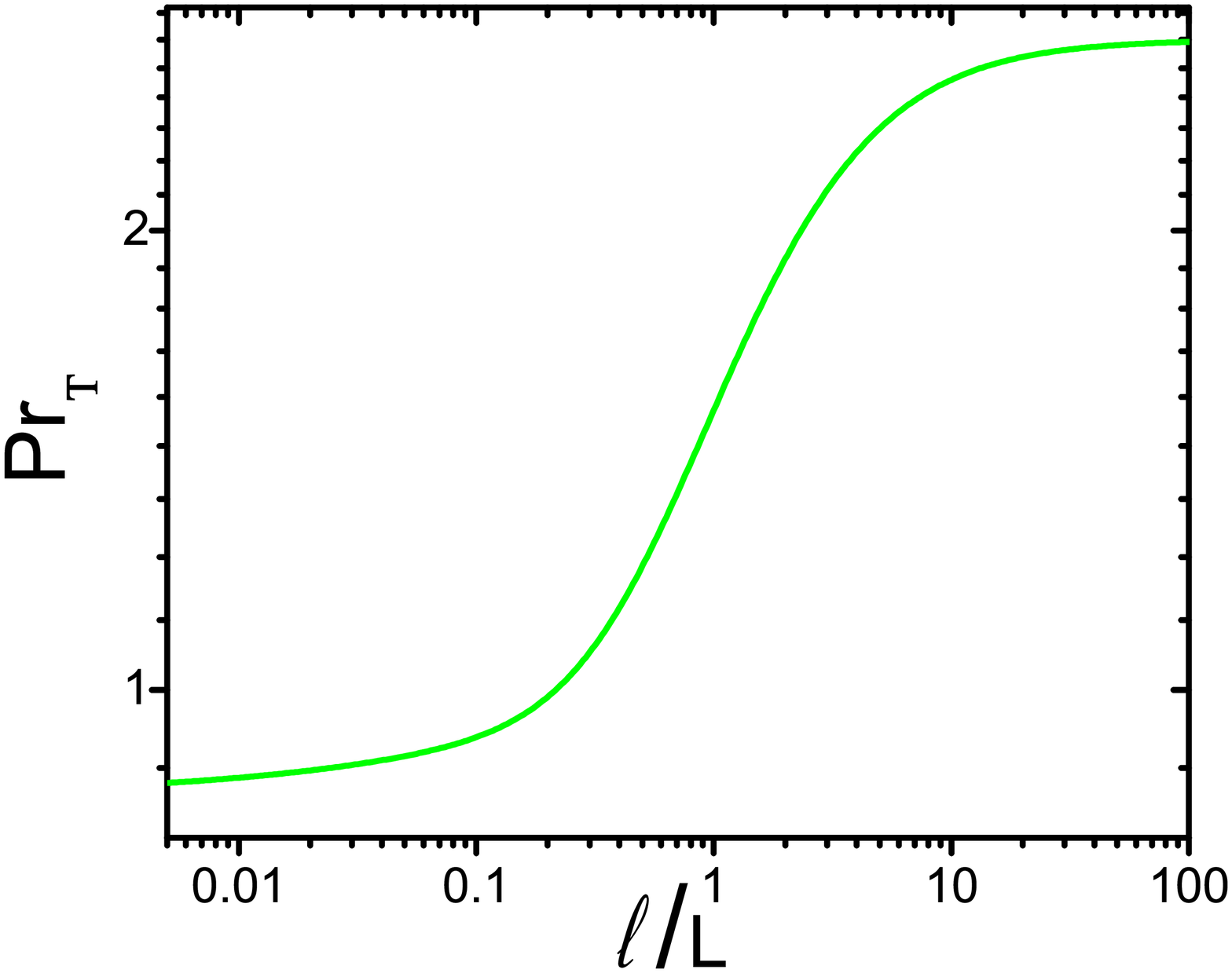}
 \includegraphics[width=0.465  \textwidth]{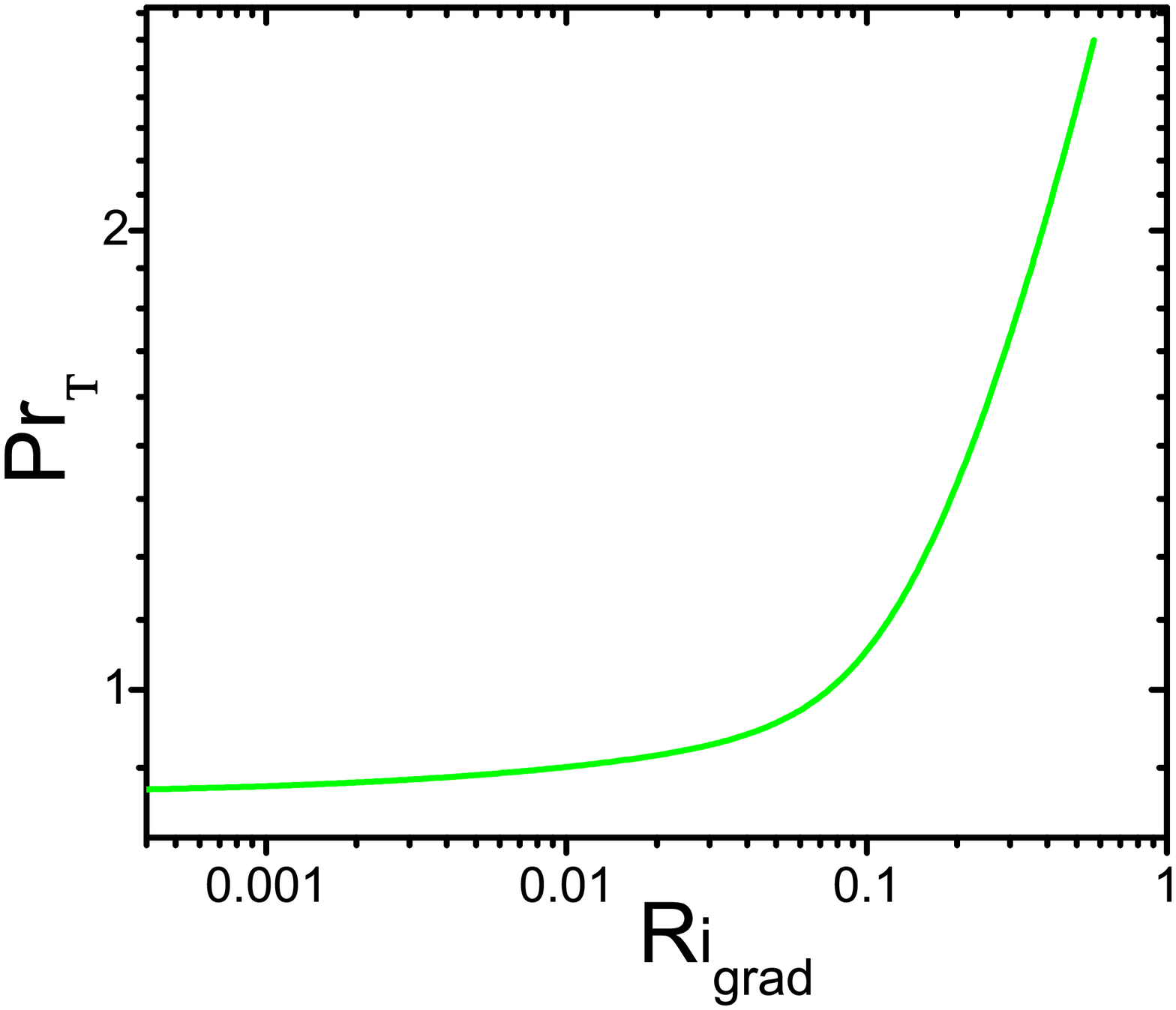}
 \includegraphics[width=0.45  \textwidth]{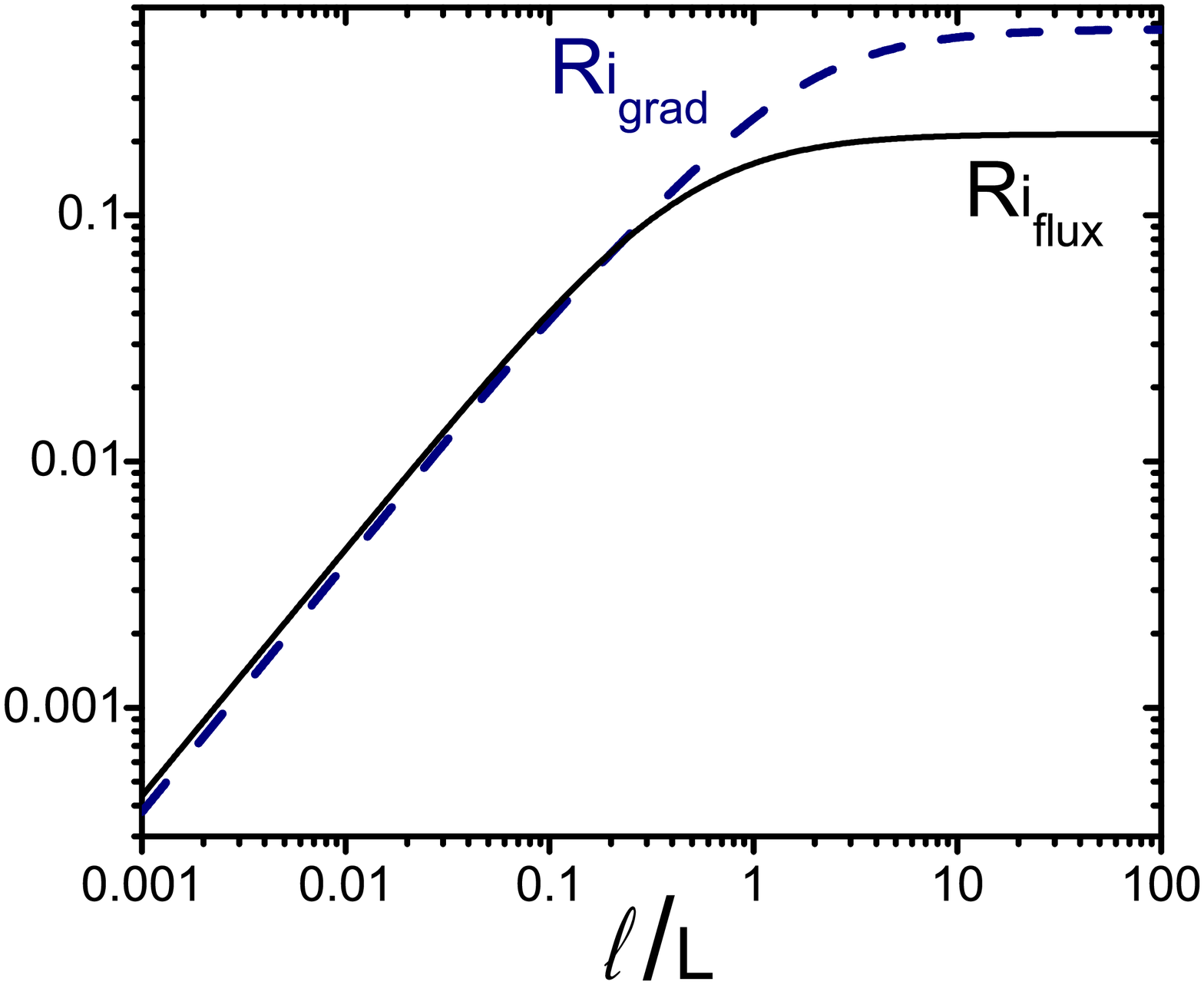}
  \includegraphics[width=0.47  \textwidth]{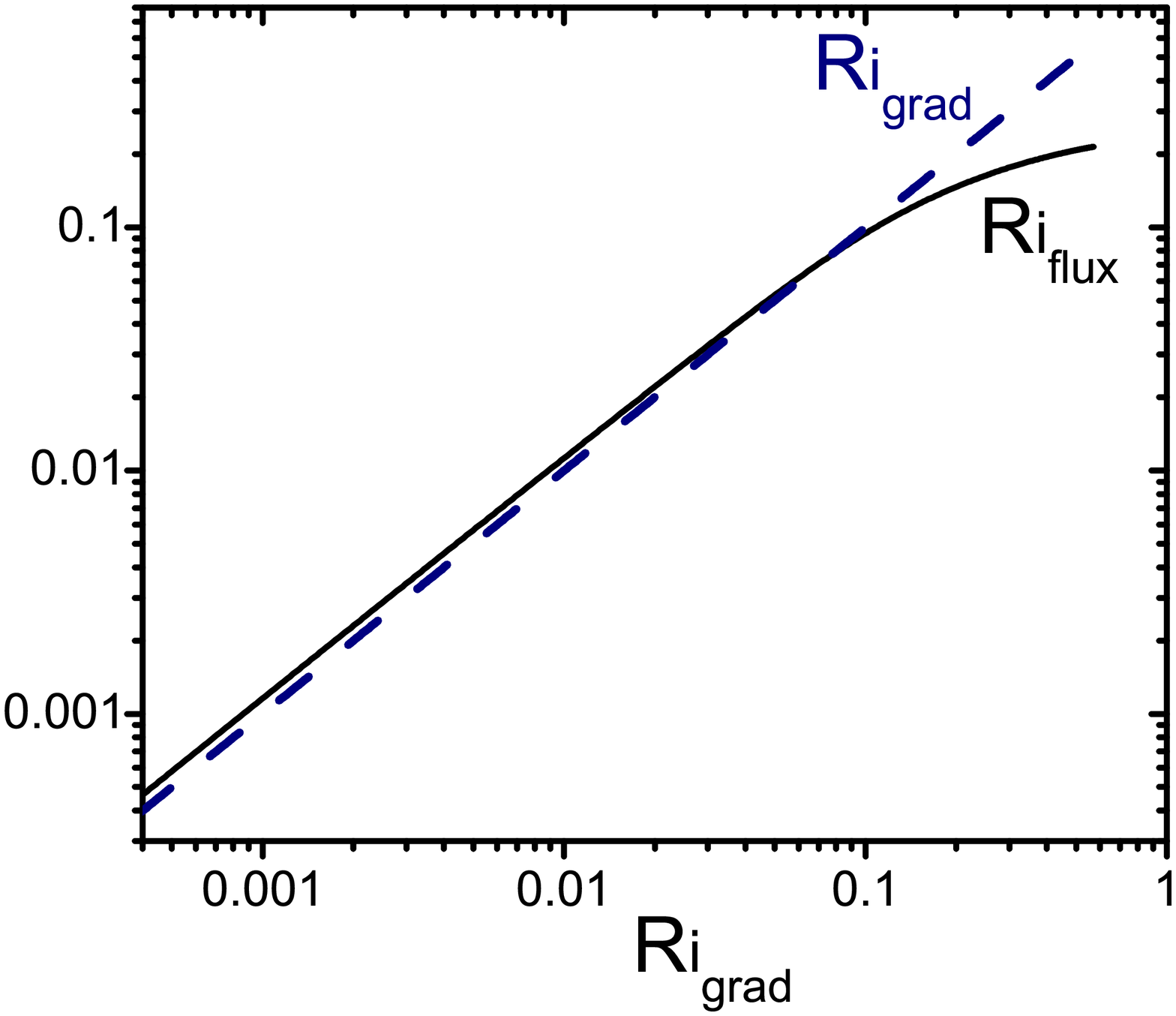}

 \caption{\label{f:res}  Color online.
 Log-log plots of ``down-gradient coefficient-functions"
    $C_\nu$ (solid blue lines) and
   $C_\chi$ (red dashed lines) --  upper panels;
    turbulent Prandtl number Pr$\Sb T$  (green lines on middle panels)
   and $\Rif$ (black solid lines),  $\Rig$ (black dashed lines) -- on lower panels
   as function of  $\ld=\ell/L$ (left panels)
   and vs. $\Rig$ (right panels). Notice, that the presented dependencies
   have qualitative character, and the choice of constants $C_{...}$
   depends on the actual functional form $\ell\(z\)$. For simplicity,
   we took $\ell\(z\) = z$.The region $\ell \gtrsim L$ may not be
  realized in the Nature. In this case it has only
  methodological character.  }
\end{figure*}
 As we mentioned in the Introduction, the concept of the
 down-gradient transport assumes that the momentum and thermal
 fluxes are proportional to the mean velocity and temperature
 gradients, see \Eqs{dt}:
  \BE\label{appr}
 \tau_{xz}=-\nu\Sb T  S_{_U}\,,\quad
 F_z=-\chi\Sb T  S\Sb \Theta \,,
 \EE
 where $\nu\Sb T$ and $\chi\Sb T$ are effective turbulent viscosity
 and thermal conductivity, that can be estimated  by dimensional
 reasoning. Equations \eq{dt}, giving this estimates,  include
 additional physical arguments that vertical transport parameters
 should be estimated via vertical turbulent velocity,
 $\sqrt{\tau_{zz}}$, and characteristic vertical scale of turbulence,
 $\ell_z$. The relations between the scales $\ell_j$ in different
 $j$- directions in anisotropic turbulence can be found in the
 approximation of time-isotropy, according to which%%
 \BE\label{ti}
 \frac{\sqrt{\tau_{xx}}}{\ell_x}=\frac{
 \sqrt{\tau_{yy}}}{\ell_y}= \frac{\sqrt{\tau_{zz}}}{\ell_z}\=
 \gamma\  \Rightarrow \ \gamma_{uu}\ .
\EE%%
 Here $\gamma$ is a characteristic isotropic frequency of turbulence,
 that for concreteness can be taken as the kinetic energy relaxation
 frequency $\gamma_{uu}$.
The approximation~\eq{ti} is supported by experimental data, according
to which in anisotropic turbulence the ratios $\ell_i/\ell_j$ ($i\ne
j$) are larger then the ratios $\ell_i\, \sqrt{\tau_{jj}}\,
/\ell_j\, \sqrt{\tau_{ii}}$ that are close to unity. With this
approximations $\nu\Sb T$ and $\chi\Sb T$ can be estimated as
follows:%%
\BE\label{est1} \nu\Sb T= C_\nu  {\tau_{zz}}/{\gamma_{uu}}\,, \quad
\chi \Sb T = C_\chi {\tau_{zz}}/{\gamma_{uu}}\,, \EE %%
where, according to the approximation of down-gradient transport,
the dimensionless parameters $C_\nu$ and $C_\chi$ are taken as
constants, independent of the level of stratification.

In order to check how the approximation~\eq{appr},~\eq{est1} works
in the stratified TBL  for both fluxes, one can consider \Eqs{appr}
as the \emph{definitions} of  $\nu\Sb T$ and $\chi\Sb T$ and
\Eqs{est1} as the  \emph{definitions} of  $C_\nu$ and   $C_\nu$.
This gives%%
\BSE\label{defC} \BEA\label{defCM} %%
C_\nu&\=& -\frac{\tau_{xz}}{\tau_{zz}}\, \frac{\gamma_{uu}}{S_{_U}}=
\frac{\gamma_{uu}^+}{ \tau_{zz}^+ S_{_U}^+} \,, \\  \label{defCH}
C_\chi&\=& -\frac{F_x}{\tau_{zz}}\, \frac{\gamma_{uu}}{S\Sb \Theta}=
\frac{\gamma_{uu}^+}{ \tau_{zz}^+ S\Sb \Theta^+} \ .  \EEA\ESE
Recall, that in this paper the down-gradient approximation is not
used at all. Instead, we are using exact balance equations for all
relevant second order correlation, including $\tau_{xz}$ and $F_x$.
Substituting our results in the RHS of the definitions~\eq{defC} we
can find, how $C_\nu$ and $C_\chi$ depend on $\ld=\ell/L$, that
determines the level of stratification in our approach.

The resulting plots of the ratios  $C_\nu(\ld)/C_\nu(0) $ and
$C_\chi(\ld)/C_\chi(0)$ are shown in the left-upper panel in
\Fig{f:res}. One sees that the $C_\nu(\ld)$ and $C_\chi(\ld)$ can be
considered approximately as constants only for $\ell \le 0.2\, L$.
For larger $\ell/ L$ both $C_\nu(\ld)$ and $C_\chi(\ld)$ rapidly
decrease, more or less in the same manner, diminishing by an order of magnitude
already for $\ell \approx 2\, L$. For larger $\ell/L$ one can use the
asymptotic solution~\eq{sol3}, \eq{sol2} for $\ld\gg 1$ according to
which %%
\BE \label{est2}%%
 S_{_U}^+\simeq \frac1{L^+}\,,\  \gamma_{uu}\simeq
\frac{\sqrt{E\Sb K^+}}{\lp}\simeq \frac{\ld^{1/3}}{\lp}\,, \
\tau_{zz}\simeq \ld^{2/3}\ . %%
\EE%%
This means that  both functions vanish as $1/\ld^{4/3}$: %%
 \BE\label{as}
C_\nu(\ld )\simeq 0.01 \( \frac{L}{ \ell} \)^{4/3}\!\!  ,  \ \
C_\chi(\ld)\simeq 0.003   \( \frac{L} { \ell} \)^{4/3} \!\!  ,
 \EE %%
 where numerical prefactors account for the accepted values of
 the dimensionless fit parameters.

The physical reason for the strong dependence  of $ C_\nu$ and
$C_\chi$ on stratification is as follows: in the   RHS of \Eq{txz}
for the momentum flux and  \Eq{Sim-F-z} for the vertical heat flux
there are two terms. The first ones, proportional to $\tau_{zz}$ and
velocity (or temperature) gradients
  correspond to the   approximation~\eq{appr}, giving
(in our notations) $C_\nu =$const and $C_\chi =$const, in agreement
with the down-gradient transport concept. However, there are second
contributions to the vertical momentum flux $\propto F_x$ and to the
vertical heat flux,  that is proportional to $ \b E_\theta$. In our
approach both contributions are negative, giving rise the
\emph{counter-gradient fluxes}.   What follows from our   approach,
is that these counter-gradient fluxes cancel (to the leading order) the
down-gradient contributions in the limit $\ld\to\infty$. As a
result, in this limit the effective turbulent diffusion  and thermal
conductivity   vanish, making the down-gradient approximation for
them (with constant $C_\nu$ and $C_\chi$) irrelevant even
qualitatively for   $\ell \gtrsim L$.

 In our picture
of stable temperature-stratified TBL, the turbulence exists at any
elevations, where one can neglect the Coriolis force. Moreover, the
turbulent kinetic and temperature energies increase as
$(\ell/L)^{2/3}$ for $\ell > L$, see left middle and lower panels in
\Fig{f:S}.   At the same time, the mean velocity and potential
temperature change the $(\ell/L)$-dependence from logarithmic lo
linear, see \Fig{f:means} and (modified) log-linear interpolation
formula~\eq{ll-aprV2}. Correspondingly, the shear of the mean
velocity and \emph{the mean temperature gradient} saturate at some
elevation (and at some $\ell/L$), and $\Rig$\emph{ saturates as
well}. This predictions agree with large eddy simulation by
Zilitinkevich and Esau (2006), see Fig. 5 in ZEKR-paper, where
$\Rig$ can be considered as saturating around 0.4 for $z/L\approx
100$.

Nevertheless, our analytical result of saturating $\Rig$ disagree
both with the ZEKR model and with various observational,
experimental and numerical data, collected in ZEKR-paper, see their
Figs. 1, 2, upper panel of Figs. 3 and 4, where various data are
plotted as functions of the gradient Richardson number in the
interval $(0,100)$. The conditions at which these data were obtained
do not correspond to the the situation considered in this paper.
Nevertheless, one cannot completely ignore the fact, shown in Fig.1
of ZEKR-paper, that various data concentrate around a linear
dependence Pr$\Sb T\sim \Rig$ in the two-decade interval $1<\Rig <
100$ (not withstanding the high degree of scatter).

 Notice that the turbulent closures of kind used above
cannot be applied for strongly stratified flows with $\Rig\gtrsim 1$
(may be even at  $\Rig\sim 1$). There are two reasons of that.  The
first one was mentioned in the Introduction. Namely, for
$\Rig\gtrsim 1$ the Brunt-V\"ais\"al\"a frequency $ N\=\sqrt{\b S\Sb
\Theta}$, $ N^+= \sqrt{\frac{S\Sb \Theta^+}{L^+}}$,
 is   larger then the eddy-turnover frequency  and
 therefore   there are weakly decaying Kelvin-Helmholts internal
 gravity
 waves which, generally speaking has to be accounted in the momentum
 and energy balance in TBL.

 The second reason, that makes the results very sensitive to the
 contribution of internal waves follows from the fact that
 vortical  turbulent fluxes vanish (at fixed velocity and temperature
 gradients). Therefore even relatively small contributions of a
 different nature to the momentum and thermal fluxes may be important.

 The final conclusion is that the TBL modeling at large level of
 stratification requires accounting for turbulence of the internal
 waves together with the vortical turbulence, and  the analysis of
 available  data calls for serious revision. Definitely, new
 observations, laboratory and numerical experiments with control of
 internal wave activity are very likely.

%%%%%%%%%%%%%%%%%%%%%%%%%%%%%%%%%%%%%%%%%%%%%%%%%%%%%%%%%%%%%%%%%%%%%%%%%%%%%%%

\acknowledgements This work has been inspired by discussions with
the authors of ZEKR-paper: Sergej Zilitinkevich, Tov Elperin, Natan
Kleeorin and Igor Rogachevskii.   We follow similar  strategy and
employ the same concept of conservation of total mechanical energy
of turbulence, proceeding further in analysis of the Reynolds
stress, thermal flux and potential energy budgets. This work has
been supported in part by the US-Israel Binational Science
Foundation and, for OR, by the Transnational Access Programme 
at RISC-Linz, funded by the European Commission Framework 6 Programme 
for Integrated Infrastructures Initiatives under the project SCIEnce 
(Contract No. 026133).

\appendix

%%%%%%%%%%%%%%%%%%%%%%%%%%%%%%%%%%%%%%%%%%%%%
\section{\label{s:approx}Obereck-Boussinesq approximation and conservation laws}
\subsection{\label{ss:basic}Basic hydrodynamic equations}

The system of hydrodynamic equations describing a fluid in which the temperature
is not uniform  consists of the
Navier-Stokes equations for the fluid velocity, $\BC U(\B r,t)$, a continuity
equation for the space and time dependent (total) density of the fluid, $\rho
(\B r,t)$, and of the heat balance equation for the (total) entropy
per unit mass, $\C S(\B r,t)$,~Landau and Lifshitz, 1987 : %%
\begin{subequations}\label{full}%%
\begin{eqnarray} %%
 \rho \frac{\C D\, \BC U }{\C D t} &=&  - \B {\nabla}
 p  - \B g\, \rho
 +  \B \nabla \cdot \mu\, \B \nabla\BC U\;,
\label{B1} \\ %%
  \frac{\partial{\rho}}{\partial{t}}&\!\!+\!\!&
\B \nabla\cdot \( \rho \, \BC U\)  =  0\,,\label{B2}\\ \label{LD}
  \rho \frac{\C D\, \C S  }{\C D t}   &=& \B \nabla  \cdot \kappa\, \B \nabla \C S
  \,,
 \label{B3}\\ \label{LD1} \frac{\C D}{\C D t}&\=&  \frac{\partial }{
\partial t} + \BC U \cdot \B {\nabla}\ .
\end{eqnarray}
\end{subequations}%%
 Here $\C D\,   /\C D t$  is the convective (Lagrangian)
derivative, $p$ is the pressure, $\B g=-\z g$ is the vertical
acceleration due to gravity, $\mu$ and $\kappa$ are the (molecular)
dynamical viscosity and heat conductivity.

These equations are considered with boundary conditions that
maintain the solution far from the equilibrium state, where $\BC
U=\C S=0$. These boundary conditions are $\BC U=0$ at zero
elevation, $\BC U=const$ at a high elevation of a few kilometers.
This reflects the existence of a wind at high elevation, but we do
not attempt to model the physical origin of this wind in any detail.
The only important condition with regards to this wind is that it
maintains a momentum flux towards the ground that is prescribed as a
function of the elevation. Similarly, we assume that a stable
temperature stratification is maintained such that the heat flux
towards the ground is prescribed as well. In the entropy  balance
\Eq{B3} we have already neglected the viscous entropy production
term, $\propto \mu \B |\nabla \BC U|^2$, assuming that the
temperature gradients are large enough such that the thermal entropy
production term on the RHS of \Eq{B3} dominates. Actually, this
assumption is very realistic in our applications. For simplicity of
the presentation we restrict ourselves by relatively small
elevations and   disregard in \Eq{B1} the Coriolis force (for more
details, see Wyngaard, 1992).

On the other hand we assume that the temperature and density
gradients  in the entire turbulent boundary layer are sufficiently
small to allow employment of local thermodynamic equilibrium. In
other words, we assume the validity of the equation of state, and
that the entropy $\C S$ is a state function of the local values of
the density and pressure:
 \begin{equation}\label{B4}%%
 \rho =\rho(T , p)\,,\quad
\C S=\C S (\rho, p)\ . %%
\end{equation}%%
In the same manner we will neglect the temperature
dependence of the dissipation parameters $\nu$ and $\kappa$.

Pressure fluctuations caused by  turbulent velocity fluctuations $\B
u$ propagate in a compressible medium with the sound velocity $c\sb
s$,  causing time dependent density fluctuations of the order of
$(u/c\sb s)^2 \rho_0$, where $\rho_0$ is the mean density. Assuming
that the square of the turbulent Mach number $M\Sb T^2 \= (u/c\sb
s)^2$ is small compared to unity, we can neglect in \Eq{B2} the
partial time derivative: %%
 $ \B \nabla\cdot \( \rho \, \BC U\)=0$, see e.g. Landau and Lifshitz, 1987.
  Even in tropical hurricanes of category five the mean wind
velocity $U$ is below 300 Km/h. Usually, the turbulent velocity
fluctuations $u$ are less then $U/10$, i.e. even in these extreme
conditions $u< 30$ Km/h and $M\Sb T^2< 10^{-3}$ (with $c\sb
s\simeq 1200$ Km/h). Therefore the incompressibility approximation $
\B \nabla \cdot\( \rho \, \BC U\)=0$  is  well justified in atmospheric physics. In the ocean where
the sound velocity is even larger and water velocities even smaller,  this
approximation  is quite excellent.
%%%%%%%%%%%%%%%%%%%%%%%%%%%%%%%%%%%
\subsection{\label{ss:Isen}Isentropic basic reference state }

In quite air, without turbulence, the pressure and the density
depend on the elevation $z$ simply due to gravity. For example, in
full thermodynamic equilibrium the temperature is uniform,
$z$-independent, and the density decreases exponentially with the
elevation. However, this equilibrium  model of the atmosphere is not
realistic, and cannot be used as a reference state about which the
actual dynamics is considered. A much better reference state is a
situation in which the entropy is space homogeneous. In this model
the thermal conductivity (leading to the temperature homogeneity) is
neglected with respect to heat transfer due to the vertical
adiabatic mixing of air, leading to a $z$-independent entropy.
Following tradition, we refer to the isentropic  model as a ``basic
reference state" and denote this state of the system with a
subscript `` $\sb b$": %%
\begin{equation}\label{A3}
 \C S\sb
b=\C  S(\rho\sb{\, b}, p\sb {\, b} )={\rm const}\,, \quad
 \rho\sb{\, b} =\rho(T\sb b, p\sb {\, b} )\ . %%
\end{equation} %%
The first of \Eq{A3} relates the gradients of the pressure and
density in this state:%%
\begin{subequations}\label{rel1}\begin{equation}\label{rel1a} 0=\B \nabla \C S\sb b
= \(\frac{\p \C S\sb b}{\p \rho\sb b}\)_p \B \nabla \rho\sb b+
\(\frac{\p \C S\sb b}{\p p\sb b}\)_\rho  \B \nabla p\sb b \ .
\end{equation} Another relation between $\rho\sb b$ and $p\sb b$
follows from the condition of hydrostatic equilibrium:%%
 \begin{equation}\label{heq} %%
 \B \nabla  p\sb b=\B g \rho\sb b\ . %%
 \end{equation} \end{subequations} %%
 Equations \eq{rel1} together with the first of Eqs.~\eq{A3} determine the
density, pressure and temperature profiles in the isentropic basic reference state.
%%%%%%%%%%%%%%%%%%%%%%%%%
 \subsection{\label{ss:mod} Hydrodynamic equations in generalized
 Obereck--Boussinesq approximation }
\subsubsection{Equations of motion}

Denote the deviations of the total density, pressure, temperature and entropy
from the basic reference state as follows:%%
 \begin{eqnarray}\label{dev}  \hat\rho \= \rho-\rho\sb{\,
b}\,, && \hat p\=p-p\sb{\, b}\,,  \br \hat T \= T-T\sb{\, b}\,, &&
\hat{\C S}\=\C S-\C S\sb{\, b}\,. \end{eqnarray} Following
Obereck (1879) and Boussinesq (1903), assume that these
deviations are small: %%
 $
\hat  \rho \ll \rho\sb{\, b}\,, \quad
\hat  p \ll p\sb{\, b}$.
Then  one   simplifies the full system of
 hydrodynamic Eqs.~\eq{full} and  rewrites them in terms of the
 fluid velocity $\BC U$ and the small deviations   $\hat p$ and $\hat {\C S}$
 (instead of $\hat \rho$). The first step is very simple:
 because of \Eq{heq}%%
 \begin{subequations}\label{step1}\begin{equation}\label{step1a}%%
  \B \nabla  p-  \B g \rho=
   \B \nabla \hat p -  \B g\, \hat \rho \ .
 \end{equation}
Next we should relate the deviations  $\hat p$, $\hat \rho$ and
$\hat {\C  S}$. In the linear approximation \Eq{B4} yields:%%
\begin{equation}\label{step1b} \hat{\C S}=\( \p \C S \sb b\big / \p p\sb {\, b}\)_\rho
\hat p+ \( \p \C S \sb b\big / \p \rho \sb {\, b} \)_p \hat \rho\ .
\end{equation} %%
With the help of Eqs.~\eq{rel1} this gives: %%
\begin{equation}\label{step1c} %%
\B g \, \hat \rho =  \frac {\B \nabla \rho\sb b } {\rho\sb b}\,\hat
p  - \B  \beta\,  \frac{\rho\sb   b }{c\sb p}\, T\sb b \, \hat{\C
S}\ . %%
\end{equation}%%
Here $\B \beta\= \B g \, \~ \beta$ is the buoyancy parameter, $\~\beta$
is the thermal expansion coefficient and $c\sb p$ is the isobaric
specific heat (heat capacity per unit mass at constant pressure) in
the basic reference state:%%
\begin{equation} \label{beta} %%
\~\beta  \=  - \frac{1}{\rho\sb b} \(\frac{\p \rho\sb b}{\p T\sb
b}\)_p= -\frac {c\sb p }{\rho\sb b\, T\sb b} \(\frac{\p \rho \sb b}
 {\p \C S\sb b}\)_p\ .
 \end{equation} %%
 Now \Eq{step1a} in the linear approximation yields:
\begin{equation}\label{step1e}%%
  \B \nabla  \hat p-  \B g \rho=
   \rho\sb b\left[\B \nabla \( \frac {p\sb b}{\rho\sb b }\) + \B \b\,
   \frac{  T\sb b}{c_p} \, \hat{\C S}
  \right ]\ .
 \end{equation}
\end{subequations}%%
 Then Eqs.~\eq{B1} can be approximated as: %%
 \label{B6} \begin{equation}\label{B6a} \frac{\C D\, \BC U }{\C D t}   = -  \B {\nabla} \(\frac{ \hat
p}{ \rho\sb b}\) - \B \b\,\frac{  T\sb b}{c_p} \, \hat {\C S}
 + \frac{1}{\rho_b}\B \nabla \cdot \mu\,\B \nabla \BC U\ .
\end{equation} %%
%%%%%%%%%%%%%%%%%%%%%
\subsubsection{Generalized potential temperature}

To proceed, we generalize the notion of potential temperature
$\Theta$ which is traditionally defined as the temperature that a
volume of \emph{dry air} at a pressure $p(z)$ and temperature $T(z)$
would attain when adiabatically compressed to the pressure $p_*$
that exists at zero elevation $z=0$. This potential temperature can
be explicitly computed for an ideal gas with the result
\begin{equation}\label{Thetaa} \Theta(z)\= T_* \big(p_*/p(z)\big)
^{(\gamma-1)/\gamma}\ ,\end{equation} where $\gamma$ is the ratio of
isobaric to isochoric specific heats, $\gamma\=c\sb p/c\sb v$, and
$T_*$ is the temperature at zero elevation.

We want to generalize the notion of potential temperature for an
arbitrary stratified fluid requiring that in the isenotropic basic
reference state it would be a constant $\Theta_*=T_*$. A second
requirement is that the definition will agree with Eq. (\ref{Thetaa})
for an ideal gas. Accordingly we define %%
\begin{equation}
 \Theta(z)= T_* \exp\left[\(\C S(z)-\C S_{\rm b} \)/c\sb p\right]\ .
 \label{defTheta}
  \end{equation}%%
For more details see also Hauf and H\"oller (1987). Indeed, if we
employ the equation of state and the equation for the entropy of an
ideal gas, i.e.
 \begin{equation}\label{ideal}%%
 p = \rho\, T \,,\quad
\C S =    \ln \Bigg (\frac{p^{c\sb v }}{\rho^{c\sb p} }\Bigg)+{\rm const} \ ,%%
\end{equation} %%
one can easily check that Eq. (\ref{Thetaa}) is recaptured.

For small  deviations of   $\Theta$ from the basic reference state
value $T_*$, \Eq{defTheta}  gives up to linear order:%%
\begin{equation} \label{ThetaD} %%
\Theta\sb{\, d}\=  \Theta -T_* =  T_*\,
 \hat {\C S}\big / c\sb p \ . %%
 \end{equation}
Now we can present Eqs.~\eq{B1}, \eq{B2} (with $\p \rho \sb{\, t}/\p
t=0$, as explained) and \eq{B3} as follows: %%
\begin{subequations}\label{FS} %%
\begin{eqnarray}\label{FSa} %%
 \frac{\C D\, \BC U }{\C D t} & =&   %%
-  \B {\nabla} \(\frac{p}{ \rho\sb b}\) - \B \b \,  \Theta\sb {\, d}
 + \nu\, \Delta\, \BC U \,,  \\
\label{FSb}  %%
  && \hskip -1 cm \B \nabla \cdot  \(\rho\sb b\,  \BC U \)  =  0\,, \\
\label{FSc}%%
\frac{\C D\, \Theta\sb {\, d}}{\C D t}  &=& \chi\, \Delta\,
\Theta\sb {\, d} \ .
\end{eqnarray}\end{subequations}%%
The dissipative terms are important only in the narrow region of the
viscous sublayer, where we can safely neglect the $z$-dependence of
$\rho \sb b$, $\mu$ and $\kappa$, and   consider the dynamical
viscosity  $\nu = \mu / \rho_b$ and dynamical thermal conductivity
$\chi = \kappa / \rho_b$ as some $z$-independent constants.

Note that for turbulence in liquids (water, etc.) one can simplify these equations
further. There one can neglect the effect of adiabatic cooling (together with
the compressibility), and simply use another reference state with constant
temperature and density:
\begin{equation}\label{BRS2} %%
T= T_*\,, \quad \rho\sb b= \rho_*\,, \quad p\sb b=p_*+ g \rho_* z\ .
\end{equation}%%
 For this reference state standard reasoning (see,
e.g.~Landau and Lifshitz, 1987) yields the same equations as \Eqs{FS} in which again
$\~\beta$, is given by \Eq{beta} and it is a parameter characterizing
a particular fluid. In this case   $\rho\sb b=\rho_*$, independent
of $z$,  and the potential temperature $\Theta=T$, such that
$\Theta\sb {\, d}= T\sb {\, d}$ is a deviation of the total
temperature $T$ from its ground (bottom, or whatever) level $T_*$.

In our \Eq{FS} the situation is more general since  we do not assume
that reference state has the simple form~\eq{BRS2} (with $\rho\sb
b=\,$const). Importantly, on the RHS of \Eq{FSa} the density
$\rho\sb b(z)$ is operated on by the gradient, and the buoyancy term
$-\B \beta \Theta\sb {\, d}$ involves $\Theta\sb {\, d}\ne  T\sb {\,
d}$, the deviation of the \emph{potential temperature} defined by
\Eq{ThetaD}. This definition for liquids has nothing in common with
the standard meteorological definition~\eq{Thetaa}. Notice also that
for an ideal gas  $\~\beta=1/T$, and \Eq{FSa} coincide with that
suggested  in the book Kurbatsky (2000) and used in ZEKR-paper.

It is important to  realize that the \emph{approximate} Eqs. \eq{FS}
 conserve \emph{exactly} an \emph{approximate} expression for the
total mechanic energy of the system in the dissipation-less limit.
Consider the sum of the kinetic, $\C E\Sb K$, and the potential energy
$\C E\Sb P$ (calculated in the basic reference state):
$ \C E\sb t=\C E\Sb K + \C E\Sb P$, %%
\begin{equation} \label{TotMechEn}%%
 \C E\Sb K\=\int d \B r\,\rho_b\,   \frac{|\BC
U|^2}{2}\,, \quad   \C E\Sb P\equiv \int d \B r \rho_b~\B \beta\cdot \B r~ \Theta_d \  .%%
\end{equation}%%
 One can check by direct substitution that this sum of energies is conserved by of motion
 \Eqs{FS} when $\nu=\chi=0$.
 %%%%%%%%%%%%%%

\subsection{\label{a:1} Potential energy in stratified TBL}

  In this Appendix we show that the potential energy of a stratified turbulent
  flow $\C E\Sb P$,  \Eq{TotMechEn}, can be presented  as the sum of
  the (time-independent) potential energy of the basic reference
  state, $\overline{\C E}\Sb P$ and a ``turbulent" potential energy,
  associated with temperature fluctuations, $\~{\C E}\Sb P$
  with the density per unit mass $E\Sb P = {\beta} E_\theta/{S_\Theta}$:
 \begin{equation}
\~{\C E}\Sb P=\int d\B r \(\rho\Sb b \beta E_{\t} / S\Sb \Theta \) \ .
 \end{equation}
Actually, we want to discuss this issue in a more general case, when
the stratification is caused by some ``internal" parameter of the
fluid,  $\xi$, not necessarily the potential temperature. It can be
the salinity of water in a sea, the humidity of the air, the
concentration of particles co-moving with the fluid as Lagrangian
tracers, etc.

In the general case  then the equation for the potential energy of a
stratified fluid has the form:%%
 \begin{subequations} \label{def-PE}%%
 \begin{equation}\label{def-PEa} \C E\Sb P= g\int \r (\B
r)   z \, dx \, dy\, dz\,,%%
 \end{equation}
In the BRS the  potential energy reaches its
minimum value referred to as  the basic  potential energy,
$\overline{\C E} \Sb P $:
\begin{equation}\label{BPE}%%
\overline{\C E} \Sb P = g\int \r \Sb b (z)   z \, dx \, dy\, dz\,,
\end{equation}%%
where $\r \Sb b(z)$ is the density profile in BRS.  Clearly,
in equilibrium the fluid density $\r \Sb b (z)$ decreases with the
elevation: $d\rho \Sb b / dz<0$.

In the turbulent state the density deviates from $\r  \Sb b (z)$:  $\r(\B
r,t)= \r \Sb b (z)+ \~ \r (\B r,t)$ and the mean  potential energy $\< \C
E\Sb P\>$ exceeds $\overline{\C E} \Sb P $. Now we
can write
\begin{equation}\label{apeb}
  \< \C E\Sb P \>  = \overline{\C E} \Sb P + \~{\C E} \Sb P \ .
\end{equation}\end{subequations}%%
We compute the turbulent potential energy  , $\~{\C E}\Sb P $, in
the case when the internal parameter $\xi$ (temperature, etc.) is
co-moving with the fluid element as a Lagrangian marker.  In the
Lagrangian approach   we can consider $\r$ as the Lagrangian marker
and introduce  the variable $z(\r,t )$, which is understood as an
elevation of the fluid element with the density $\r$.  Noticing that
$zdz=\frac 12 dz^2$, and integrating  \Eq{def-PEa}  by parts with
respect of $z^2$,  we can present $\<\C E\Sb P\>$   in the
Lagrangian approach as:%%
 \begin{equation}\label{PE} \<\C E\Sb P\> = -
\frac g2\int \<[z(\r,t )]^2 \>   dx \, dy\, d  \r \ . \end{equation}
 As a result of turbulent motion, the elevation $z(\r,t )$ at given $\r$ fluctuates and  can be decomposed into the mean and fluctuating parts:%%
\begin{equation}\label{dec1} z(\r,t )=z \Sb b (\r)+ \~ z(\r,t )\,, \quad\<\~ z(\r,t
)\>=0\ . \end{equation} Substituting $z^2(\r,t )=z^2 \Sb b +2\, z \Sb b \~z
+ \~z^{\; 2} $ in \Eq{PE} we have three contributions to  the potential
energy. The first one (originating from $z^2 \Sb b $) describes the basic
 potential energy, \Eq{BPE}. The second contribution, which is  linear in $\~ z$, disappears
because $\<\~z\> =0$. The last one describes  the turbulent potential
energy:%%
 \begin{equation}\label{PE1}
 \~{\C E}\Sb P = - \frac g2\int \<[\~ z (\r,t)]^2\>    \, dx \, dy\, d\r \ . \end{equation}
Relating  the density fluctuations $\~\r$,  around the BRS density
profile $\r \Sb b (z)$ (caused by the fluctuations $\~ z$) with $\~ z$
 \begin{equation}\label{def-d}
\d \r \= \frac{d \r \Sb b (z)}{d\, z}\, \d z\,, \end{equation} and
returning back to the Eulerian description in \Eq{PE1}
  one has:
\begin{equation}\label{tr1b}  \~{\C E}\Sb P =  - \frac g2\int   \< \~\r ^{\,2}
\> \Big[\frac{d\, \r \Sb b (z) }{d z  }\Big]^{-1}   dx \, dy\, dz \ .
\end{equation} Here we used the transformation formula, similar to \Eq{def-d}:
$d \r =  [d \r _0 (z)\big / d\, z ]\,  d z$. Equation~\eq{tr1b}
allows one to introduce a local density of  turbulent potential
energy per unit mass,
\begin{subequations}\label{def1}
\begin{equation}\label{def1a}
 E\Sb P  =  - \frac g2  \,  \Big[\frac{d\, \r \Sb b
 (z) }{d z  }\Big]^{-1} \frac{\< {\~\r\,}^{2}\>}{\rho\Sb b}  \,,
\end{equation}
 such that
 \begin{equation}\label{def1b}\~{ \C E}\Sb P  =  \int
\rho\Sb b E\Sb P\, dx \, dy\, dz \ .
 \end{equation} \end{subequations}

For the particular case of temperature stratification %%
\BE\label{case}%%
\frac{d\, \r \Sb b
 (z) }{d z  } = \rho\Sb b \~\b\, \frac{d \Theta}{d z} \,,\quad \~\r=\rho\Sb b \~\b \theta\,, \quad \b=g \~\b\,,
 \EE
where $\~\b$ is the thermal expansion coefficient, related, as
shown,  to the buoyancy parameter $\b$. With \Eqs{case} and
definition of the temperature energy $E_\theta =
\frac12\<\theta^2\>$, \Eq{def1a} goes to $E\Sb P={\beta}
E_\theta/{S_\Theta}$, as declared.

%%%%%%%%%%%%%%%%%%%%%%%%%%%%%%%%%%%%%%%%%%%%%%%%%%%%%%%%%%%%%%%%%%%%%%%%%%%%%%%%%%%%%%%%%%%%%%%%%%%%%%%%%%%%%%%%%%%%%%%%%%%%%%%%%%%%%%%%%%%%%%%%%%%%
%%%%%%%%%%%%%%%%%%%%%%%%%%%%%%%%%%%%%%%%%%%%%%%%%%%%%%%%%%%%%%%%%%%%%%%%%%%%%%%%%%%%%%%%%%%%%%%%%%%%%%%%%%%%%%%%%%%%%%%%%%%%%%%%%%%%%%%%%%%%%%%%%%%%
\section{\label{ss:gen} Analysis of the Minimal-Model  Balance Equations  }%%
\subsection{\label{ss:plus} Wall unites and $\ddag$-representation}
The Eqs.~\eq{simple} in the wall units take the form:
\begin{subequations}\label{eqs-plus}
\begin{eqnarray}\nn %%
\tau^+_{xx} &=& E^+\Sb K -\frac{ F^+_z}{2\,L^+
 \gamma^+_{uu}}\,,\quad%%
\tau^+_{yy} = \frac{E^+\Sb K}2\,,\quad \\ %%
\tau^+_{zz} &=& \frac{E^+\Sb K}2 +\frac{ F^+_z}{2\,L^+
 \gamma^+_{uu}}\,, %%
\\
 \gamma^+_{uu} E^+\Sb K &=&  \frac{F^+_z}{L^+}  -\tau^+_{xz}
S^+_{_U} \,, \\ %%
4\,\~C\Sb{RI}\,\gamma^+_{uu} \tau^+_{xz} &=&  \frac{F^+_x}{L^+}
-\tau^+_{zz}\,S^+_{_U} \,, \\%%
C_{\theta\theta}\,\gamma^+_{uu} E^+_\theta &=&
-F^+_z\,S^+_{_\Theta}\,, \\%%
%%%%%%%%%%%%%%%%%%%%%%%%%%%%%%%%%
C_{u\theta}\,\gamma^+_{uu} F^+_x &=& - \tau^+_{xz}S^+_{_\Theta}
-C\Sb{SU}F^+_z\, S^+_{_U} \,,  \\%%
%%%%%%%%%%%%%%%%%%%%%%%%%%%%%%%%%
C_{u\theta}\,\gamma^+_{uu} F^+_z &=&
 - \tau^+_{zz}  S^+_{_\Theta}
 -2\,\frac{C\Sb{E\Theta\,} E^+_{\theta}  }{L^+}  \,,%%
\end{eqnarray}%%
with%%
\begin{equation}%%
\gamma^+_{uu} = c_{uu} {\sqrt{E^+\Sb K} \over \ell^+}\,, \quad \tau^+_{xz}  = F^+_z = -1 \ .%%
\end{equation}\end{subequations}

Outside of the viscous region, the problem has only one
characteristic length, the Monin-Obukhov scale  $ L$.
Correspondingly, one expects that the only dimensionless parameter
that governs the turbulent statistics in this region should be the
ratio of the outer scale of turbulence, $\ell$, to the Monin-Obukhov
length-scale $L$, which we denote as $ \ell^\ddag\=
 {\ell}/{L}=   \ell^+ / L^+  $.
Indeed, introducing ``$\ddag$-objects": %%
\begin{equation}\label{ddagb} %%
 \ell^\ddag\=
 {\ell}/{L}\,, \quad
 S_{_U}^\ddag\= S_{_U}^+\ell^+\,, \  S_{_\Theta}^\ddag\=
S_{_\Theta}^+\ell^+\,,  \end{equation} %%
one rewrites the balance Eqs.~\eq{eqs-plus}
as follows: %%
\begin{subequations} \label{MMA}%%
\begin{eqnarray} \label{9-eqAA}%%
&& \hskip -1.3 cm \tau^+_{xx} =  E^+\Sb K + {\ell^\ddag  } /2
c_{uu}
\sqrt{E^+\Sb K}\,, \quad \tau^+_{yy}  =  {E^+\Sb K}/2\,,\\
\label{9-eqBA}%%
&&  \hskip  -1.3 cm 2\, \tau^+_{zz}= {E^+\Sb K}  -{\ell^\ddag
 } / c_{uu} \sqrt{E^+\Sb K}\,,~~~~ \\ \label{9-eqCA}
 &&  \hskip  -1.3 cm c_{uu}  { E^+\Sb K}^{3/2} =  \ell^\ddag F^+_z -\tau^+_{xz}
S^\ddag_{_U} \,, \\ \label{9-eqDA}%%
\label{not-usedA}%%
&&  \hskip  -1.3 cm 4\,\~C\Sb{RI}\,c_{uu} \sqrt{E^+\Sb K}
\tau^+_{xz} =\ell^\ddag F^+_x -\tau^+_{zz}\,S^\ddag_{_U} \,, \\
\label{9-eqEA}%%
&&  \hskip  -1.3 cm  C_{\theta\theta}\,c_{uu} \sqrt{E^+\Sb K}
E^+_\theta = -F^+_z\,S^\ddag_{_\Theta}\,, \\ \label{9-eqFA}%%
&&  \hskip  -1.3 cm C_{u\theta}\,c_{uu} \sqrt{E^+\Sb K} F^+_x = -
\tau^+_{xz}S^\ddag_{_\Theta} -C\Sb{SU}F^+_z\, S^\ddag_{_U} \,,
\\ \label{9-eqGA}%%
&&  \hskip  -1.3 cm  C_{u\theta}\,c_{uu} \sqrt{E^+\Sb K} F^+_z = -
\tau^+_{zz}S^\ddag_{_\Theta} - 2\,C\Sb{E\Theta}\,\ell^\ddag
E^+_{\theta} \,,%%
\end{eqnarray}\end{subequations}%%

\subsection{\label{ss:poly}Polynomial~formulation~of~balance~Eqs.~\eq{MMA}}

In Sect. \ref{ss:Ri-infty} we justify the simple choice
\begin{equation} \label{CEtCtt} %%
C\Sb{E\Theta} \approx -2\,  C_{\theta\theta} \big/ 3\,,%%
\end{equation} %%
which is used presently.  With it the equations~~\eq{MM} [except of
(\ref{not-used})] yield expressions for $S_{_U}^\ddag$, $S\Sb
\Theta^\ddag$, $E_\theta^+$ and $F_x^+$ as functions of $\ell^\ddag$
and $X$:   %%
\begin{subequations} \label{sols} %%
\begin{eqnarray}\label{X}%%
X &\=&  \sqrt{E\Sb K^+(\ell^\ddag) } \,, \\ %%
\label{sols-a}%%
S^\ddag_{_U} (\ell^\ddag) &=& {\ell^\ddag F^+_z -c_{uu}
X^3}\over{\tau^+_{xz}}\,, \\ %%
\label{sols-b}%%
S^\ddag\Sb \Theta(\ell^\ddag) &=& -\frac{6\,c _{uu}^2
C_{u\theta}\, X^2 F^+_z}{3\,c_{uu}  X^3 +11\,\ell^\ddag F^+_z}
\,,\\ %%
\label{sols-d}%%%%
E^+_\theta (\ell^\ddag)&=& -\frac{S^\ddag\Sb
\Theta}{c_{uu}C_{\theta\theta}X}F^+_z,
\\ %%
\label{sols-e}%%%%
F^+_x (\ell^\ddag)&=& -\frac{\tau^+_{xz} S^\ddag\Sb \Theta
+C\Sb{SU}S^\ddag_{_U}F^+_z}{c_{uu}\,C_{u\theta}X}\ .
 \end{eqnarray}\end{subequations}%% %%

Substituting Eqs.~(\ref{sols})   into the remaining \Eq{not-used}
one gets an equation for $E\Sb K$ which can be represented as the
$9^\textrm{th}$ order  polynomial for $X\=\sqrt{E\Sb
K^+(\ell^\ddag)}$: %%
\begin{eqnarray}\nn  %%
&&-12\, c_{uu}^2 C_{u\theta}\,{\tau^+_{xz}}^2 \ell^\ddag F^+_z X^2
+\(3\,c_{uu}X^3 +11\,{\ell^\ddag}F^+_z\)~~~~~~~\\ \label{poly}
&&\times\Big\{8\,c_{uu}^2 \~C\Sb{RI} C_{u\theta}\,{\tau^+_{xz}}^2
X^2 -  \(c_{uu}X^3 -{\ell^\ddag}F^+_z\)\br &&\times \big[ \(2\,
C\Sb{SU}+ C_{u\theta}\)\ell^\ddag F^+_z +c_{uu} C_{u\theta}X^3
\big]\Big\} =0 \ . \end{eqnarray} %%
This equation can be solved numerically to find the function
$X(\ell^\ddag)$. After that Eqs.~\eq{sols} allow to find all the
required correlations as functions of $\ell^\ddag$, see red solid
lines in  \Fig{f:S}.

The following subsections are dedicated to finding of an
approximate analytical solutions for all correlations that will
describe their $\ell^\ddag$-dependence with reasonable accuracy.
First we will find in the next Sec.~\ref{ss:Ri0} the solution of
Eqs.~\eq{sols} and \eq{poly} for $\ell^\ddag=0$ , corrected up to
linear order in $\ell^\ddag$. Then, in Sec.~\ref{ss:Ri-infty} we
will find the asymptotic solution for $\ell^\ddag \to \infty$ with
 corrections, linear in the small parameter $\d =2(c_{uu}/\ell^\ddag)^{4/3}$.

\subsection{\label{ss:Ri0}Solution for neutral and weak stratification}

For neutral stratification, when  $\ell^\ddag=0$, \Eq{poly}
trivially yields: %%
\begin{subequations}\label{sol1} %%
\begin{equation}\label{sol1a} %%
X=X_0= (8\, \~C\Sb{RI})^{1/4} \,, \quad E\Sb K^+={E\Sb
{K,\scriptstyle{0}}^+} \=2\sqrt {2\,  \~ C\Sb{RI} } \ . %%
\end{equation} %%
With Eqs.~\eq{sols} this gives: %%
\begin{eqnarray}\label{sols-a-0}%%
S^\ddag\Sb{U,{\scriptstyle{0}}} &=& -{{c_{uu}
X^3_0}\over{\tau^+_{xz}}}\,,\\ %%
\label{sols-b-0}%%
S^\ddag\Sb {\Theta,{\scriptstyle{0}}} &=& -\frac{2\,c _{uu}
C_{u\theta}\,F^+_z}{X_0}\,,\\ %%
\label{sols-d-0}%%%%
E^+_{\theta,{\scriptstyle{0}}} &=&
-\frac{S^\ddag\Sb{\Theta,{\scriptstyle{0}}}F^+_z}{c_{uu}C_{\theta\theta}X_0}\,,
\\ %%
\label{sols-e-0}%%%%
F^+_{x,0} &=&  -\frac{\tau^+_{xz} S^\ddag\Sb
{\Theta,{\scriptstyle{0}}}
+C\Sb{SU}S^\ddag\Sb{U,{\scriptstyle{0}}}F^+_z}{c_{uu}\,C_{u\theta}X_0}\
.
 \end{eqnarray}\end{subequations}%%
 Remembering that $ S\Sb{U,\scriptstyle{0}} ^\ddag =
   1/ \kappa$ and using
experimental values in the lag-law region $E^+\Sb
{K,{\scriptstyle{0}}}\approx
 3.42$ and $\kappa\approx 0.436 $  one finds:
\begin{equation}\label{est} \~ C\Sb{RI}\approx 1.46\,,
\quad c_{uu}\approx 0.36\
. \end{equation}%%
 Using the relationship between $E\Sb {K,{\scriptstyle{0}}}^+$ and
$\kappa$ as suggested in L'vov~et~al.~(2006), $E\Sb
{K,{\scriptstyle{0}}}^+= 18\k^2$, one finds a relationship between
these two constants, %%
\begin{equation}\label{rel3} %%
4\sqrt 2 \, c_{uu}\~C \Sb{RI}=3\,, %%
\end{equation} %%
which agrees with the known values~\eq{est} with an accuracy that
is better then   1\%.

In the case of weak stratification
 one can develop a  perturbative approach to \Eq{poly}, using
  $ \ell^\ddag \ll 1$ as a small
parameter, to develop a solution in the form of  a Taylor series
in powers of $ \ell^\ddag $:%%
\begin{subequations}\label{exp}
  \begin{equation}\label{expX} X=X_0+X_1 {\ell^\ddag}+X_2
{\ell^\ddag}^2+\dots \end{equation} %%
For our purpose only the linear term in $\ell^\ddag$ is required:
\begin{equation} \label{expa} X_1= - F_z^+
\frac{2\,{\tau^+_{xz}}^2C_{u\theta}
+C\Sb{SU}X^4_0}{2\,c_{uu}C_{u\theta}X^6_0}  \,,
 \end{equation}\end{subequations}
where $X_0$ is given by \Eq{sol1a}. Using now the exact
relations~\eq{sols} with $E\Sb K^+=X^2$  one can immediately find
the linear terms in $\ell^\ddag$ in the expansions for $E\Sb
K^+(\ell^\ddag)$, $S_{_U}^\ddag$, $S\Sb \Theta^\ddag$,
$E_\theta^+$ and $F_x^+$:  %%

\begin{subequations}\label{exp1} \begin{eqnarray} %%
E\Sb K^+ &=&  E^+\Sb{K,{\scriptstyle{0}}}+
E^+\Sb{K,{\scriptstyle{1}}}\ell^\ddag
  +\dots \ ,\\
S_{_\Theta}^\ddag &=&  S^\ddag_{_{\Theta, {\scriptstyle{0}}}}+
S^\ddag_{_{\Theta,{\scriptstyle{1}}}} \ell^\ddag  +\dots \ , \quad
\mbox{etc.}\end{eqnarray}\end{subequations} %%
where%%
 \begin{subequations}\label{small-Ri} \begin{eqnarray}%%
E\Sb{K,{\scriptstyle{1}}} ^+ &=& -\frac{F^+_z}{c_{uu} E^{+\,1/2}
\Sb{K,{\scriptstyle{0}}}}\( 2\frac{{\tau^+_{xz}}^2}{E^{+\,2}
\Sb{K,{\scriptstyle{0}}}} +\frac{C\Sb{SU}}{C_{u\theta}}\),\\ %%
E_{\theta,{\scriptstyle{1}}} ^+ &=& \frac{2\,{F^+_z}^2} {3\,c_{uu}
E^{+\,5/2} \Sb{K,{\scriptstyle{0}}}}
\(6\frac{{\tau^+_{xz}}^2}{E^{+\,2} \Sb{K,{\scriptstyle{0}}}} -11
+3\frac{C\Sb{SU}}{C_{u\theta}}\),\\ %%
S^+\Sb{U,{\scriptstyle{1}}} &=& \frac{F^+_z}{\tau^+_{xz}}
\(1+3\frac{{\tau^+_{xz}}^2}{E^{+\,2}
\Sb{K,{\scriptstyle{0}}}} +\frac32 \frac{C\Sb{SU}}{C_{u\theta}}\)\,, \\
S^\ddag\Sb {\Theta, {\scriptstyle{1}}} &=&
-\frac{{2\,C_{u\theta}F^+_z}^2} {3\,E^{+\,2}
\Sb{K,{\scriptstyle{0}}}}
 \(3\,\frac{{\tau^+_{xz}}^2}{E^{+\,2}
 \Sb{K,{\scriptstyle{0}}}} -11
 +\frac32 \frac{C\Sb{SU}}{C_{u\theta}}\),~~~~~~~\\ %%
F_{x,{\scriptstyle{1}}}^+ &=& -\frac{\tau^+_{xz}
 S^\ddag\Sb {\Theta,{\scriptstyle{1}}}
 +C\Sb{SU} F^+_z S^\ddag\Sb{U,{\scriptstyle{1}}}}{c_{uu}
  C_{u\theta}\sqrt{E^{+} \Sb{K,{\scriptstyle{0}}}}} %%
\br && +\frac{\tau^+_{xz} S^\ddag\Sb {\Theta,{\scriptstyle{0}}}
+C\Sb{SU} F^+_z S^\ddag\Sb{U,{\scriptstyle{0}}}}{2\,c_{uu}
C_{u\theta}\sqrt{E^{+}\Sb{K,{\scriptstyle{0}}}}}
\frac{E^{+}\Sb{K,{\scriptstyle{1}}}}{E^{+}\Sb{K,{\scriptstyle{0}}}}\
. \end{eqnarray}
 \end{subequations}

 %%%%%%%%%%%%%%%%%%%%%%%%%%%%%%%%%%%%%%%%%%%%%%%%%%%%%%%%%%%%%%%%%%%%%%%%
\subsection{\label{ss:Ri-infty}Solution for  strong sratification}

To analyze Eqs.~\eq{eqs-plus} for large $ \ell^\ddag $, introduce
new variables   for the second order correlations $e\Sb K$,
$e_\theta$, and $f_x$, according to %%
\begin{eqnarray}\nn  %%
E^+\Sb K &=&   {\(e\Sb K{\ell}^\ddag
 \)}^{2/3}, \quad %%
E^+_\theta = \frac{e_\theta}{\sqrt[3]{e}\Sb K}\, {{\ell }^\ddag
 }^{2/3}= \frac{e_\theta \ell^\ddag}{\sqrt{E\Sb K^+}}\,,  \\
 \label{sol3}
F^+_x &=& \frac{f_x}{\sqrt[3]{e}\Sb K}\, {{\ell }^\ddag
 }^{2/3} = \frac{ f_x\ell ^\ddag}{\sqrt{E\Sb K^+}} \ . %%
\end{eqnarray}
In these variables \Eqs{eqs-plus} take  the  form:
\begin{subequations}\label{stress5}\begin{eqnarray} \label{stress5a}%%
 c _{uu}e\Sb K &=&  {L^+S\Sb U^+} -1  \,,  \\ %%
\label{stress5b}%%
4\,\d \,\~ C\Sb {RI} \,(c_{uu}\, e\Sb K )^{2/3} &=& \(c_{uu} e\Sb
K -1\) {L^+S\Sb U^+}  -2\,c_{uu}\,f_x, ~~~~~~~~~~ \\ %%
\label{stress5c} C_{\theta\theta}\,c _{uu}e _\theta &=& {L^+S\Sb
\Theta ^+} \,,
\\ \label{stress5d}%%
%%%%%%%%%%%%%%%%%%%%%%%%%%%%%%%%%
C_{u\theta}\,c_{uu} f_x &=&  L^+(S\Sb \Theta ^+ +C\Sb{SU}\, S\Sb
U^+ ) \,, \br%%
%%%%%%%%%%%%%%%%%%%%%%%%%%%%%%%%%
\d \,C_{u\theta}\,( c_{uu} \, e\Sb K )^{2/3} &=& %%
L^+S\Sb{\Theta}^+\(c_{uu} e\Sb K -1\)%%
\\ \label{stress5e} &&
+4\,C\Sb{E\Theta}\,c_{uu} e_{\theta} \,,%%
\end{eqnarray} \end{subequations}%%
with a small parameter $\d$
\begin{equation}\label{d} \d \=
  2\,
  \Big(\frac{ c_{uu}}{\ell^\ddag}\Big)^{4/3} \ , \end{equation}
  appearing in the LHS of the equations for the vertical momentum and thermal fluxes.
%%%%%%%%%%%%%%%%%%%%%%%%%%%%%%%%%%%%
\subsubsection{\label{ss:Ri-infty}Asymptotic solution for
 $ \ell^\ddag \to \infty$}

In the limit $ \ell^\ddag \to \infty$, one takes $\d=0$ in
\Eqs{stress5}  and gets an asymptotic solution (denoted by
 ``$^\infty$") %%
\begin{subequations}  \label{sol2} %%%%%%%%%%%%%%%%%%%%%
\begin{eqnarray} \label{sol2-1}%%
e^{\infty}\Sb K &=& \frac{1}{c_{uu}}\!\(1
 -4 \frac{C\Sb{E\Theta}}{C_{\theta\theta}}\), \\
\label{sol2a}%%%
S\Sb U^{+\infty} &=& \frac 2{L^+}\(1 -2\frac{C\Sb{E\Theta}}{C_{\theta\theta}}\), \\
  \label{sol2b}%%
S\Sb\Theta ^{+\infty} &=& - S\Sb U^{+\infty} \Big(C\Sb{SU}
+2C_{u\theta}\frac{C\Sb{E\Theta}}{C_{\theta\theta}}\Big), \\  %%
\label{sol2d} %%
e_\theta ^\infty &=& \frac{ L^+S\Sb\Theta
^{+\infty}}{c_{uu}C_{\theta\theta}} \,,  \\  %%
\label{sol2e}%%
f_x  ^\infty &=& -2\frac{C\Sb{E\Theta}}{C_{\theta\theta}}\frac{
L^+ S\Sb U^{+\infty}}{c_{uu}}\ . %%
\end{eqnarray} \end{subequations}%%
The physics of the problem requires  that  ${S\Sb U^+}$ $S_{_
\Theta}^+$, $e\Sb K$ and  $e_\theta$  be positively definite. This
constraints the possible values of the fitting constants. With
$C\Sb{E\Theta} < 0$, \Eq{CEtCtt}, we have:%%
\begin{equation}%%
c_{uu} > 0\,, \quad C_{\theta\theta} > 0\,, \quad  C_{u\theta} >
C\Sb{SU}\, C _{\theta\theta}/ 2|C_{_{E\theta}}|  \ .%%
\end{equation}

Notice, that   the maximal value of Ri$\sb {flux}$, [defined in
\Eq{numRig}],  which we denote as  Ri$\sb {f,\,max}$, corresponds to
$ \ell^\ddag \to \infty$ and equals to $1/L^+S\Sb U^{+\infty} $.
Taking for $C_{E\theta}=-2/3$ according to the
estimate~$C_{_{E\theta}}\approx -2/3$, and, following ZEKR-paper,
accept $C_{\theta\theta}=1$, one gets Ri$\sb {f,\,max}=3/14 \approx
0.21$ in agreement with the experimental value (which is between
0.20 and 0.25). With the choice%%
\begin{equation}\label{choice2} C_{E\theta}= -2/3\,,\quad  C_{\theta\theta}=1\ .
\end{equation}
 Equations \eq{sol2a} give the following predictions:%%
\begin{subequations} \label{sol5}  %%
\begin{eqnarray} \label{sol3-part}%%
e\Sb K ^\infty \!\! &=&\!\! \frac{11}{3\,c_{uu}}   \approx 10.1\,,
\ \Rightarrow\ E\Sb K^+\approx 4.7\, {\ld}^{2/3}\,,
\\ %%
 S\Sb U^{+ \infty}\!\! &=&\!\! \frac{14}{3\, L^+}\,,  \ \ \Rightarrow \ \ \max
\Rif
  = \frac3{14}  \approx0.21\,,~~~~~~~ \\ %%
S\Sb\Theta ^{+\infty}\!\! &=&\!\! -S\Sb U ^{+\infty}\(C\Sb{SU}
-\frac43
C_{u\theta}\) ,  \\
e_\theta ^\infty\!\! &=&\!\! L^+\! S\Sb\Theta
^{+\infty}/{c_{uu}}\,,  \\ %%
f_x  ^\infty \!\! &=&\!\! \frac{4}{3 } \frac{L^+ S\Sb U
^{+\infty}}{c_{uu}}
 \approx 17.2\,, \  \Rightarrow \ F_x^+\approx 6.7\, \ld^{2/3}.~~~~ %%
\end{eqnarray} %%
\end{subequations}%

\subsubsection{\label{ss:Ri-infty}Expansion around
 $ \ell^\ddag\to \infty $}

The small parameter in \Eqs{stress5}  allows one  to find the
solution as Taylor series in $\d\ll 1$:%%
\begin{subequations} \label{pert}  %%
\begin{equation}\label{perta} %%
{S\Sb \Theta ^+}= {S\Sb \Theta ^+}^{ \infty } +  S\Sb {\Theta,
1}^{ +\infty }\d + S\Sb {\Theta, 2} ^{ +\infty }\d^2+\dots\,, \
\mbox{etc.} %%
\end{equation} %%
For our purposes we need to know only the linear terms in $\d$:
\begin{eqnarray}\label{pertb}%%  %%   %%   %%   %%   %%  %%%  %%%   %%%
e\Sb{K,1} ^\infty &=& \frac{3}{11}C_{u\theta}\frac{e\Sb K
^\infty}{S^{+\infty}\Sb\Theta L^+}\,,\\ %%
S^{+\infty}_{_U,_1} &=& \frac{C_{u\theta}}{S^{+\infty}\Sb\Theta {L^+}^2}\,, \\
S^{+\infty}\Sb {\Theta,1} &=&
\frac{C_{u\theta}}{L^+}\(\frac{11\,C_{u\theta}
-3\,C\Sb{SU}}{{3\,S\Sb \Theta ^{+\infty}}L^+} -2\~C\Sb{RI}\), \\
e_{\theta,_1} ^\infty &=& \frac{S^{+\infty}\Sb
{\Theta,1}L^+}{c_{uu}}\,,\\ %%
f_{x,_1}^\infty &=& C_{u\theta}\frac{e\Sb K ^\infty}{S\Sb \Theta
^{+\infty}L^+} -2\frac{\~C\Sb{RI}}{c_{uu}}. \end{eqnarray}
 \end{subequations}
%%%%%%%%%%%%%%%%%%%%%%%%%%%%%%%%%%%%%%%%%%%%%%%%%%%%%%%%%%%%%%%%%%%%%%%%%%

\section{\label{ss:onClosure} On the closure problem of triple  correlations via second
order correlations}

Let us look more carefully at the approximation~\eq{freqs}, which is
\begin{eqnarray}\label{App-Appr}%%
\g_{uu}&=&c_{uu} \sqrt{E\Sb K} \big / \ell\,, \quad \g\Sb {RI}= C\Sb
{RI} \g_{uu}\,,  \br \~ \gamma\Sb {RI}& = &  \~ C\Sb {RI} \gamma\Sb
{RI}\,, \gamma_{\theta\theta}  = C_{\theta\theta} \g_{uu}\,, \quad
\gamma\Sb{RD}= C_{u\theta}\g_{uu}\ .%%
\end{eqnarray}%%
The dimensional reasoning that
leads to this approximation is questionable for problems having a
dimensionless parameter $\ld$.  Generally speaking, all
``constants" $c_{...}$ and $C_{...}$ in \Eq{App-Appr} can be any
functions of $\ld$.  Presently we just  hope that a possible $\ld$
dependence of these functions is relatively weak and does not
affect the qualitative picture of the phenomenon.

Moreover, even the assumption  \eq{distau1a} that the dissipation
of the thermal flux $\e_i$ is proportional to the thermal flux and
the assumption \eq{distau1b} that the dissipation of $E_\theta$,
$\ve\propto E_\theta$ are also questionable. Formally speaking,
one cannot guarantee that the tripple cross-correlator $ \<\theta
uu \>^+$ that estimates $\e^+$, can be (roughly speaking)
decomposed like $\<u\theta\> \sqrt {\<uu\>}$, i.e  really
proportional to $F=\<u\theta\>$ as it stated in   \Eq{distau1a}.
Theoretically, one cannot exclude the decomposition $\<\theta uu
\>\sim \<uu\> \sqrt {\<\theta\theta\>}$, i.e. a contribution to
$\e\propto E \Sb K$. Similarly, the dissipation $\ve$ in the
balance~\eq{corrc} of $E_\theta$, that is determined by the
correlator~\eq{distau1b}, is $\propto \<\theta\theta u \>$, as it
follows from the decomposition $\<\theta\theta u \>\sim
\<\theta\theta\> \sqrt {\<uu\>} $ and is stated in \Eq{distau1b}.
This correlator allows, for example, the decomposition
$\<\theta\theta u \>\sim \<\theta u\> \sqrt {\<\theta\theta\>} $,
i.e. contribution to $\ve\propto F$. This discussion demonstrates,
that the situation with the dissipation rates is not so simple, as
one may think and thus requires careful theoretical analysis that
is in our agenda for future work. Our preliminary analysis of this
problem shows that all fitting constants are indeed functions of
$\ld$. Fortunately, they vary within finite limits in the entire
interval $0\le \ld < \infty$. Therefore we propose that the
approximations used in this paper preserve the qualitative picture
of the phenomenon. Once again, the traditional down-gradient
approximation does not work even qualitatively because
corresponding ``constants" $C_\nu$ and $C_\chi$ vanish in the
limit $\ld\to\infty$.

%%%%%%%%%%%%%%%%%%%%%%%%%%%%%%%%%%%%%%%%%%%%%%%%%%%%%%%%%%%%%%%%%%%%%%%%%%%%%%%%%%

%

\end{document}